\documentclass[prd,eqsecnum,twocolumn,nofootinbib,superscriptaddress,floatfix]{revtex4-1}

\usepackage{amsfonts}
\usepackage{amsmath}
\newcommand{\sgn}{\operatorname{sgn}}
\usepackage{amssymb}
\usepackage{amsthm}
\usepackage{bm}
\usepackage{dcolumn}
\usepackage{epsfig}
\usepackage{graphicx}
\usepackage{graphics}
\usepackage[latin1]{inputenc}
\usepackage{latexsym}
\usepackage{rotating}
\usepackage{subfigure}
\usepackage{tensor}
\usepackage{hyperref}
\graphicspath{{pdf_figures/}}
\begin{document}

\title{Exciting black hole modes via misaligned coalescences:\texorpdfstring{\\ II. The mode content of late-time coalescence waveforms}{}}

\author{Halston Lim} \affiliation{Department of Physics and MIT Kavli Institute, Massachusetts Institute of Technology, Cambridge, MA 02139}

\author{Gaurav Khanna} \affiliation{Department of Physics, University of Massachusetts, Dartmouth, MA 02747}

\author{Anuj Apte} \affiliation{Department of Physics and MIT Kavli Institute, Massachusetts Institute of Technology, Cambridge, MA 02139}

\author{Scott A.\ Hughes} \affiliation{Department of Physics and MIT Kavli Institute, Massachusetts Institute of Technology, Cambridge, MA 02139}

\begin{abstract}
Using inspiral and plunge trajectories we construct with a generalized Ori-Thorne algorithm, we use a time-domain black hole perturbation theory code to compute the corresponding gravitational waves.  The last cycles of these waveforms are a superposition of Kerr quasinormal modes.  In this paper, we examine how the modes' excitations vary as a function of source parameters, such as the larger black hole's spin and the geometry of the smaller body's inspiral and plunge.  We find that the mixture of quasinormal modes that characterize the final gravitational waves from a coalescence is entirely determined by the spin $a$ of the larger black hole, an angle $I$ which characterizes the misalignment of the orbital plane from the black hole's spin axis, a second angle $\theta_{\rm fin}$ which describes the location at which the small body crosses the black hole's event horizon, and the direction $\sgn(\dot\theta_{\rm fin})$ of the body's final motion.  If these large-mass-ratio results hold at less extreme mass ratios, then measuring multiple ringdown modes of binary black hole coalescence gravitational waves may provide important information about the source's binary properties, such as the misalignment of the orbit's angular momentum with black hole spin.  This may be particularly useful for large mass binaries, for which the early inspiral waves are out of the detectors' most sensitive band.
\end{abstract}
\maketitle
\section{Introduction}
\label{sec:intro}

In a companion paper (Ref.\ {\cite{ah2019}}, hereafter Paper I), we introduced a model describing the worldline followed by a small body that orbits a Kerr black hole and is driven by gravitational-wave (GW) emission to inspiral until it encounters a dynamical instability and plunges into the hole's horizon.  Our model describes the transition from inspiral to plunge for orbits that are misaligned from the larger black hole's equatorial plane, generalizing earlier work by Ori and Thorne {\cite{ot00}} which did this for equatorial orbits.  Our goal now is to use this model in order to study the GWs produced by such misaligned plunges.

Our motivation for this study is to understand how a coalescing binary's final GW cycles depend on its orbital geometry as the binary enters its final plunge and merger.  Especially for binaries with total masses greater than $50\,M_\odot$ or so (which are amply represented in the sample that LIGO and Virgo have discovered {\cite{gwtc1}}), the early inspiral waves are emitted at low frequencies for which ground-based GW detectors have relatively poor sensitivity.  The inspiral richly encodes information about the system's masses and spins.  However, if it is not in the detector's band, we cannot measure these waves well, and we do not benefit from this rich encoding.  By contrast, for systems with $M \gtrsim 50\,M_\odot$, the late merger and final ringdown waves are generated at frequencies which are nearly ideal for ground-based detectors.  Especially as detectors' mid- and high-frequency noise is reduced in future upgrades {\cite{einsteintel, thirdgensens1, dwyeretal, thirdgensens2, rana, yulowfreq}}, we can expect these final merger cycles to be measured ever more precisely.  These final cycles will also be important components of the waves that are measured by the space-based detector LISA {\cite{lisa}}, which will measure processes involving black holes of tens of thousands to tens of millions of $M_\odot$.

Our particular goal here is to characterize how strongly different ringdown modes are excited as a function of the inspiral and plunge geometry, as well as on the spin of the binary's larger black hole.  Past work (e.g., Refs.\ {\cite{bhspectro1,bhspectro2, carullo2018}} for recent examples) has examined the measurement of multiple ringdown modes.  Such work has typically focused on the fact that, for a Kerr black hole, each mode's frequency and damping time depends on the final merged remnant hole's mass and spin in a unique way.  Measuring two such modes and assuming the Kerr spacetime thus suffices, at least in principle, to measure the merged remnant's mass and spin.  Measuring more than two modes makes it possible to test the Kerr hypothesis.

To date, not as much attention has been given to what can be learned by measuring the amplitudes of ringdown modes (though see Ref.\ {\cite{london2018}} for an important recent example to the contrary).  The relative excitation of different modes depends on the geometry of the system as it approaches its final state.  This geometry, in turn, depends upon the astrophysical history of the binary, with the spins and orbit expected to be substantially aligned for compact binaries that form {\it in situ} from stellar binaries, and with these angular momentum directions largely randomly oriented for binaries that form dynamically through multi-body interactions (e.g., in globular clusters); see Refs.\ {\cite{mandelfarmer2018,ligoastro2016,gerosaetal2018}} for recent discussion and review.

For a coalescence in which the orbital angular momentum is nearly parallel to the large black hole's spin, the $(\ell, m) = (2, 2)$ ringdown mode is likely to be the most strongly excited.  If the orbit is substantially misaligned from the larger hole's spin axis, then modes with $(\ell, m) = (2,2)$, $(2,1)$, or $(2,0)$ might be excited by roughly the same amount.  One must model {\it all} of the binary's GW modes and multipoles to completely assess what can be learned from their measurement {\cite{londonetal2018}}.  For the ringdown, a careful analysis is needed to understand how the excitation of these modes depends on the coalescence geometry.

Our goal is to begin developing such an analysis.  We use black hole perturbation theory to provide an easily parameterized framework for studying how different modes are excited by binary black hole coalescence.  Strictly speaking, our results thus only describe the limit in which one member of the binary is far more massive than the other.  We expect, however, that insight from this limit will carry over to coalescences with general mass ratio at least qualitatively, and perhaps even provide good quantitative understanding for mass ratios larger than some threshold.  Numerical relativity will be needed to explore late mode excitation by highly misaligned coalescences for general mass ratios.

We find that mode excitation varies in a predictable fashion as a function of certain parameters that describe the final coalescence geometry.  Given the spin $a$ of the larger black hole, we find the relative excitation of black hole modes is determined by two angles --- an angle $I$ describing the inclination of the binary's orbit relative to the black hole's equatorial plane, and an angle $\theta_{\rm fin}$ which describes the polar location at which the smaller body crosses the larger black hole's event horizon --- and its angular direction $\sgn(\dot\theta_{\rm fin})$ as it crosses.  (Because we work in perturbation theory and linearize all deviations from Kerr in the small body's mass $\mu$, the {\it absolute} excitation of ringdown modes also depends on this mass.)

This suggests that, if measurements by GW detectors can accurately measure multiple ringdown modes, it may be possible to use the relative amplitude of these modes to learn about the binary's spin-orbit misalignment.  This may be particularly valuable for ground-based measurements of high mass systems for which the inspiral waves (which encode spin-orbit misalignment through amplitude and frequency modulation of the waveform) are poorly measured.  In other words, it may be possible to get information about the system's spin-orbit misalignment from the late ringdown signal.  At the very least, the ringdown may provide information that complements spin-orbit constraints obtained from earlier in the waveform, improving our ability to make inferences about the nature of a measured binary black hole.

We begin with a synopsis of how we compute GWs in Sec.\ {\ref{sec:computinggws}}.  We review the results of Paper I in Sec.\ {\ref{sec:worldline}}, describing how we build the worldline which the smaller member of the binary follows as it inspirals and then plunges into the larger black hole.  As described in Sec.\ {\ref{sec:tdt}}, we then use this worldline to build the source term for the time-domain Teukolsky equation, which allows us to compute GWs produced by a small body following that worldline.  The last several to several dozen cycles of the waveforms that we compute consist of ringing modes of the larger black hole.  We characterize this mode content in Sec.\ {\ref{sec:modes}}.  We review the properties of Kerr black hole quasinormal modes in Sec.\ {\ref{sec:qnmoverview}}, discuss important issues regarding the angular bases used to described these modes and our data in Sec.\ {\ref{sec:spheroidvsphere}}, and then describe our algorithm for extracting the mode content of our waveforms in Sec.\ {\ref{sec:modeextract}}.  In most of the cases that we examine, the ringdown content we find is accurately described as entirely due to superpositions of the ``fundamental'' quasinormal black hole modes.  In some cases, we believe that we may be able to discern the presence of the first overtone mode.  Extracting this mode from our waveform data requires some care; our procedure for doing this extraction is described in Sec.\ {\ref{sec:overtone_results}}.

Section {\ref{sec:checks}} describes how we parameterize ringdown, a describes checks we instituted to make sure our results are reasonable.  We begin in Sec.\ {\ref{sec:parameterization}} by describing how the plunges which produce ringdown waveforms can be characterized using four parameters: the black hole spin $a$, an angle $I$ describing the inclination of the orbital plane, an angle $\theta_{\rm fin}$ that describes where the plunge terminates at the black hole, and the direction of the final plunge's angular motion, $\sgn(\dot\theta_{\rm fin})$.  As expected, we find that these parameters work well to describe the ringdown waveform.  We next discuss in Sec.\ {\ref{sec:modesymmetry}} certain symmetries which ringdown waveforms should inherit from the Kerr spacetime, and verify that our results respect these symmetries.  Finally, we describe a comparison to previous results for the equatorial case in Sec.\ {\ref{sec:equatorialcomp}}.  As we describe, we do not expect perfect agreement because of differing methodologies, but we find good agreement over the regime where agreement is expected.

We show results in Sec.\ {\ref{sec:results}}.  We begin with a catalog of modes with spheroidal index $\ell = 2$ in Sec.\ {\ref{sec:catalog}}; additional modes are shown in Appendix {\ref{app:catalog_more}}.  This catalog indeed demonstrates that the excitation of each mode depends cleanly and predictably on the geometry of the final plunge, suggesting that the inverse problem --- inferring the properties of the plunge geometry from a spectrum of measured modes --- may be feasible.  We describe in more detail interesting features we find in this catalog in Secs.\ {\ref{sec:universal}} and {\ref{sec:overtone_results}}, and discuss the impact of numerical errors in Appendix.\ {\ref{app:resolution_comparison}}.  In Sec.\ {\ref{sec:universal}}, we report interesting universality behavior that appears to emerge, at least at shallow inclination angle, as we examine mode excitation.  In brief, across a wide range of spins, excitation of the fundamental $(\ell, m)$ mode appears to follow a universal functional form that depends only on $\ell - m$ or $\ell + m$.  In Sec.\ {\ref{sec:overtone_results}}, we present results which suggest that the first overtone mode can be discerned in our results for rapid spin and shallow inclination angle, and provide further evidence in Appendix\ {\ref{app:overtone_test}}.

Our concluding discussion is given in Sec.\ {\ref{sec:conclude}}.  In addition to summarizing our findings, we suggest directions for future work that may improve our ability to model ringdown from misaligned binaries and to use these models to learn about coalescing black holes from their measured gravitational waves. A concise discussion of our key findings and their implications can be found in a companion Letter \cite{modesletter}.

\section{Gravitational waves from large-mass-ratio inspiral and plunge}
\label{sec:computinggws}

To model the GWs from a large-mass-ratio binary system, we first compute the worldline that a small body follows as it adiabatically evolves through a sequence of geodesic orbits, and then compute the transition to a plunging trajectory that carries it into the larger black hole.  We use this worldline to build the source function to the time-domain Teukolsky equation {\cite{td1,td2}}, solve this equation {\cite{td3,td4}} to compute the GWs that arise from the small body's motion on this worldline, and then characterize the quasinormal modes (QNMs) that terminate the waveform we find.  In this section, we briefly summarize the key steps in this analysis in order to set up how we characterize the QNM content of these GWs.

\subsection{Computing the inspiral, transition, and plunge worldline: A brief synopsis}
\label{sec:worldline}

Paper I discusses in detail how we construct the worldline describing how the small member of the binary inspirals and then plunges into its companion black hole.  Here we recap the main points of Paper I, emphasizing aspects of that analysis that are important for this paper:

\begin{enumerate}

\item[(1)] The small body begins on an inclined circular Kerr geodesic orbit, parameterized by initial radius $r_0$ and inclination angle $I$.  All orbit properties vary smoothly over the domain $I = 0^\circ$ (prograde equatorial) to $I = 180^\circ$ (retrograde equatorial).  The initial orbit is taken to adiabatically {\it inspiral} due to GW emission, shrinking in radius while keeping $I$ nearly constant {\cite{dh06}}.  Inspiral describes the system until the small body comes near the separatrix separating stable from unstable orbits.

\item[(2)] As the small body approaches the separatrix, inspiral accelerates until its inward motion is no longer adiabatic.  We compute the {\it transition} connecting inspiral to the final plunge by following the principles developed by Ori and Thorne {\cite{ot00}}, who showed how to do this for equatorial orbits.  Our ``generalized'' Ori-Thorne procedure lifts this restriction, allowing us to compute the transition for arbitrarily inclined circular orbits.

\item[(3)] At very late times, the small body's motion is well approximated by a plunging geodesic (with constant parameters) that crosses the event horizon.  In Boyer-Lindquist coordinates, this {\it plunge} terminates the small body's motion at some $\theta_{\rm fin}$ on the horizon.  This is an artifact of Boyer-Lindquist time, and reflects the fact that the event horizon is a surface of infinite redshift.  As we discuss in Sec.\ {\ref{sec:tdt}}, this behavior ensures that the time-domain Teukolsky equation goes over to its homogeneous form at late times, which in turn ensures that the final GW cycles we compute are QNMs, coherently joined in phase to the preceding waveform.

\end{enumerate}

Our worldline model requires us to make three {\it ad hoc} choices.  The first is how to define the end of ``inspiral'' and the beginning of ``transition.''  The physics governing the transition picks out a range of times for this moment.  We show in Paper I that the worldlines we develop vary very little over that range.  The second choice is how to model the evolution of the orbit's parameters $E$, $L_z$, and $Q$ during the transition.  In Paper I, we refine the Ori-Thorne procedure to eliminate unphysical discontinuities present in their model, but note that there are many ways to implement this refinement.  The two which we have investigated barely differ from one another.  We suspect this would be the case for any reasonable model.

The third {\it ad hoc} choice describes when ``transition'' ends and ``plunge'' begins.  As with the first choice, the physics of the transition picks out a range of times for this moment.  Disconcertingly, we find that the worldlines we develop depend on this parameter in a non-negligible way (see Sec.\ V\,B of Paper I, especially Fig.\ 7).  This could raise concerns that our conclusions will not be robust, but will depend upon how we make this choice.

Fortunately, we find that the {\it ringdown modes} we find are robust with respect to this choice, even though the plunge worldlines are not.  As we have outlined in the Introduction, the ringdown modes we find depend on the spin $a$ of the larger black hole, the inclination angle $I$, and an angle $\theta_{\rm fin}$ which defines where the smaller body plunges into the black hole.  As we vary the end of transition, the relation between $\theta_{\rm fin}$ and the worldline's initial conditions can change by quite a bit.  However, the {\it dependence of the ringdown waves on $\theta_{\rm fin}$ does not depend on this choice}.  As long as we parameterize our modes using the parameter set $\boldsymbol{(}a, I, \theta_{\rm fin}, \sgn(\dot{\theta}_{\rm fin})\boldsymbol{)}$, our conclusions are robust against how we choose to end the transition.  See Appendix {\ref{app:worldline_robust}} for detailed discussion and results.

Although it is a relief that our physical conclusions are not impacted by how we choose these parameters, it is a fundamental flaw of the generalized Ori-Thorne model that these {\it ad hoc} choices exist.  Work to improve this, or at least to better inform how these choices should be made, would be salubrious.

\subsection{Solving the time-domain Teukolsky equation}
\label{sec:tdt}

Following the procedure of Paper I summarized in the previous section, we make the worldline which the small body follows as it inspirals and plunges into the large black hole.  We then use this worldline to build the source term for the time-domain Teukolsky equation.  This equation describes scalar, vector, and tensor field perturbations to the spacetime of a rotating black hole.  In Boyer-Lindquist coordinates, it takes the form \cite{teuk}
\begin{eqnarray}
\label{teuk}
&&
-\left[\frac{(r^2 + a^2)^2 }{\Delta}-a^2\sin^2\theta\right]
        \partial_{tt}\Psi
-\frac{4 M a r}{\Delta}
        \partial_{t\phi}\Psi \nonumber \\
&&- 2s\left[r-\frac{M(r^2-a^2)}{\Delta}+ia\cos\theta\right]
        \partial_t\Psi\nonumber\\  
&&
+\,\Delta^{-s}\partial_r\left(\Delta^{s+1}\partial_r\Psi\right)
+\frac{1}{\sin\theta}\partial_\theta
\left(\sin\theta\,\partial_\theta\Psi\right)+\nonumber\\
&& \left[\frac{1}{\sin^2\theta}-\frac{a^2}{\Delta}\right] 
\partial_{\phi\phi}\Psi + 2s \left[\frac{a (r-M)}{\Delta} 
+ \frac{i \cos\theta}{\sin^2\theta}\right] \partial_\phi\Psi  \nonumber\\
&&- \left(s^2 \cot^2\theta - s \right) \Psi = -4\pi\left(r^2+a^2\cos^2\theta\right)T   ,
\end{eqnarray}
where $M$ is the mass of the black hole, $a$ is its angular momentum per unit mass, $\Delta = r^2 - 2 M r + a^2$, and $s$ is the ``spin weight'' of the field.  For $s = -2$, this equation describes the radiative degrees of freedom of the gravitational field, and is related to the Weyl curvature scalar as $\Psi = (r - ia\cos\theta)^4\psi_4$.   At future null infinity,
\begin{equation}
    \psi_4 = \frac{1}{2}\frac{d^2}{dt^2}\left(h_+ - ih_\times\right)\;.
\end{equation}
With this quantity in hand, $h_{+}$ and $h_{\times}$ can be easily computed by a double time-integration.

The source $T$ in Eq.\ (\ref{teuk}) is computed from the small body's energy-momentum tensor,
\begin{equation}
T_{\alpha\beta} 
= \mu\, \frac{u_\alpha u_\beta}{\Sigma\, u^t\,\sin\theta}\,
\delta\left[r - r(t)\right]\,
\delta\left[\theta - \theta(t)\right]\,
\delta\left[\phi - \phi(t)\right]\;,
\label{source_stress}
\end{equation}
where $\Sigma = r^2 + a^2\cos^2\theta$ and $u^\alpha$ denotes components of the small body's 4-velocity along its worldline.  To construct $T$, project $T_{\alpha\beta}$ onto certain legs of the Kinnersley tetrad, and then operate upon the resulting quantity with a second-order differential operator.  See Ref.\ \cite{teuk} for detailed discussion.

We solve Eq.\ (\ref{teuk}) in the time domain, dynamically providing information about the small body's worldline to make the source $T$ for our inspiraling and plunging body; Fig.\ {\ref{fig:examplewaveform}} shows a representative example of the waveform produced by this procedure.  Details of our approach have been extensively described in past literature~\cite{td1, td2, td3, td4}, so we do not repeat this discussion here, modulo one remark that is significant for the purposes of this current work.  Notice that the stress-energy tensor $T_{\alpha\beta}$ is inversely proportional to $u^t = dt/d\tau$.  This factor converts between time $\tau$ along the worldline and time $t$ as measured by a distant observer.  As the small body approaches the event horizon, $dt/d\tau \to \infty$.  The source term thus ``redshifts away,'' smoothly converting the Teukolsky equation to its homogeneous form and connecting the GWs from the small body's plunge phase to the Kerr black hole's QNMs very naturally.

\section{Characterizing a waveform's quasinormal mode content}
\label{sec:modes}

\begin{figure}
    \centering
    \includegraphics[width=.45\textwidth]{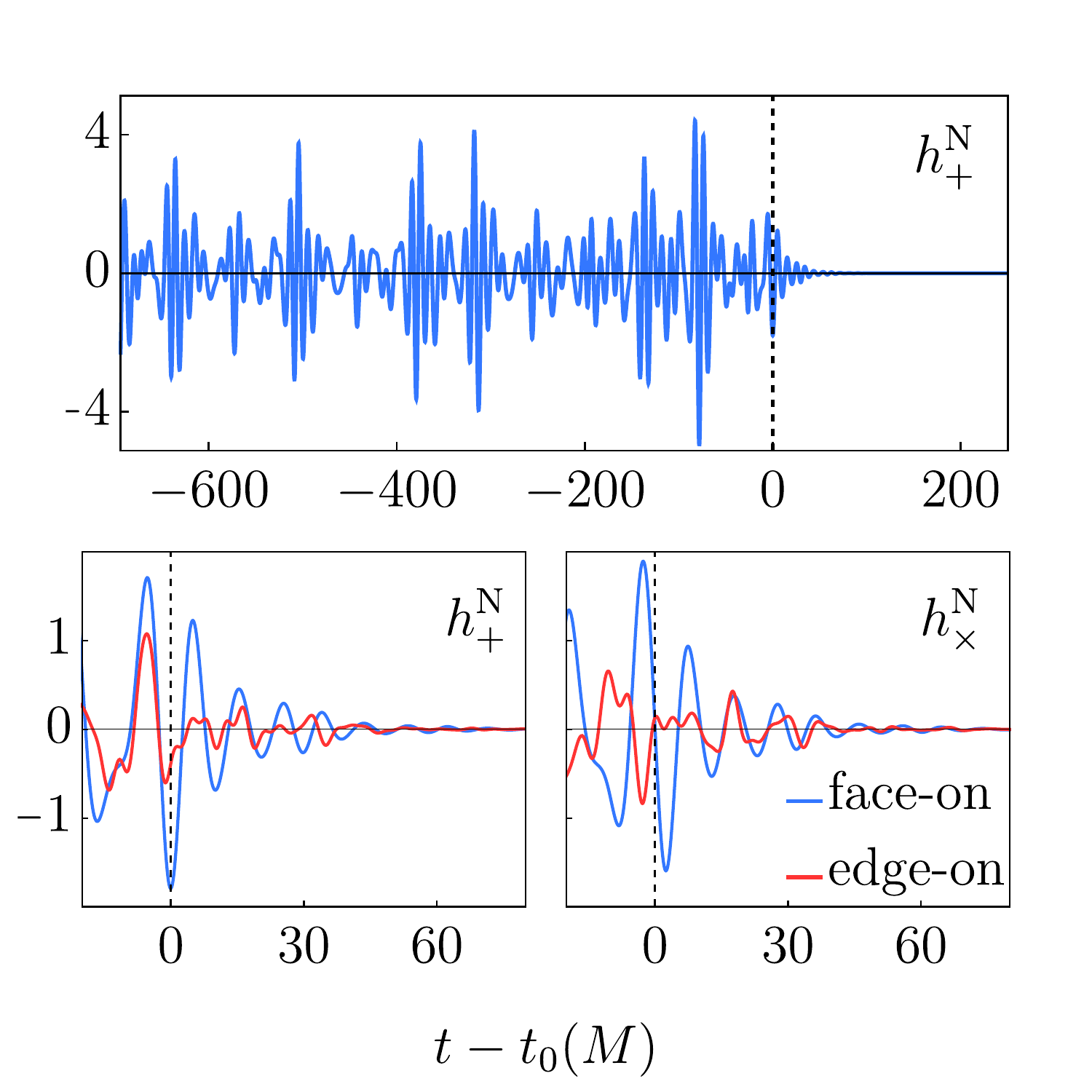}
    \caption{Gravitational waves from a small body on an inclined ($I=60^\circ$) orbit that plunges into a spinning ($a=0.9M$) black hole at a final polar angle of $\theta_{\rm fin} = 138.5^\circ$ and angular direction $ \dot{\theta}_{\rm fin} < 0$.  The small body reaches the radius of the prograde (equatorial) photon orbit at $t = t_0$.  Top panel shows the numerical waveform we find following the calculation described in Sec.\ \ref{sec:computinggws}.  Bottom shows the ringdown radiation produced by the small body plunging into the black hole.  In both panels, the waveform is normalized by a factor $\mu/D$, where $\mu$ is the small body's mass and $D$ is distance to the source. Notice that the nature of the final ringdown waves varies significantly depending on the relative orientation of the binary and the system.  The ``face-on'' waves (blue trace) show the waveform as measured by an observer who looks down the large black hole's spin axis ($\theta = 0^\circ)$; the ``edge-on'' waves (red trace) show the waves for an observer in the black hole's equatorial plane ($\theta = 90^\circ$).}
    \label{fig:examplewaveform}
\end{figure}

The analysis presented in Sec.\ {\ref{sec:computinggws}} yields the waveform produced by a small body that inspirals and plunges into a black hole; an example is shown in Fig.\ {\ref{fig:examplewaveform}}.  The final cycles of these waveforms can be modeled as a linear superposition of QNMs.  In this section, we describe how we characterize and extract the QNM content that describes the final cycles of inspiral and plunge waveforms like that shown in Fig.\ {\ref{fig:examplewaveform}}.

\subsection{quasinormal modes}
\label{sec:qnmoverview}

Teukolsky showed {\cite{teuk}} that Eq.\ (\ref{teuk}) separates after decomposing in the frequency domain, yielding ordinary differential equations which govern the $\theta$ and $r$ dependence of the field $\Psi$.  The solution for a given mode with (possibly complex) frequency $\sigma$ then becomes
\begin{equation}
    \Psi = \sum_{\ell = 2}^\infty\sum_{m = -\ell}^\ell\sum_{n = 0}^\infty {_s}{R}^{a\sigma}_{\ell m n}(r){_s}S^{a\sigma}_{\ell m n}(\theta,\phi)e^{-i\sigma t}\;.
    \label{eq:sep_teuk_soln}
\end{equation}
The functions ${_s}S^{a\sigma}_{\ell m n}(\theta,\phi)$ are known as spin-weighted spheroidal harmonics; when $a = 0$, they reduce to the spin-weighted {\it spherical} harmonics ${_{s}}Y_{\ell m}(\theta,\phi)$ {\cite{Goldberg67}}.  We discuss these functions and their properties in further detail below.  See Ref.\ {\cite{qnmreview}} for a discussion of the radial function ${_s}R^{a\sigma}_{\ell m n}(r)$, as well as for additional details regarding the spin-weighted spheroidal harmonics.

For certain frequencies, the modes (\ref{eq:sep_teuk_soln}) satisfy physical boundary conditions: they describe radiation that is purely ingoing on the black hole's event horizon, and purely outgoing at null infinity.  Such solutions are the black hole's QNMs.  Frequencies for which such solutions hold are written $\sigma_{\ell m n}$, each labeled by the mode's spheroidal harmonic indices plus an overtone index $n = 0, 1, \ldots$ {\cite{teuk2, qnmreview}}.  These frequencies are complex, and can be written
\begin{equation}
    \sigma_{\ell mn} = \omega_{\ell mn} - i/\tau_{\ell mn}\;.
\end{equation}
Given mode indices $(\ell,m,n)$ and assuming that the endstate of the merged system is a Kerr black hole, these frequencies depend only on the black hole's mass $M$ and spin parameter $a$.  Code for computing $\omega_{\ell mn}$ and $\tau_{\ell mn}$ for ringing modes of Kerr black holes, as well as tables describing the results, are provided by Berti {\cite{qnmreview,qnmlisa}}. 

The Teukolsky equation's symmetry properties dictate that for each mode $(\ell, m,n)$, the separated angular equation is actually satisfied by two eigenfunctions: $_sS^{a\sigma}_{\ell mn}(\theta,\phi)$ and $_sS^{a\sigma}_{\ell -mn}(\pi-\theta,\phi)^*$, where $^*$ denotes complex conjugation.  A given QNM is thus actually characterized by pairs of frequencies $\omega_{\ell mn}$ and $-\omega^*_{\ell -mn}$ {\cite{qnmlisa}}.  Specializing to spin weight $s = -2$ and bearing this symmetry in mind, we follow Ref.\ {\cite{qnmlisa}} [cf.\ their Sec.\ II\,A, especially Eq.\ (2.9) and nearby text] and write the gravitational waveform for a mode as
\begin{eqnarray}
h(t) &=& \sum_{kmn} \bigg[\mathcal{A}_{kmn}e^{-i[\sigma_{kmn}(t-t_0) - \phi_{kmn}]} {_{-2}}S^{a\sigma_{kmn}}_{kmn}(\theta,\phi)
\nonumber\\ 
&+& \mathcal{A}'_{kmn} e^{i[\sigma^*_{kmn}(t-t_0)+\phi'_{kmn}]}{_{-2}}S^{a\sigma_{kmn}}_{kmn}(\pi-\theta,\phi)^*\bigg]\;.
\nonumber\\
\label{eq:spheroidaldecomp}
\end{eqnarray}
(We explain why we have shifted the spheroidal index from $\ell \to k$ in the next section.)  Here ($\mathcal{A}_{kmn},\mathcal{A}'_{kmn})$ and $(\phi_{kmn},\phi'_{kmn})$ are the mode amplitude magnitudes and phases\footnote{Our notation, which associates the QNM amplitudes labeled by $(k,m)$ with $\mathcal{A}_{kmn},\mathcal{A}'_{kmn},\phi_{kmn},\phi'_{kmn}$, is aligned with Refs.~\cite{qnmlisa,tshift4}, but differs from Ref.~\cite{tbkh14}[cf.Eq.(5)]. See Sec.~\ref{sec:equatorialcomp} for discussion on converting between these different conventions.} and $t_0$ marks the time at which quasinormal ringing begins. Each mode's absolute excitation depends in addition on the small body's mass $\mu$ and the distance $D$ to the binary: ${\cal A}_{\ell mn} \propto \mu/D$.  In the results we present, we set this factor to 1; one can then scale the amplitudes by multiplying by $\mu/D$.

We will imagine a waveform that is ringdown dominated and thus is accurately described by 
Eq.\ (\ref{eq:spheroidaldecomp}) for $t \ge t_0$.  However, for perturbations sourced by a plunging body, it is not possible to absolutely determine the start of the ringdown {\cite{tshift1,tshift2}}.  Previous studies considering comparable mass binaries have associated $t_0$ with various phenomenological indicators such as the peak of the gravitational-wave amplitude \cite{tshift4,tshift3} or the peak orbital frequency \cite{tbkh14}.  Such associations are not well adapted to inclined, extreme mass ratio orbits since the radiation and orbital frequency do not always exhibit a clear peak.  In the analysis we describe below, we will take $t_0$ to be the time at which the small body reaches a radius equal to that of the prograde equatorial photon orbit (see Refs. \cite{lightring,lightring2} for further discussion).  Although the mode amplitudes $\mathcal{A}_{kmn},\mathcal{A}'_{kmn}$ and phases $\mathcal{\phi}_{kmn},\mathcal{\phi}'_{kmn}$ that we find then depend on $t_0$ (see Ref.~\cite{tshift1} for further discussion on the time-shift problem), our discussion of how the mode excitation depends on $\boldsymbol{(}I,a,\theta_{\rm fin},\sgn(\dot\theta_{\rm fin})\boldsymbol{)}$ will be independent of $t_0$. 

\subsection{Spherical and spheroidal expansions}
\label{sec:spheroidvsphere}

The numerical code which we use to solve Eq.\ (\ref{teuk}) {\cite{td1,td2,td3,td4}} decomposes the gravitational radiation as\begin{equation} \label{eq:sphericaldecomp}
\tensor*[]{h}{^{\rm N}}(t)=\sum_{\ell ,m}\tensor*[]{h}{_{\ell m}^{\rm N}}(t)\,{_{-2}}Y_{\ell m}(\theta,\phi)\;,
\end{equation}
where $h_{\ell m}^{\rm N}(t) = h_{\ell m,+}^{\rm N}(t) -i h_{\ell m,\times}^{\rm N}(t)$ is the $(\ell ,m)$ {\it spherical} multipole component. The superscript N emphasizes that each component is output from our numerical code.  Since we model the ringdown in the spheroidal basis, we must take into account spherical-spheroidal mode mixing \cite{BertiKlein, cz14, lf-j_mixing}.  This occurs because basis functions with the same degree $m$ overlap:
\begin{equation}
\label{eq:angdecomp} 
{_{-2}}S^{a\sigma_{\ell mn}}_{kmn}(\theta,\phi) = \sum_\ell\mu^*_{m\ell kn}(a\sigma_{\ell mn})\, {_{-2}}Y_{\ell m}(\theta,\phi)\;,
\end{equation}
where $k$ is the spheroidal harmonic order and $\ell$ is the spherical harmonic order.  In the Schwarzschild limit, $\mu_{m\ell k n}$ collapses to $\delta_{k\ell}$.  Our definition of the overlap coefficient coincides with that used in Ref.\ \cite{BertiKlein}, cf.\ their Eq.\ (5).  To avoid confusion, we use $k$ to label spheroidal harmonic decomposed QNMs, and label decompositions based on spherical harmonics with $\ell$.  Our code for computing the coefficients $\mu_{m\ell kn}$ is based on the algorithm described in Appendix A of Ref.\ {\cite{hughes2000}} (see also Ref.~{\cite{hughes2000err}}).  It has been validated to numerical precision using comparison data kindly provided by Cook, and agrees with Berti's online data tables (modulo an unimportant overall phase).

Equating the left-hand side of Eq.\ (\ref{eq:sphericaldecomp}) to the left-hand side of (\ref{eq:spheroidaldecomp}), multiplying both sides by $_{-2}Y^*_{\ell m}(\theta,\phi)$, integrating over the sphere and using Eq.\ (\ref{eq:angdecomp}), we find
\begin{equation} \label{eq:multipole}
\tensor*[]{h}{_{\ell m}^{\rm N}}(t) = \!\!\sum_{k=k_{\rm min}}^{\infty}\sum_{n = 0}^\infty\left[a_{m\ell kn}(t)\, \mathcal{C}_{kmn} + a_{-m\ell kn}'(t)\, \mathcal{C}'_{k-mn}\right]\;,
\end{equation}
where $k_{\rm min} = {\rm max}(2, |m|)$.  In practice, the sums are truncated at some finite maximum index; we discuss this truncation in detail in the next section.  The mode amplitudes and phases have been absorbed here into complex amplitudes,
\begin{equation} 
\mathcal{C}_{kmn} \equiv \mathcal{A}_{kmn} e^{i \phi_{kmn}}\;,\quad
\mathcal{C}'_{kmn} \equiv \mathcal{A}'_{kmn} e^{i\phi'_{kmn}}\;,
\end{equation} 
and the time dependent coefficients are given by
\begin{eqnarray}
a_{m\ell kn}(t) &=& \mu^*_{m\ell kn}(a\sigma_{kmn}) e^{-i\sigma_{kmn}(t-t_0)}\;,
\nonumber\\
a'_{m\ell kn}(t) &=& (-1)^\ell \mu_{m\ell kn}(a\sigma_{kmn}) e^{i\sigma^*_{kmn} (t-t_0)}\;.
\label{eq:as}
\end{eqnarray}
To derive these results, we have used the fact that
\begin{equation}
    _{-2}Y_{\ell m}(\pi-\theta,\phi)^* = (-1)^\ell\,_{-2}Y_{\ell-m}(\theta,\phi)\;.
    \label{eq:spherical_reflection}
\end{equation}
It is important to note that the spherical multipole $h^{\rm N}_{\ell m}$ contains, in general, contributions from spheroidal modes of both $m$ and $-m$.

\subsection{Mode extraction}
\label{sec:modeextract}
\begin{figure*}[ht]
    \centering
    \includegraphics[width=.32\textwidth]{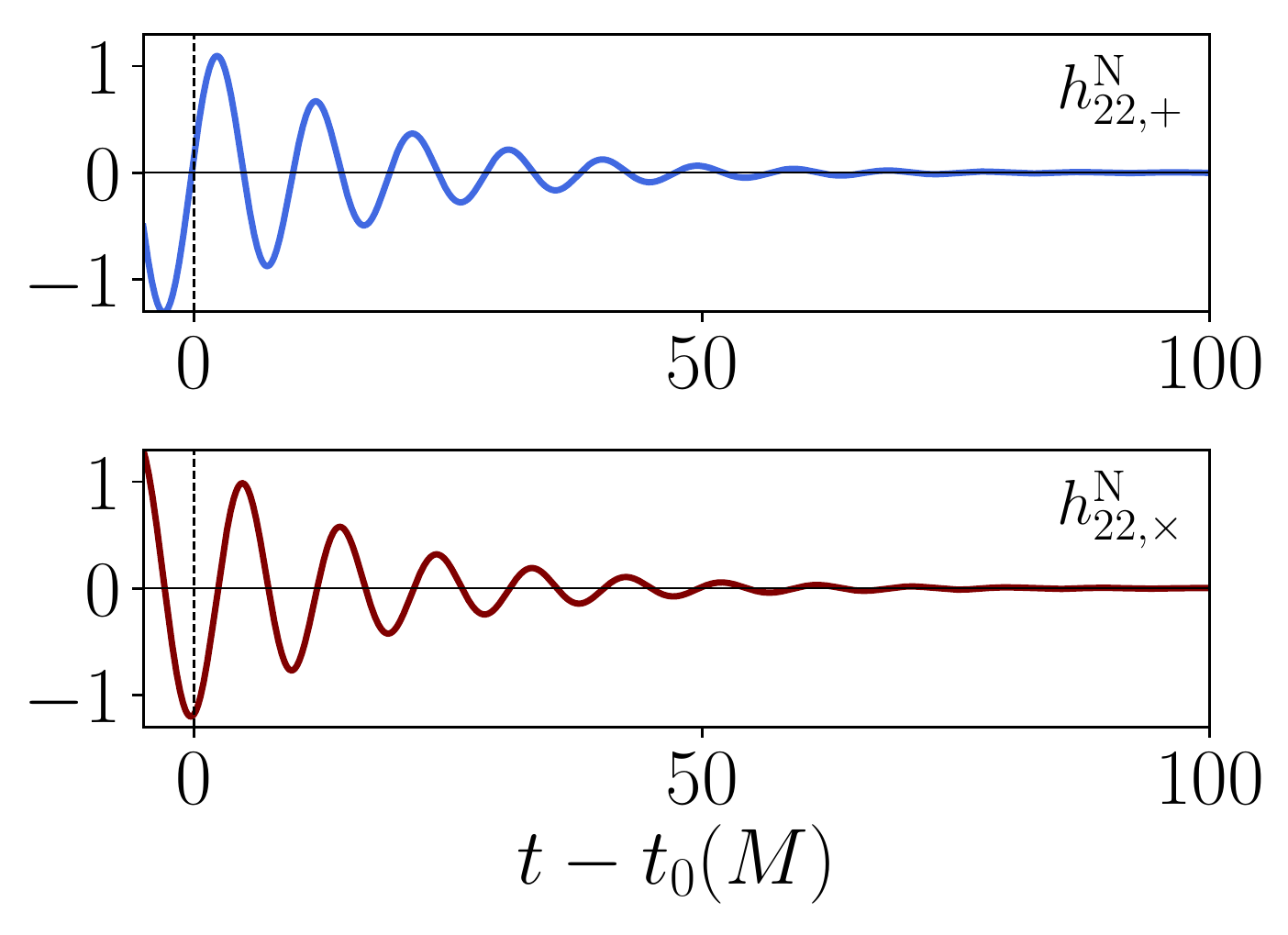}
    \includegraphics[width=.31\textwidth]{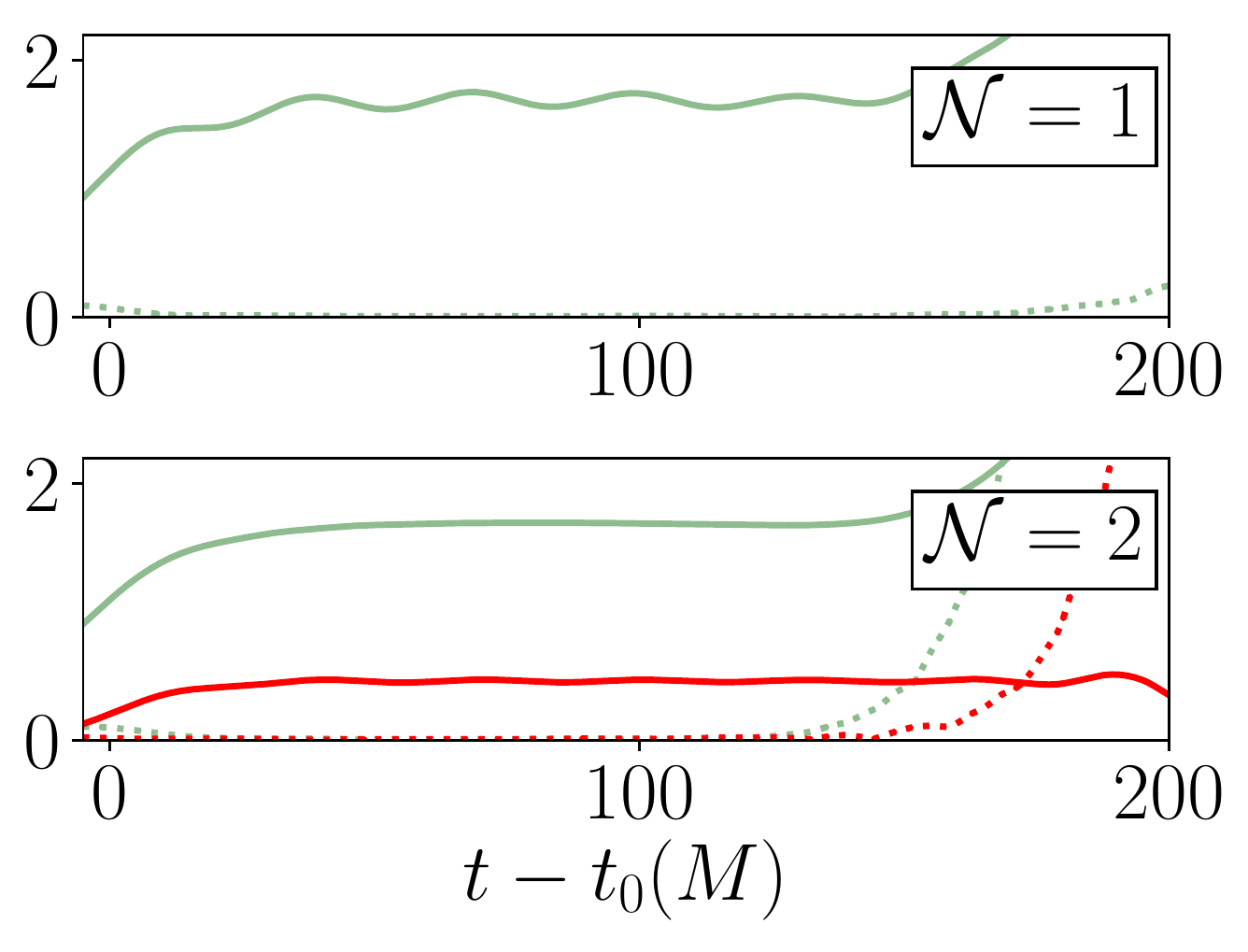}
    \includegraphics[width=.32\textwidth]{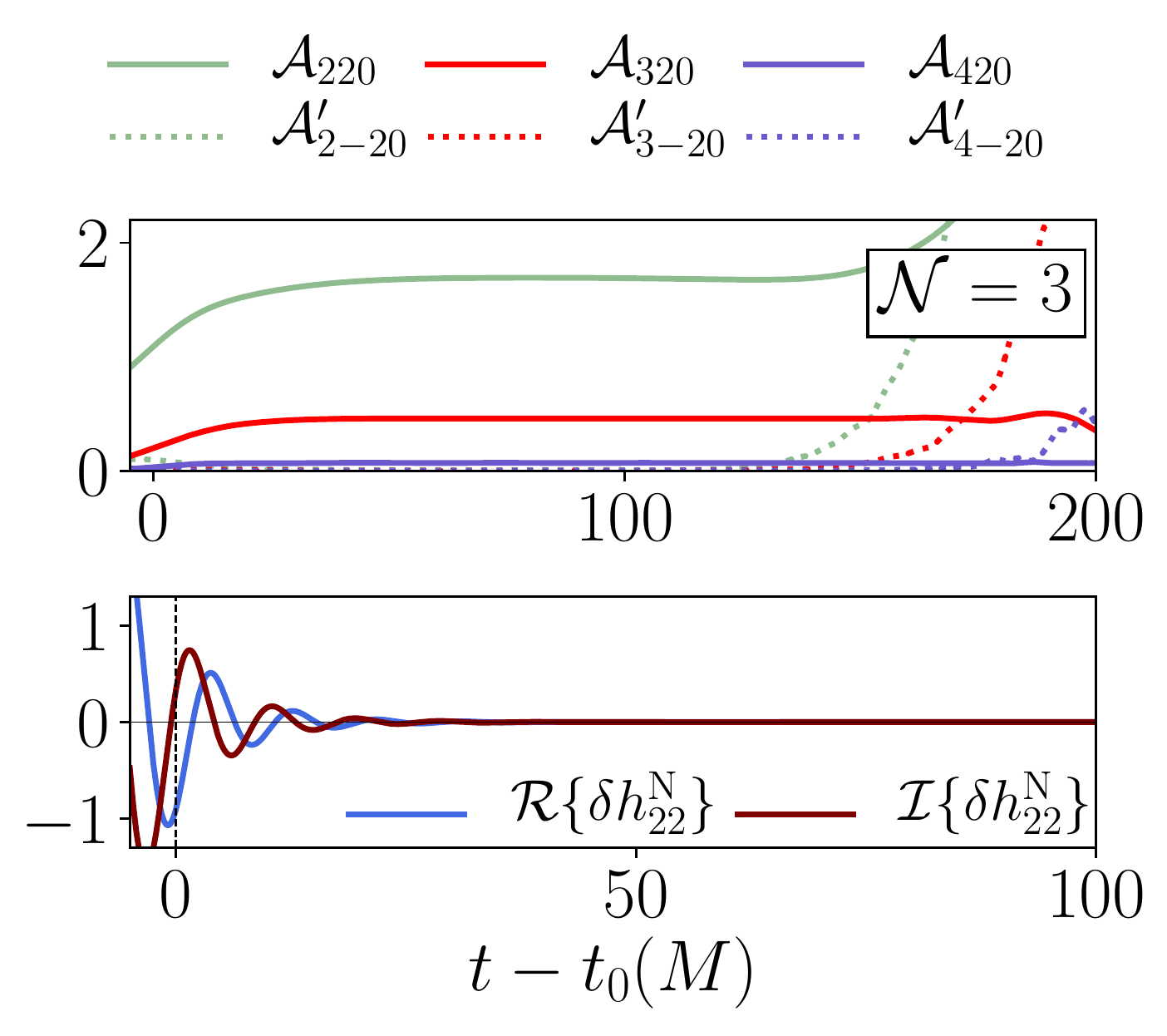}
    \caption{Extraction of mode amplitudes, following the algorithm described in Sec.\ {\ref{sec:modeextract}}.  We show results for a small body on an inclined ($I=20^\circ$) orbit that plunges into a black hole with $a = 0.9M$, crossing the horizon at polar angle $\theta_{\rm fin} = 72.1^\circ$ and with $\dot{\theta}_{\rm fin} < 0$.  The two left panels show the numerical mode amplitudes for $m = 2$ versus time [see Eq.\ (\ref{eq:multipole})].  The two middle panels and the top-right panel illustrate how our method of fitting to the ringdown converges as more spheroidal multipoles are included in the fit.  In the top middle, we consider fits using only $k = 2$ modes, $\mathcal{N}=1$.  The extracted amplitudes (green curves) never settle to a constant, indicating that our model has not captured the waves' full mode content.  The extracted amplitudes are closer to constants when we fit both $k = 2$ and $k = 3$ modes (bottom middle; green curves show $k = 2$ amplitudes, red show $k = 3$), $\mathcal{N}=2$, though we still find oscillations especially for $k = 3$.  We finally find good behavior when we fit $k = 2$, $k = 3$, and $k = 4$ (violet curves showing the $k = 4$ amplitudes), $\mathcal{N}=3$.  The extracted amplitudes stabilize to constants in the interval $25M \lesssim t - t_0 \lesssim 120M$.  At earlier times, the numerical waveform is not yet ringdown dominated, and at later times, the wave has decayed away, and the extraction is noise dominated.  The bottom right panel shows the residuals $\delta h^{\rm N}_{22} = h^{\rm N}_{22} - h^{\rm RD0}_{22}$ of the $\mathcal{N}=3$ ringdown model $h^{\rm RD0}_{22}$.}
%
    \label{fig:modeextraction}
\end{figure*}
Our work to characterize mode amplitudes can be considered an extension of Taracchini \textit{et al.} \cite{tbkh14}, whose analysis was limited to equatorial trajectories ($I=0^\circ$ and $180^\circ$). Initially, we extended their algorithm to include higher-order modes. However, with this algorithm the calculated mode amplitudes, especially sub-dominant modes, are rather sensitive to the choice of fitting interval. This makes it essentially impossible to draw robust conclusions about the waveform's mode content. Other studies in the literature that rely on fits to damped sinusoids in either the time or frequency domain have dealt with this ambiguity by choosing the fitting interval with lowest error (\cite{tshift1,kellybaker,lsh_overtone,tshift4}). This choice leads to discontinuities in the mode amplitudes as a function of $\theta_{\rm fin}$, which is a general symptom of the fact that the fit error (as a function of chose fit interval) suffers from multiple local minima. Oscillations in fit error, similar to those discussed in Ref.~\cite{othertshift}, lead to jumps in the calculated mode amplitude as $\theta_{\rm fin}$ is incremented.

Our aim also differs from that of numerical relativity analyses. Since we are working within black hole perturbation theory, we know the BH spin and mass \textit{a priori} and can assume that the ringdown comprises those Kerr QNM frequencies. In addition, numerical relativity analyses have primarily focused on nonprecessing comaprable mass systems, where the effect of higher-order mode mixing [specifically beyond the (2,2) and (3,2) QNMs] can be neglected. It remains to be determined whether the same trend holds for highly precessing systems with larger mass ratios. Instead of neglecting mode mixing, our algorithm specifically addresses how to disentangle mode mixing by factoring in the higher-order spherical waveform modes, for example in utilizing $\psi^{\rm N}_{2 2},\psi^{\rm N}_{3 2}$ and $\psi^{\rm N}_{4 2}$ simultaneously. An exception is Ref.~\cite{lsh_overtone}, which adopts an algorithm to measure higher-order modes and overtones. However, this algorithm is based on a fit to a frequency domain waveform, which leads to mode amplitudes that still depend on the choice of fitting interval.

Before describing our mode extraction algorithm in detail, we introduce some simplifications to the general framework we have described.  First, we will typically neglect overtone modes, $n\ge 1$.  The damping times $\tau_{kmn}$ tend to decrease fairly rapidly with $n$ {\cite{Leaver85}}.  The fundamental mode $n = 0$ thus dominates the ringdown, especially for ``late'' times $t - t_0 \gtrsim 25M$.  We thus mostly focus on $n = 0$, and drop the sum over $n$ (though see the following section for a discussion about including the first overtone).

Second, for the situations that we consider, the coefficients $\mu_{m\ell kn}$ tend to peak at $k = \ell$, falling off rapidly in magnitude away from this peak.  As such, we truncate the sum over $k$ at $k_{\rm max} = \ell + K_\ell$, where $K_\ell$ is found by determining how many spheroidal modes must be included to accurately fit the spherical mode numerical data $h^{\rm N}_{\ell m}(t)$; we elaborate on this below.  Equation (\ref{eq:multipole}) becomes
\begin{equation} \label{eq:multipole2}
\tensor*[]{h}{_{\ell m}^{\rm N}}(t) =\sum_{k=k_{\rm min}}^{\ell + K_\ell}\left[a_{m\ell k0}(t)\, \mathcal{C}_{km0} + a_{-m\ell k0}'(t)\, \mathcal{C}'_{k-m0}\right]\;.
\end{equation}
With this framework in hand, our goal now is to calculate the mode amplitudes $\mathcal{C}_{km0}$ and $\mathcal{C'}_{km0}$ given $\tensor*[]{h}{_{\ell m}^{\rm N}}(t)$, for which we have developed the following algorithm:
\begin{itemize}

    \item Consider a set of ${\cal N}$ spherical multipoles $(\ell_i,m)$.  Choose $k_{\rm min}=\ell_1$ where $\ell_1 = {\rm max}(2,|m|)$, $\ell_2 = \ell_1 + 1$, $\ldots$, $\ell_{\cal N} = \ell_1 + {\cal N} - 1$.  Choose the factor $K_\ell$ appearing in Eq.\ (\ref{eq:multipole2}) such that $K_{\ell_1} = {\cal N} - 1$, $K_{\ell_2} = {\cal N} - 2$, $\ldots$, $K_{\ell_{\cal N}} = 0$.
    
    \item Consider a moment $t = t_j$; evaluate Eq.\ (\ref{eq:multipole2}) and its time derivative at this moment.  This yields $2{\cal N}$ linear equations which can be inverted for the $2{\cal N}$ unknown mode amplitudes ${\cal C}_{km0}$ and ${\cal C}'_{k-m0}$.
    
    \item Check consistency of the algorithm by computing the mode amplitudes at multiple times.  If the ringdown model is consistent with the data, then we find that each ${\cal C}_{km0}$ and ${\cal C}'_{k-m0}$ settles down to a constant during the time period in which the waveform is QNM dominated.  If these amplitudes do not stabilize to a constant, then the model is not adequate.  It may be that the model needs to include more spheroidal modes (in which case we increase ${\cal N}$ and repeat the algorithm), or that the radiation is simply not QNM dominated.
    
    \item When the model is consistent with data, we denote by an overbar the values to which the amplitudes settle down: $\bar{\mathcal{C}}_{km0}$, $\bar{\mathcal{C'}}_{k-m0}$.  In practice, we determine $\bar{\mathcal{C}}_{km0}$ and $\bar{\mathcal{C'}}_{k-m0}$ by calculating a moving average with fixed size $\Delta t$.  We then associate $\bar{\mathcal{C}}_{km0}$ and $\bar{\mathcal{C'}}_{k-m0}$ with the average in the interval with the least variance.  The ringdown model becomes 
\begin{equation}\label{eq:fundamentalmodel}
    \tensor*[]{h}{_{\ell m}^{\rm RD0}}(t) =\sum_{k=k_{\rm min}}^{\ell + K_\ell} \left[a_{m\ell k0}(t)\, \bar{\mathcal{C}}_{km0} + a_{-m\ell k0}'(t)\, \bar{\mathcal{C}'}_{k-m0}\right]\;.
\end{equation}
    The ``0'' in the superscript on the right-hand side of Eq.\ (\ref{eq:fundamentalmodel}) labels the fact that this model is based on the $n = 0$ fundamental ringdown modes.

\end{itemize}

We illustrate this algorithm with an example.  The left panel of Fig.\ \ref{fig:modeextraction} shows the final several numerical GW cycles $h^{\rm N}_{22}$ arising from a small body inspiraling and plunging into a black hole with $a = 0.9M$.  The orbit is inclined at $I = 20^\circ$, and crosses the horizon at a final polar angle $\theta_{\rm fin} = 72.1^\circ$ with angular velocity $\dot \theta_{\rm fin} < 0$.  We first try to fit these waves using a model with ${\cal N} = 1$ (i.e., using only the $\ell = 2$ modes).  The middle top panel of Fig.\ \ref{fig:modeextraction} shows the coefficients ${\cal A}_{220}$ and ${\cal A}'_{2-20}$ we find in this case.  We find {\it no} span of time at which ${\cal A}_{220}$ and ${\cal A}'_{2-20}$ settle down to constants.  The choice ${\cal N} = 1$ poorly describes these data.

Consider next fits using a model with ${\cal N} = 2$ (i.e., now using the $\ell = 2$ and $\ell = 3$ modes).  As the middle bottom panel of Fig.\ \ref{fig:modeextraction} shows, the fit is improved, but we still see oscillations in the extracted amplitudes.  This indicates that there is still room for improvement in this model.  Finally, the top right-panel shows the fit for ${\cal N} = 3$ (now including modes $\ell = 2$, $\ell = 3$, and $\ell = 4$).  Here at last we find that the amplitudes have settled down to a nearly constant level, at least over the time interval $25M \lesssim t - t_0 \lesssim 120M$.  At earlier times, the signal is not yet QNM dominated; at later times, the modes have decayed away, and our fit becomes noise dominated.  The residual $\delta h^{\rm N}_{22} \equiv h^{\rm N}_{22} - h^{\rm RD0}_{22}$ shown in the lower-right panel illustrates that the fit describes the numerical data well over this time interval.

\subsection{Overtones}
\label{sec:OvertoneMethod}

The ringdown model $\tensor*[]{h}{_{\ell m}^{\rm RD0}}(t)$ only includes fundamental QNMs.  As we will show in Sec.\ {\ref{sec:checks}}, this model is consistent with data during the late ringdown.  However, in the early ringdown, overtone modes (which tend to decay much more quickly than the fundamental) may be present, and a ``fundamentals-only'' model will not capture their contributions to the radiation that arises from the small body's final plunge.  For larger spins (for which the rapid decay of the overtones is not so rapid), it may be feasible to isolate their contribution to the early ringdown.  Past work has similarly proposed techniques to isolate overtones from numerical relativity models of binary black hole coalescence {\cite{lsh_overtone}}, where the initial ``perturbation'' is very large and overtone excitation is important.

Suppose we have computed a fundamental-only ringdown model $h^{\rm RD0}_{\ell m}(t)$.  Define the residuals from this model as
\begin{equation}\label{eq:residual}
    \delta \tensor*[]{h}{_{\ell m}^{\rm N}}(t) = \tensor*[]{h}{_{\ell m}^{\rm N}}(t) - \tensor*[]{h}{_{\ell m}^{\rm RD0}}(t)\;.
\end{equation}
In principle, we could repeat the algorithm described in Sec.\ {\ref{sec:modeextract}}, but using $\delta h^{\rm N}_{\ell m}$ rather than $h^{\rm N}_{\ell m}$, and fitting with modes which have $n = 1$.  In practice, because the overtones are quite short-lived, we do not find a time domain over which their amplitudes settle down to a constant level.  In Sec.\ {\ref{sec:overtone_results}}, we describe a modification to our mode extraction algorithm which we use to account for this difficulty and to estimate overtone amplitudes.

One could imagine iterating further, yielding ever higher-order fits to the ringdown overtones.  In practice, we expect that this method will be greatly limited by numerical accuracy, and that it is likely to be quite challenging to find all but perhaps the $n = 1$ overtones.  Even in that case, overtones are most likely to be discernible only if the black hole's spin is quite rapid (so that the overtone's decay is relatively slow).  Evidence that we may be finding the first overtone with $a = 0.99M$ is presented in Sec.\ {\ref{sec:overtone_results}}.

\section{Parameterization, checks, and comparisons with past work}
\label{sec:checks}

Before discussing the results we find, we describe how we parameterize our data, examine symmetries that our results should respect, and check that our results behave as expected in the Schwarzschild limit, where spherical symmetry implies certain relations among the different amplitudes.  We also verify that our results agree with past work in the equatorial limit.  We then explore the inclined Kerr case in Sec.\ {\ref{sec:results}}.

\subsection{A clean and complete parameterization}
\label{sec:parameterization}

The ringing cycles that we wish to study are sourced by the final moments of the small body's worldline.  To describe those waves, we need a parameterization that completely characterizes the small body's final motion on its worldline as it plunges into the black hole. 

Begin by considering the circular and inclined geodesic orbits on which the small body initially moves.  Circular geodesics of Kerr black holes are generally described using the orbit's radius $r$ and some angle describing the orbit's tilt from the equatorial plane.  We use the angle $I$, defined (in radians) by
\begin{equation}
    I = \pi/2 - [\sgn(L_z)]\theta_m\;,
\end{equation}
where $\theta_m$ is the minimum value of the Boyer-Lindquist angle $\theta$ reached on an orbit.  $I$ smoothly varies from $0^\circ$ for prograde equatorial orbits to $180^\circ$ for retrograde equatorial orbits.  For Schwarzschild, $I$ corresponds exactly to the angle at which the orbit is inclined from the equatorial plane.  Although not quite amenable to such a simple interpretation for general spin, it provides a useful notion of orbit tilt for Kerr as well.
\begin{figure*}[ht]
    \centering
    \includegraphics[width=.45\textwidth]{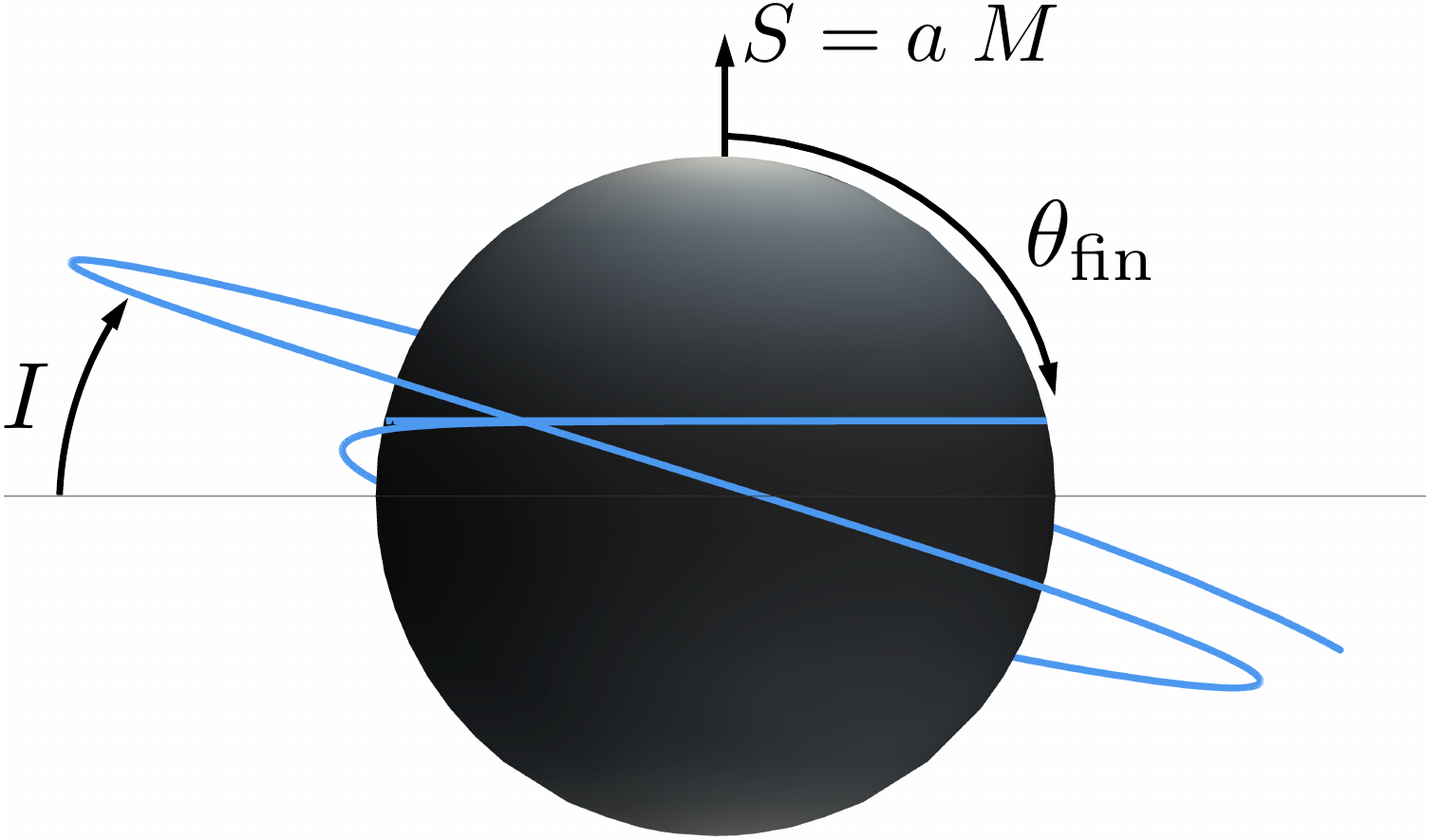}
    \includegraphics[width=.45\textwidth]{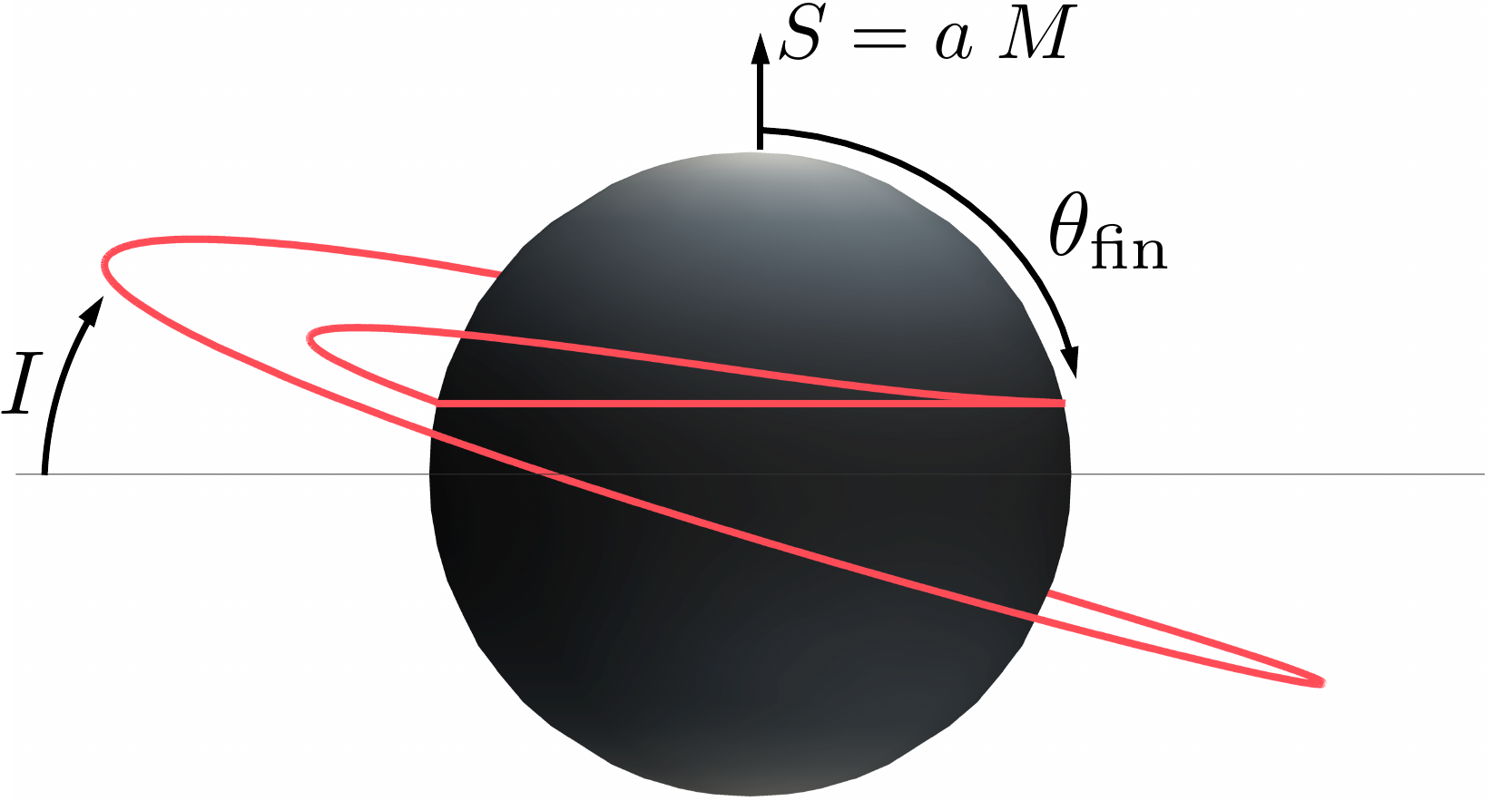}
    \includegraphics[width = 0.38\textwidth]{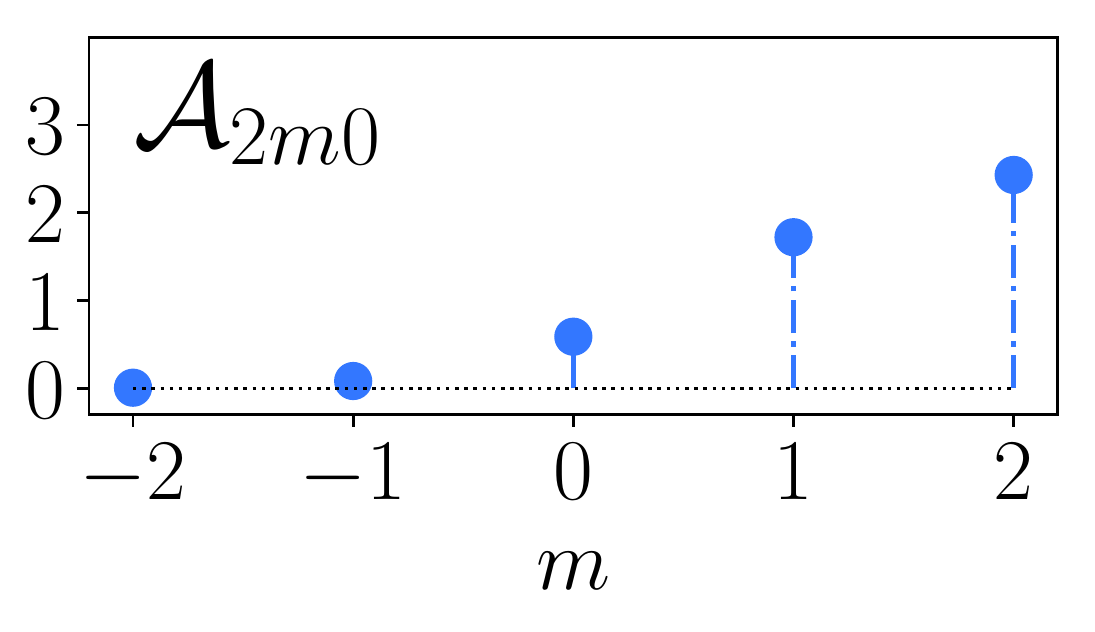}\qquad\qquad
    \includegraphics[width = 0.38\textwidth]{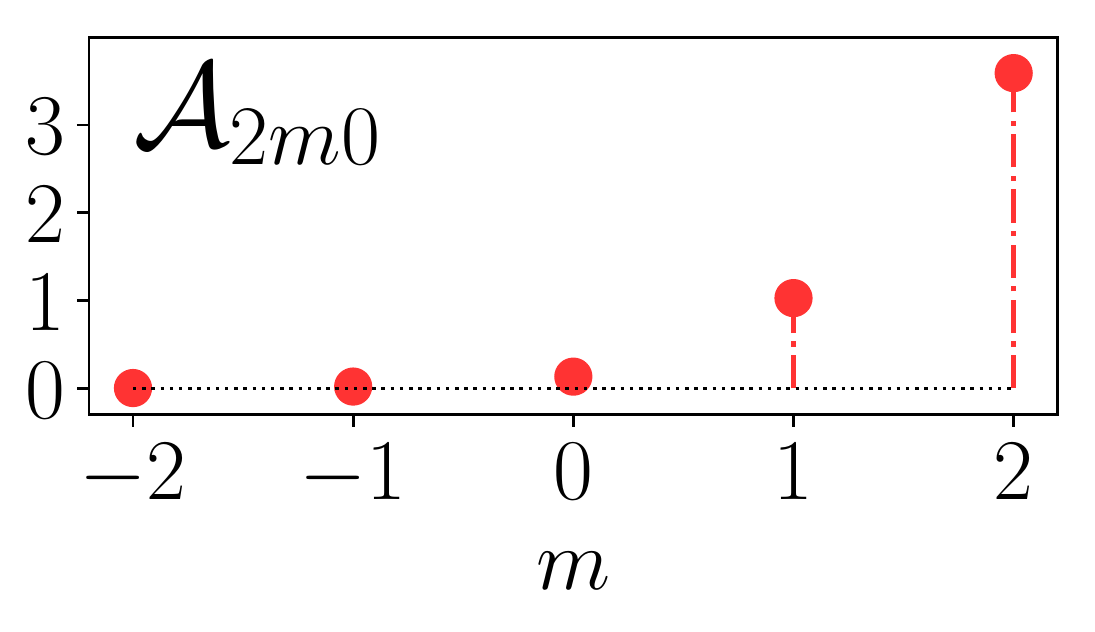}
    \caption{ Example of two plunge trajectories around a Kerr black hole ($a=0.5M$) and some of the mode content in the ringdown waves. Each trajectory shares the same inclination $I=20^\circ$ and final polar angle $\theta_{\rm fin}=77.3^\circ$. The left (blue) worldline approaches the horizon with  $\dot\theta < 0$ and the right (red) wordline approaches with $\dot\theta > 0$, which leads to different mode excitation.}
    \label{fig:3dtrajectory}
\end{figure*}
An orbit with inclination $I$ oscillates in $\theta$ in the range
\begin{eqnarray}
    90^\circ - I \le\; &\theta& \;\le 90^\circ + I\qquad\mbox{(prograde)}\;,
    \nonumber\\
    I - 90^\circ \le\; &\theta& \;\le 270^\circ - I\qquad\mbox{(retrograde)}\;.
    \label{eq:polarrange}
\end{eqnarray}
It turns out that $I$ remains nearly constant during the small body's inspiral and plunge {\cite{hughes2000}}.  This means that over the small body's worldline, its polar position oscillates over the range given by Eq.\ (\ref{eq:polarrange}) until the moment that it crosses the larger black hole's event horizon.  In Boyer-Lindquist coordinates, the small body freezes at the angle $\theta_{\rm fin}$ where it enters the event horizon.
\begin{figure}[ht]
    \centering
    \includegraphics[width=.45\textwidth]{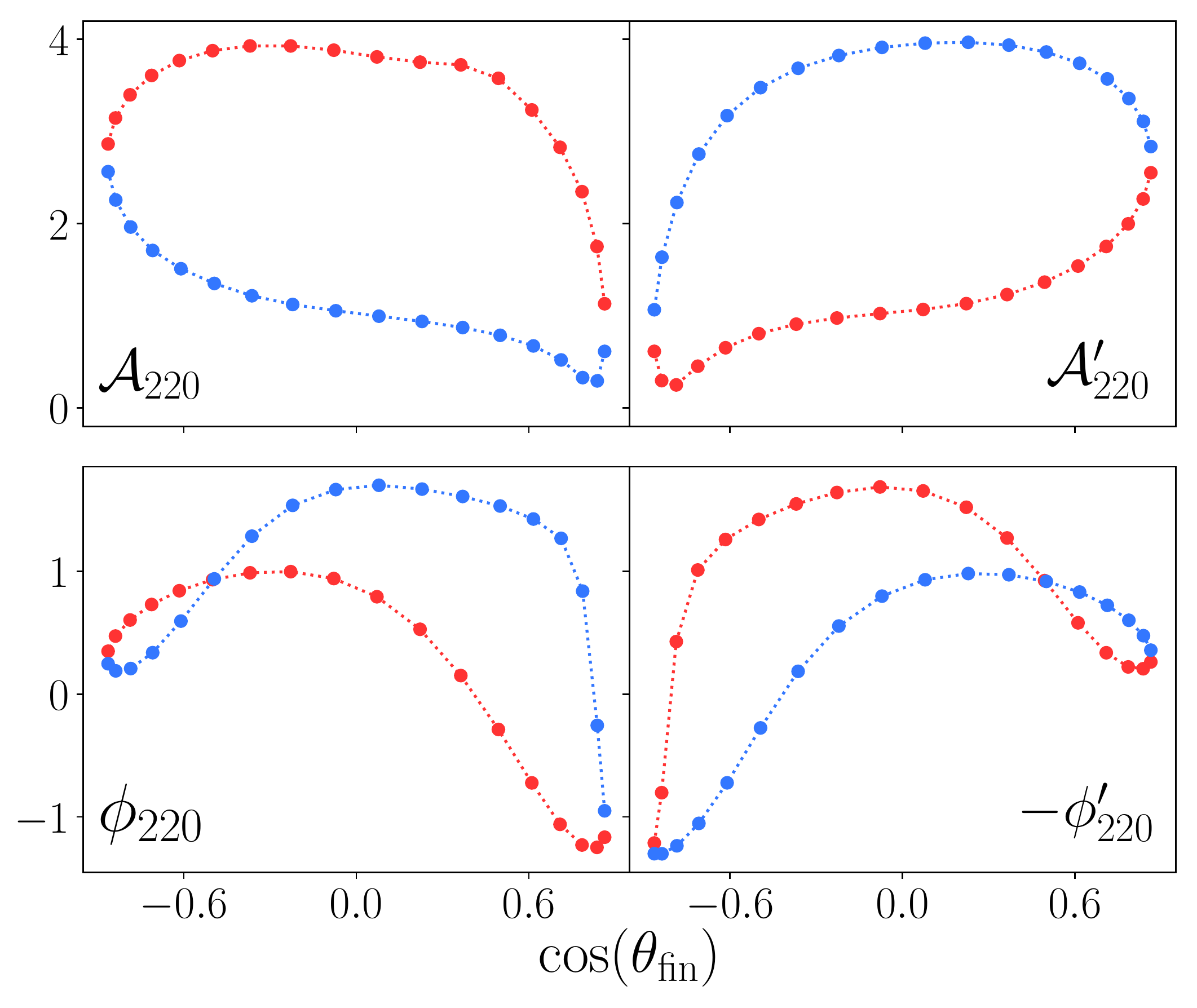}
    \caption{Example of reflection symmetry in mode excitation.  Top panels show the mode amplitudes $\mathcal{A}_{220}$ and $\mathcal{A}'_{220}$; bottom ones show the phases $\phi_{220}$ and $\phi'_{220}$.  All data are for a plunging small body with $I=60^\circ$ into a black hole with spin $a=0.5M$.  Each point corresponds to the mode excitation resulting from a worldline that terminates at a final polar angle $\theta_{\rm fin}$; red (blue) points indicate  $\dot{\theta}_{\rm fin} > 0$ ($\dot{\theta}_{\rm fin} < 0$).  In all cases, we find mode amplitudes that are consistent with the symmetry given in Eq.\ (\ref{eq:conjugate}).}
    \label{fig:symmetry}
\end{figure}

As discussed at length in Paper I, two worldlines with the same initial values of $r$ and $I$ can begin with different values of $\theta$.  Although these worldlines will follow the same trajectories in $(r,I)$ during inspiral and plunge, their worldlines will follow different trajectories in $\theta$, eventually crossing the event horizon at different final polar angles $\theta_{\rm fin}$.  An example of this is shown in Fig.\ 5 of Paper I.

Given $I$, there are in fact {\it two} worldlines which terminate at each value of $\theta_{\rm fin}$ allowed by the range (\ref{eq:polarrange}): one with $\dot\theta > 0$ during the final plunge, and one with $\dot\theta < 0$ in these moments.  This is illustrated in Fig.\ {\ref{fig:3dtrajectory}}.  Although the final state in both cases is identical, how the system reaches that final state is quite different.  The QNM signature of these two situations differs accordingly as well.

We thus find that a waveform's QNM content depends on 4 parameters which characterize its final plunging behavior: the black hole spin $a$, the orbital inclination $I$, the final polar angle $\theta_{\rm fin}$, and the sign of angular velocity in the plunge's final moments, $\sgn(\dot\theta_{\rm fin})$.  As we show in Sec. {\ref{sec:results}}, these parameters completely describe the mode amplitudes $\mathcal{C}_{kmn}$ and $\mathcal{C}'_{kmn}$:
\begin{eqnarray}\label{eq:function}
\mathcal{C}_{kmn} &=& \mathcal{C}_{kmn}\boldsymbol{(}a,I,\theta_{\rm fin},\sgn({\dot{\theta}_{\rm fin}})\boldsymbol{)}\;,
\nonumber\\
\mathcal{C}'_{kmn} &=& \mathcal{C}'_{kmn}\boldsymbol{(}a,I,\theta_{\rm fin},\sgn({\dot{\theta}_{\rm fin}})\boldsymbol{)}\;.
\end{eqnarray}

\subsection{Mode excitation symmetry characteristics}
\label{sec:modesymmetry}

The properties of black hole spacetimes imply that certain symmetries should exist in the mode excitation.  First consider two plunges that are on worldlines $u^{(1)}_\alpha$, $u^{(2)}_\alpha$ with the same orbital inclination about a given black hole (so that they share values of $a$ and $I$).  Imagine that these plunges approach the horizon such that
\begin{eqnarray}
\theta^{(2)}_{\rm fin} &=& \pi - \theta^{(1)}_{\rm fin}\;,
\label{eq:refl_angle}\\
\sgn(\dot{\theta}^{(2)}_{\rm fin}) &=& -\sgn(\dot{\theta}^{(2)}_{\rm fin})\;.
\label{eq:refl_angledot}
\end{eqnarray}
The Kerr spacetime is reflection symmetric about the equatorial plane ($\theta \rightarrow \pi - \theta$), so these infalling bodies have identical worldlines modulo a reflection which inverts in $\cos\theta$ (and possibly modulo a rotation $\phi \rightarrow \phi + \delta \phi$, which is also a continuous symmetry of Kerr). 

As the worldlines are related by a reflection, the radiation they produce, which is decomposed onto the harmonics $\tensor*[_{-2}]{Y}{_{\ell m}}(\theta,\phi)$, should also be related by a reflection.  The spin-weighted spherical harmonics transform under reflection by Eq.\ (\ref{eq:spherical_reflection}).  This implies that the primed and unprimed mode amplitudes of the two worldlines are related by the following reflection symmetry:
\begin{eqnarray}
& &\mathcal{C}'_{kmn}\boldsymbol{(}a,I,\theta_{\rm fin},\sgn({\dot{\theta}_{\rm fin}})\boldsymbol{)} =
\nonumber\\
& &\qquad\qquad
\mathcal{C}^*_{kmn}\boldsymbol{(}a,I,\pi - \theta_{\rm fin},-\sgn({\dot{\theta}_{\rm fin}})\boldsymbol{)}\;.
\label{eq:conjugate}
\end{eqnarray}
For each QNM, it suffices to only specify either $\mathcal{C}_{kmn}\boldsymbol{(}a,I,\theta_{\rm fin},\sgn({\dot{\theta}_{\rm fin}})\boldsymbol{)}$ or $\mathcal{C}'_{kmn}\boldsymbol{(}a,I,\theta_{\rm fin},\sgn({\dot{\theta}_{\rm fin}})\boldsymbol{)}$.  In all of our calculations, we find that the mode amplitudes we find are consistent with the symmetry (\ref{eq:conjugate}).  Figure {\ref{fig:symmetry}} shows this explicitly for one example (which we discuss in much greater depth in Sec.\ {\ref{sec:results}}).  Moving forward, we will generally only show one of $\mathcal{C}_{kmn}$ or $\mathcal{C}'_{kmn}$ in our results.

\begin{figure*}[ht]
    \centering
    \includegraphics[width=.45\textwidth]{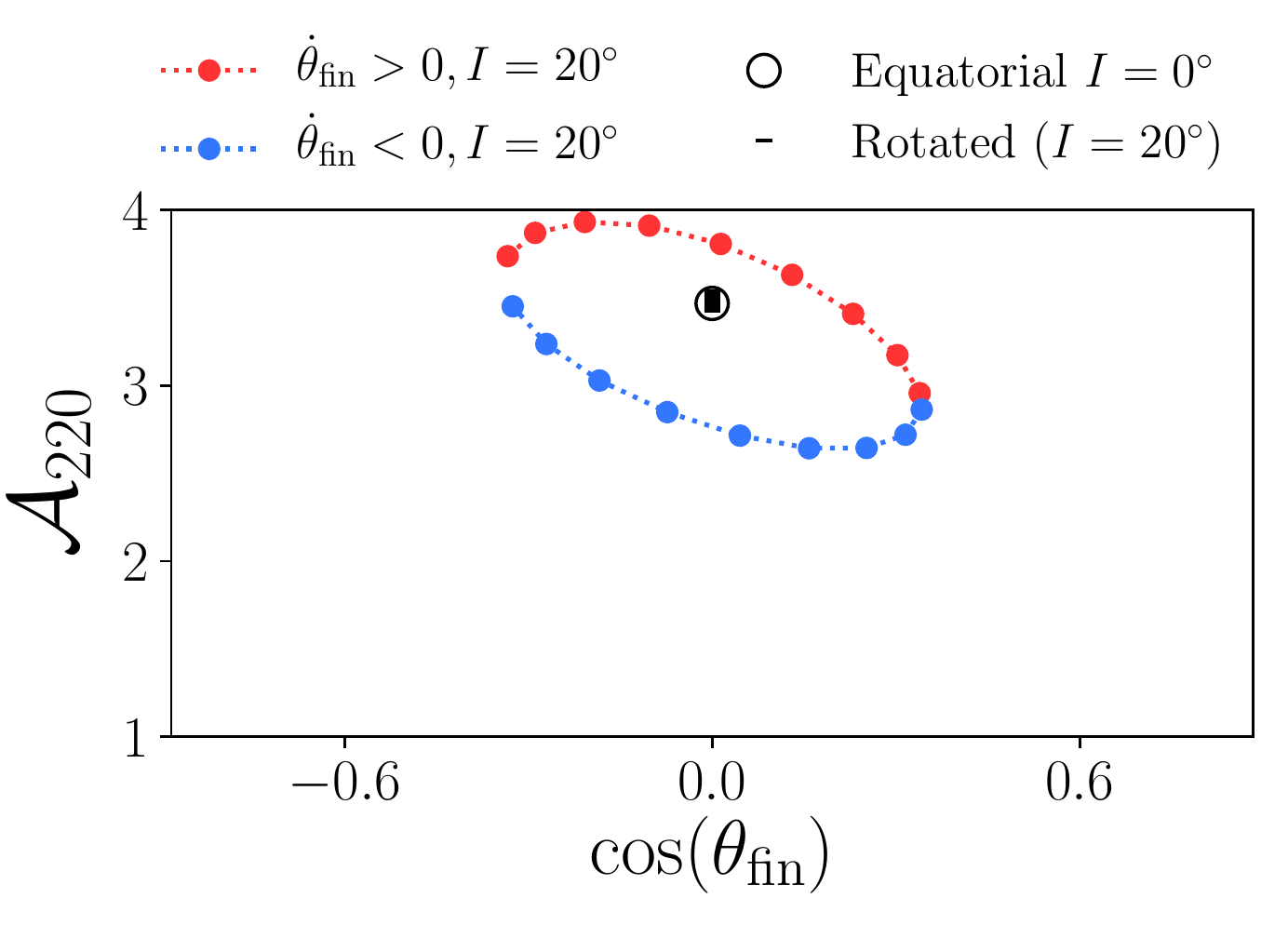}
    \includegraphics[width=.45\textwidth]{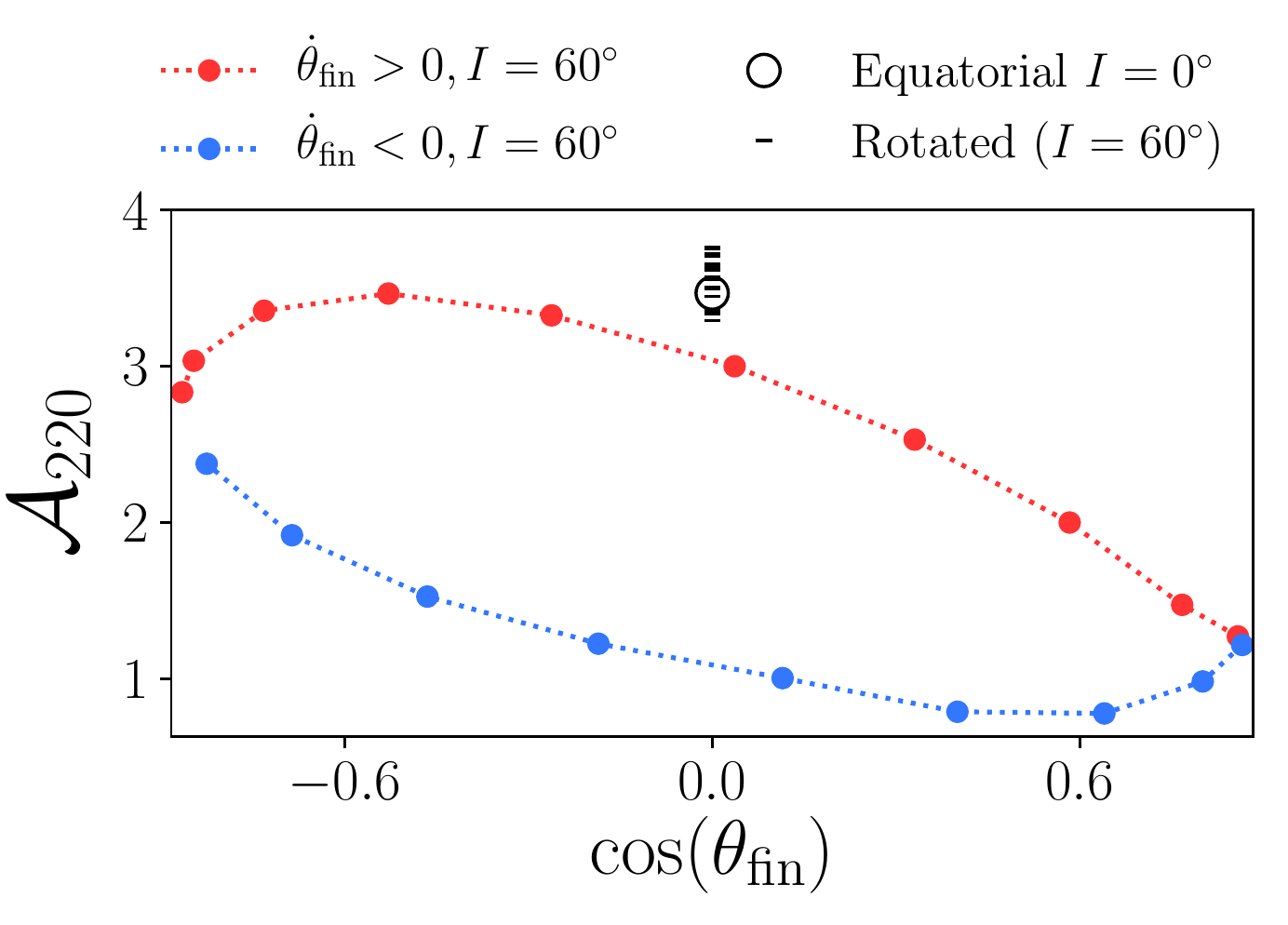}
    \caption{Mode excitation of a Schwarzschild black hole. Red and blue points plot mode amplitudes calculated from inclined trajectories ($I = 20^\circ$ on the left panel, $I = 60^\circ$ on the right panel).  The black circles show the mode amplitude calculated from an equatorial trajectory.  Mode excitations from each inclined trajectory are rotated into a new basis [cf.\ Eq.\ (\ref{eq:moderotation})] where the mode amplitudes equal those of the equatorial trajectory.  We find outstanding agreement for the shallow inclination case, $I = 20^\circ$.  The relative error is larger (several percent) for $I = 60^\circ$ due to numerical errors from the small body rapidly crossing many angular grid zones at this inclination.}
    \label{fig:schwarzschildtest}
\end{figure*}

Second, consider the Schwarzschild limit, $a = 0$.  Because of this spacetime's spherical symmetry, there is no unique notion of an ``equatorial plane,'' so any two worldlines $u^{(1)}_\alpha$ and $u^{(2)}_\alpha$ that begin from circular orbits can be related to each other by a rotation.  Consider the rotation generated by the quaternion $\mathbb{R}$ which relates the spatial components of the worldlines $\mathbf{u}^{(1)}$ and $\mathbf{u}^{(2)}$ as
\begin{equation}\label{eq:trajrotate}
\mathbf{u}^{(1)} = \mathbb{R}^{-1} \mathbf{u}^{(2)} \mathbb{R}\;.
\end{equation}
Now, decompose the radiation resulting from $u^{(i)}_\alpha$ onto the spin-weighted spherical harmonics,
\begin{equation}
\tensor*[]{h}{^{(i)}}=\sum_{\ell ,m} \tensor*[]{h}{_{\ell m}^{(i)}}\, \tensor*[_{-2}]{Y}{_{\ell m}}(\theta_i,\phi_i)\qquad (i=1,2)\;,
\end{equation}
where $(\theta_i,\phi_i)$ are the coordinates in which the angular components of $\mathbf{u}^{(i)}$ are expressed.  Then the radiation multipoles $h^{(1)}_{\ell m}$ and $h^{(2)}_{\ell m}$ can be related to each other by the basis transformation formula,
\begin{equation} \label{eq:rotation}
\tensor*[]{h}{_{\ell m}^{(1)}} = \sum_{m'} \mathfrak{D}^{\ell}_{m'm}( \mathbb{R}) \tensor*[]{h}{_{\ell m'}^{(2)}},
\end{equation}
where $\mathfrak{D}^{\ell}_{m'm}( \mathbb{R})$ is the Wigner rotation matrix, which takes on a simple form in terms of $\mathbb{R}$ \cite{swshboyle}. Using Eq.\ (\ref{eq:rotation}) and the fact that $\tensor*[_{-2}]{S}{^{a\sigma}_{k mn}} = \tensor*[_{-2}]{Y}{_{k m}}$ for $a = 0$, we can relate the mode amplitudes between any two worldlines as
\begin{equation} \label{eq:moderotation}
\mathcal{C}^{(1)}_{kmn} = \sum_{m'} \mathfrak{D}^{k}_{m'm}( \mathbb{R})\mathcal{C}^{(2)}_{km'n},
\end{equation}
provided that their trajectories satisfy Eq.\ (\ref{eq:trajrotate}). 

We use Eq.\ (\ref{eq:moderotation}) to test our mode extraction algorithm as follows.  Let $u^{(1)}_\alpha$ be the worldline of a small body on an equatorial trajectory ($I^{(1)} = 0$), and let $u^{(2)}_\alpha$ be the worldline of a small body on an inclined trajectory ($I^{(2)} > 0$).  By Eq.\ (\ref{eq:moderotation}), the mode amplitudes calculated from the $u^{(2)}_\alpha$ plunge should transform to amplitudes computed from the equatorial $u^{(2)}_\alpha$ plunge.

Figure \ref{fig:schwarzschildtest} shows our results for $I^{(2)} = 20^\circ$ and $I^{(2)} = 60^\circ$ using several different values of $\theta^{(2)}_{\rm fin}$.  The black circle at $(\theta_{\rm fin}, {\cal A}_{220}) = (90^\circ, 2.3)$ shows the mode amplitude we find for the equatorial worldline $u^{(1)}_\alpha$.  The blue and red dots show the amplitudes ${\cal A}_{220}$ we find for the various inclined worldlines $u^{(2)}_\alpha$ we consider.  The black dashes show the amplitude that we find by applying Eq.\ (\ref{eq:moderotation}) to rotate the inclined case into the equatorial plane.  At high inclination ($I^{(2)} = 60^\circ$), the rotated mode amplitudes agree with the equatorial amplitude to within $8.5\%$.  As we'll discuss in Secs.\ {\ref{sec:catalog}} and {\ref{app:resolution_comparison}}, numerical noise appears to limit our accuracy for high orbit inclination.  At lower inclination ($I^{(2)} = 20^\circ$) the numerical error is reduced, and we find that all rotated mode amplitudes agree with the equatorial amplitudes to within $2.0\%$. In Appendix\ {\ref{app:resolution_comparison}}, we further discuss the impact of numerical errors on our results. Code for computing the Wigner rotation matrix elements in terms of quaternions is provided by M.\ Boyle {\cite{quaternion}}.

\subsection{Comparison with past equatorial results}
\label{sec:equatorialcomp}

\begingroup
\squeezetable
\begin{table*}[t]
\begin{ruledtabular}
\begin{tabular}{c|ccc|ccc|ccc}
 $a/M$ & & $|A_{2-10}/A_{210}|$ & & & $|A_{220}/A_{320}|$ & & &  $|A_{3-20}/A_{320}|$ & \\ 
        &  This paper & & Ref.\ {\cite{tbkh14}}        & 
           This paper & & Ref.\ {\cite{tbkh14}}        & 
      	   This paper & & Ref.\ {\cite{tbkh14}}   \\    \hline
             
 $0.9$   &  0.0066  & & 0.0025                            &
            2.40    & & 2.11                          &
            0.044  & &                       \\ 
 $0.5$   &  0.011   & & 0.010                             &
            0.34    & & 0.34                             &    
            0.0044 & &              \\    
 $0.0$   &  0.071   & & 0.069                              &
            0.000   & &                         & 
            0.010  & & 0.010              \\ 
 $-0.5$  &  0.20    & & 0.21                             &
            0.10    & & 0.13                           &    
            0.096   & & 0.093              \\    
 $-0.9$   &  0.32   & & 0.32                              &
             0.18   & & ---                       &  
             0.29   & & ---                             \\  
\end{tabular}
\end{ruledtabular}
\caption{
    Mode amplitudes that we find compared with those presented by Taracchini \textit{et al.} \cite{tbkh14} for equatorial inspiral and plunge [cf.\ their Table II and Eq.\ (5)].  We calculate the mode amplitudes as described in Sec.\ {\ref{sec:modeextract}} using waveform data provided by the authors of Ref.\ {\cite{tbkh14}}.  In this table, we use the notation of Ref.\ {\cite{tbkh14}}, so that positive $a$ denotes prograde ($I = 0^\circ$), negative $a$ denotes retrograde ($I = 180^\circ$).  Dash entries indicate that the mode, while present, could not be reliably fitted (using the method described in Ref.\ {\cite{tbkh14}}).  Blank entries indicate that the mode is not significantly excited and thus excluded from the model. At least for modest spins ($|a| \le 0.5M$), we find fairly good agreement. Because of systematic differences in the mode extraction algorithm, we do not expect perfect agreement; and, we expect the differences to be particularly marked for $a$ large.  See text in Sec.~\ref{sec:equatorialcomp} for details of how to convert between mode amplitudes presented in Ref.~\cite{tbkh14} $A_{\ell mn}$, and mode amplitudes presented in this paper, $\mathcal{A}_{\ell mn},\mathcal{A}'_{\ell mn}$.}
\label{tab:taracchinicomparison}
\end{table*}
\endgroup

As a final check, we compare our results for equatorial inspiral and plunge ($I = 0^\circ$ and $I = 180^\circ$) with those given by Taracchini \textit{et al.}, Ref.\ {\cite{tbkh14}}.  We do not expect these two analyses to agree perfectly.  This is in part due to differences in the trajectory and waveform calculations.  Also, the method used in Ref.\ {\cite{tbkh14}} to extract QNMs from late waveforms is quite different from that which we developed.  In particular, the number of spheroidal modes that are included in the fit in Ref.\ {\cite{tbkh14}} is typically fewer than the number of modes that we use here.  Since mode mixing is strongest at large spin, we expect the largest systematic differences at large $a$.  
Given these considerations, we limit our comparison by analyzing the same waveform data from Ref.\ {\cite{tbkh14}}, which was provided by the authors (two of whom are authors of this paper). We also only model the modes which were considered in Ref.\ {\cite{tbkh14}}.
To properly compare with Ref.\ {\cite{tbkh14}}, we also have to adjust our notation slightly by relating the amplitudes defined here [appearing in Eq.\ ({\ref{eq:spheroidaldecomp}})] to those defined in Ref.\ {\cite{tbkh14}} [cf.\ their Eq.~(5)]. For prograde orbits $a \ge 0$ ($I = 0^\circ$),
\begin{eqnarray}
\left|\frac{A_{\ell-m0}}{A_{\ell m0}}\right| &=& \left|\frac{\mu_{-m\ell \ell 0}\mathcal{A}'_{\ell -m0}}{\mu_{m\ell \ell 0}\mathcal{A}_{\ell m0}}\right|\;,
\nonumber\\
\left|\frac{A_{\ell' m0}}{A_{\ell m0}}\right| &=& \left|\frac{\mu_{m \ell \ell' 0}\mathcal{A}_{\ell' m 0}}{\mu_{m \ell \ell 0}\mathcal{A}_{\ell m0}}\right|\;.
\end{eqnarray}
For retrograde orbits, the waveform multipoles in Ref.\ {\cite{tbkh14}} (which we refer to as $h^{\rm N}_{\ell m}$) were computed in a coordinate system where $L_z$ is positive.  Although this is opposite of our convention, the computed modes in this study can be related to those modes computed in Ref.\ {\cite{tbkh14}} by a $180^\circ$ coordinate rotation [see Eq.~({\ref{eq:moderotation}})]. Thus for $a < 0$,
\begin{eqnarray}
\left|\frac{A_{\ell-m0}}{A_{\ell m0}}\right| &=&  \left|\frac{\mu_{m\ell \ell 0}\mathcal{A}'_{\ell m0}}{\mu_{-m\ell \ell 0}\mathcal{A}_{\ell -m0}}\right|\;,
\nonumber\\
\left|\frac{A_{\ell' m0}}{A_{\ell m0}}\right| &=& \left|\frac{\mu_{-m \ell \ell' 0}\mathcal{A}_{\ell' -m 0}}{\mu_{-m \ell \ell 0}\mathcal{A}_{\ell -m0}}\right|\;.
\end{eqnarray}
We also adapt the conventions used in Ref.\ {\cite{tbkh14}} for choosing $t_0$.

Table {\ref{tab:taracchinicomparison}} shows the result of this comparison.  We indeed find decent agreement with their mode amplitudes, at least for $a \le 0.5M$.  The lack of perfect agreement, and disagreement at $a=0.9M$, is to be expected given the rather different methods of calculating the mode excitation. For instance, our analysis suggests that the $(k,m,n)=(3,1,0)$ QNM is excited, but this mode is not modeled in Ref.\ {\cite{tbkh14}}.  This leads to larger systematic differences between the two methods at high spin, $a = 0.9M$.

\section{Results}
\label{sec:results}

\begin{figure}
    \includegraphics[width=.45\textwidth]{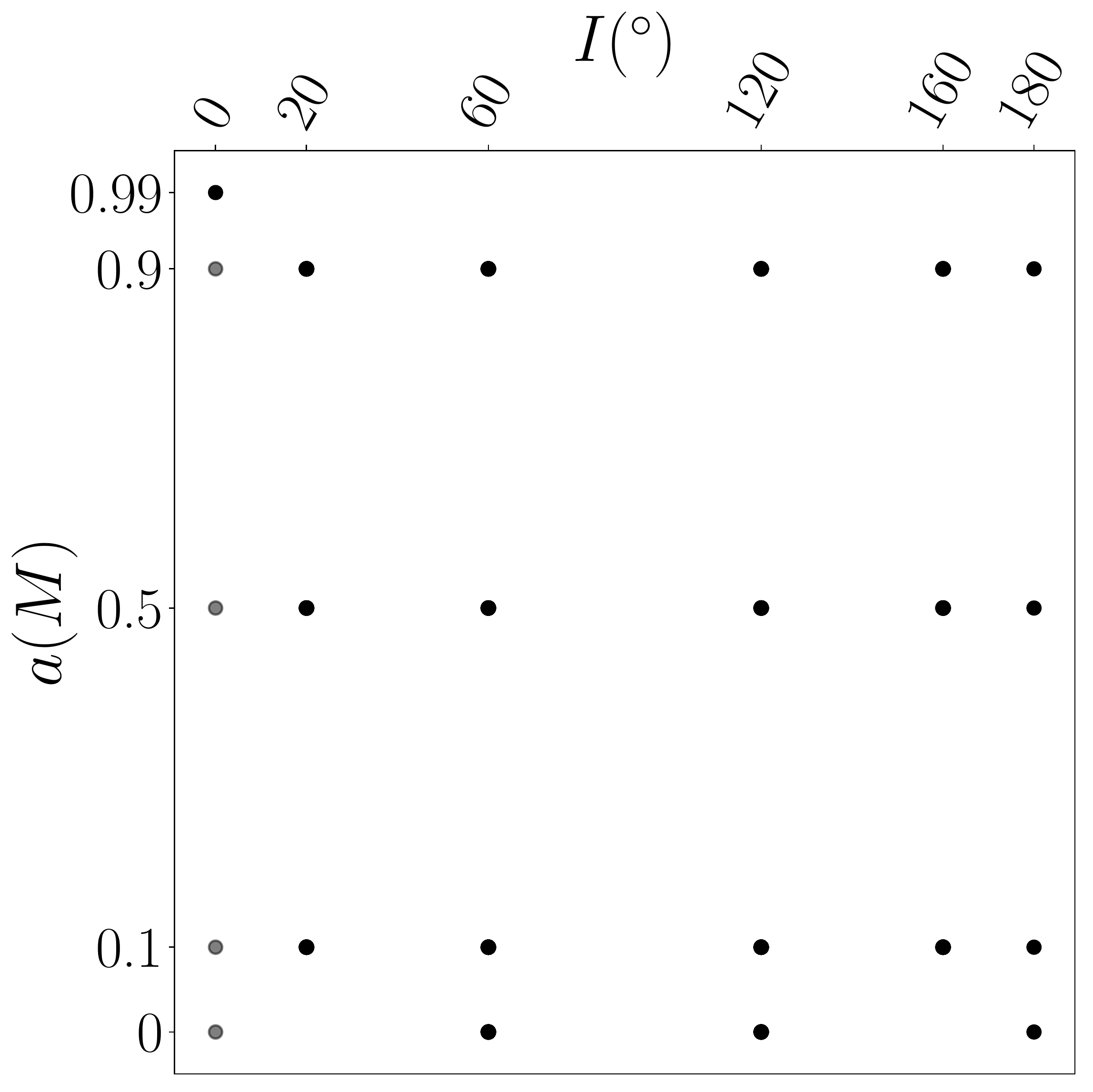}
    \caption{Spins and inclinations investigated in this study.  Each point $(a,I)$ shown here represents an orbital configuration that we have studied in detail.  For each non-equatorial case ($I \ne 0^\circ$ and $I \ne 180^\circ$), we examine 36 different trajectories, each corresponding to a different polar angle $\theta_{\rm fin}$ at which the trajectory terminates (18 with $\dot\theta_{\rm fin}$ positive, 18 with it negative).  Only one example suffices for the equatorial configurations.  The results for spin $a = 0$ are primarily used to check that our results respect Schwarzschild symmetry properties, and are discussed in Sec.\ {\ref{sec:modesymmetry}}; the results for spin $a = 0.99M$ are used to investigate whether we can ascertain the presence of overtone modes ($n \ge 1$), and are discussed in Sec.\ {\ref{sec:overtone_results}}.}
    \label{fig:cases}
\end{figure}

We now present detailed results describing the mode excitation we find for inclined plunges of rotating black holes.  Figure {\ref{fig:cases}} shows the range of cases that we examine.  Each point in this figure represents a particular choice of $a$ and $I$ that we study.  For each non-equatorial case ($I \ne 0^\circ$ and $I \ne 180^\circ$), we examine 36 different values of $\theta_{\rm fin}$, 18 for each sign of $\dot\theta_{\rm fin}$.  For the equatorial cases, we always have $\theta_{\rm fin} = 90^\circ$, and $\dot\theta_{\rm fin} = 0$.

\subsection{A mode excitation catalog}
\label{sec:catalog}

\begin{figure*}[p]
\includegraphics[width = 1\textwidth]{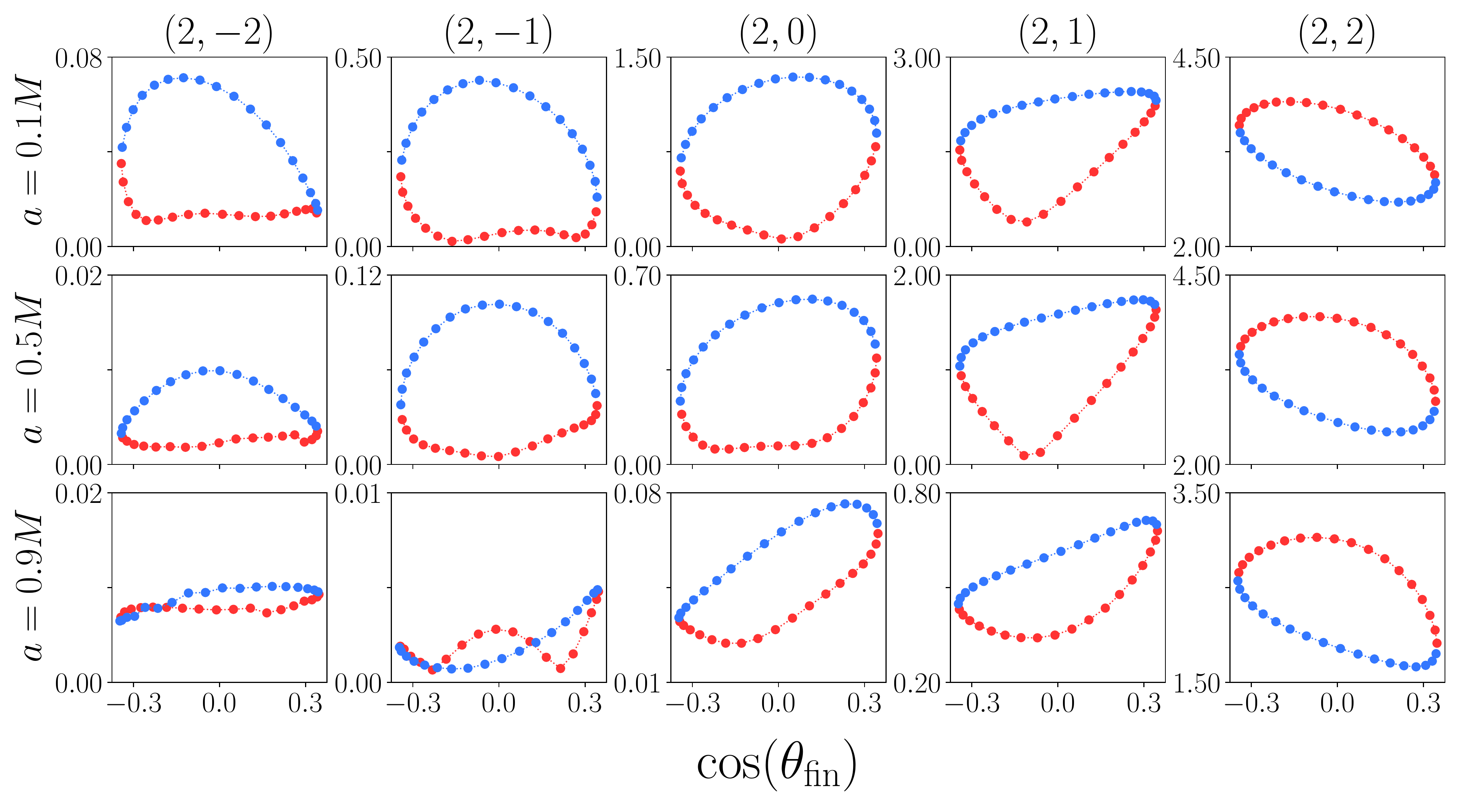}
\caption{Mode excitation magnitude ${\cal A}_{kmn}$ for spheroidal QNMs with $k = 2$, $m \in (-2,\ldots,2)$, $n = 0$ for inspiral and plunge with $I = 20^\circ$.  From top to bottom, the black hole spins vary from $a = 0.1M$ to $a = 0.5M$ to $a = 0.9M$; from left to right, $m$ varies from $-2$ to $2$.  In each plot, red (blue) dots indicate $\dot\theta_{\rm fin} > 0$ ($\dot\theta_{\rm fin} < 0$).  Note that the vertical scale varies in each panel.}
\label{fig:I020_l2}
\end{figure*}

\begin{figure*}[p]
\includegraphics[width = 1\textwidth]{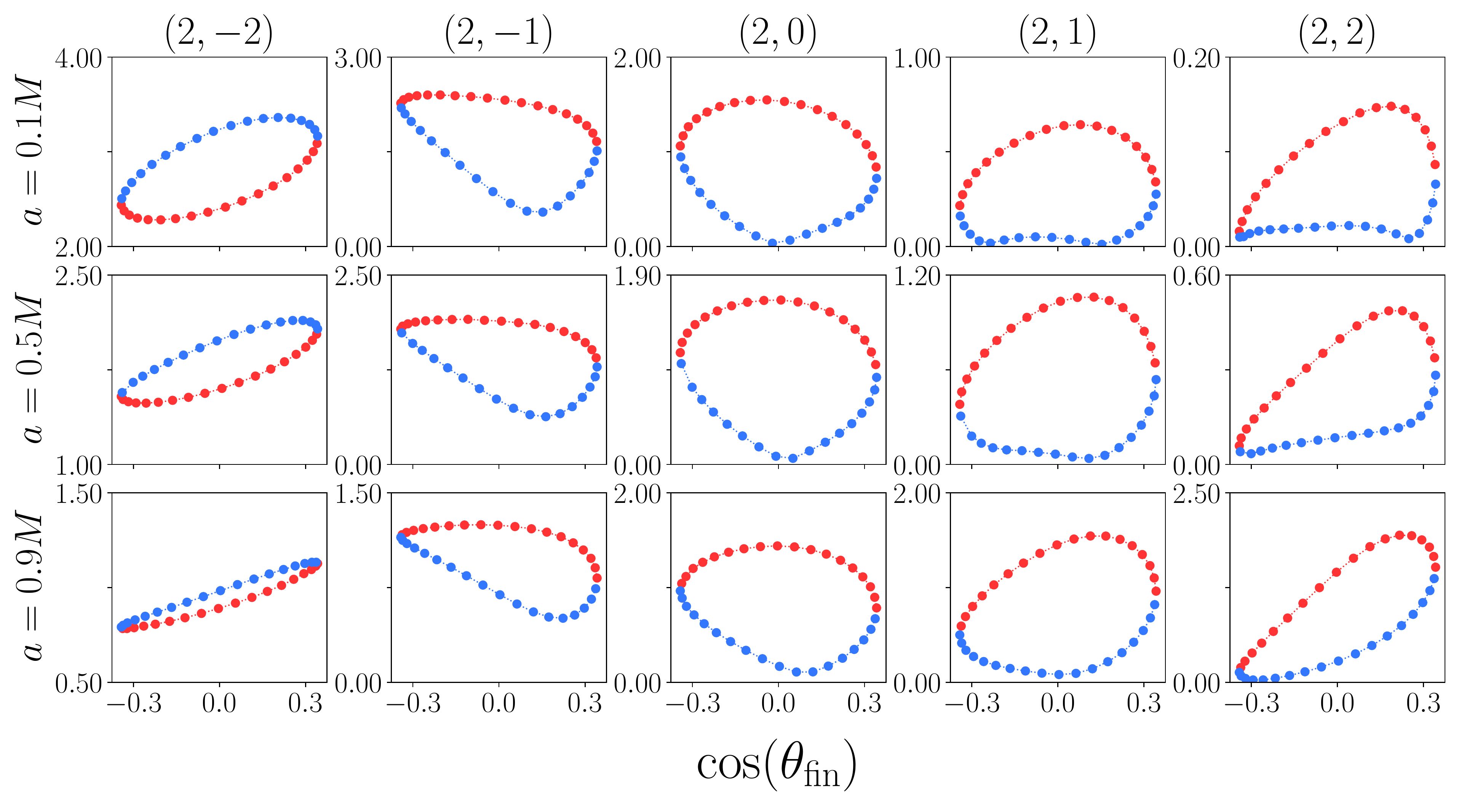}
\caption{Mode excitation magnitude ${\cal A}_{kmn}$ for spheroidal QNMs with $k = 2$, $m \in (-2,\ldots,2)$, $n = 0$ for inspiral and plunge with $I = 160^\circ$ (the retrograde complement of the case presented in Fig.\ {\ref{fig:I020_l2}}).  All other details are as in Fig.\ {\ref{fig:I020_l2}}.  }
\label{fig:I160_l2}
\end{figure*}

\begin{figure*}[p]
\includegraphics[width = 1\textwidth]{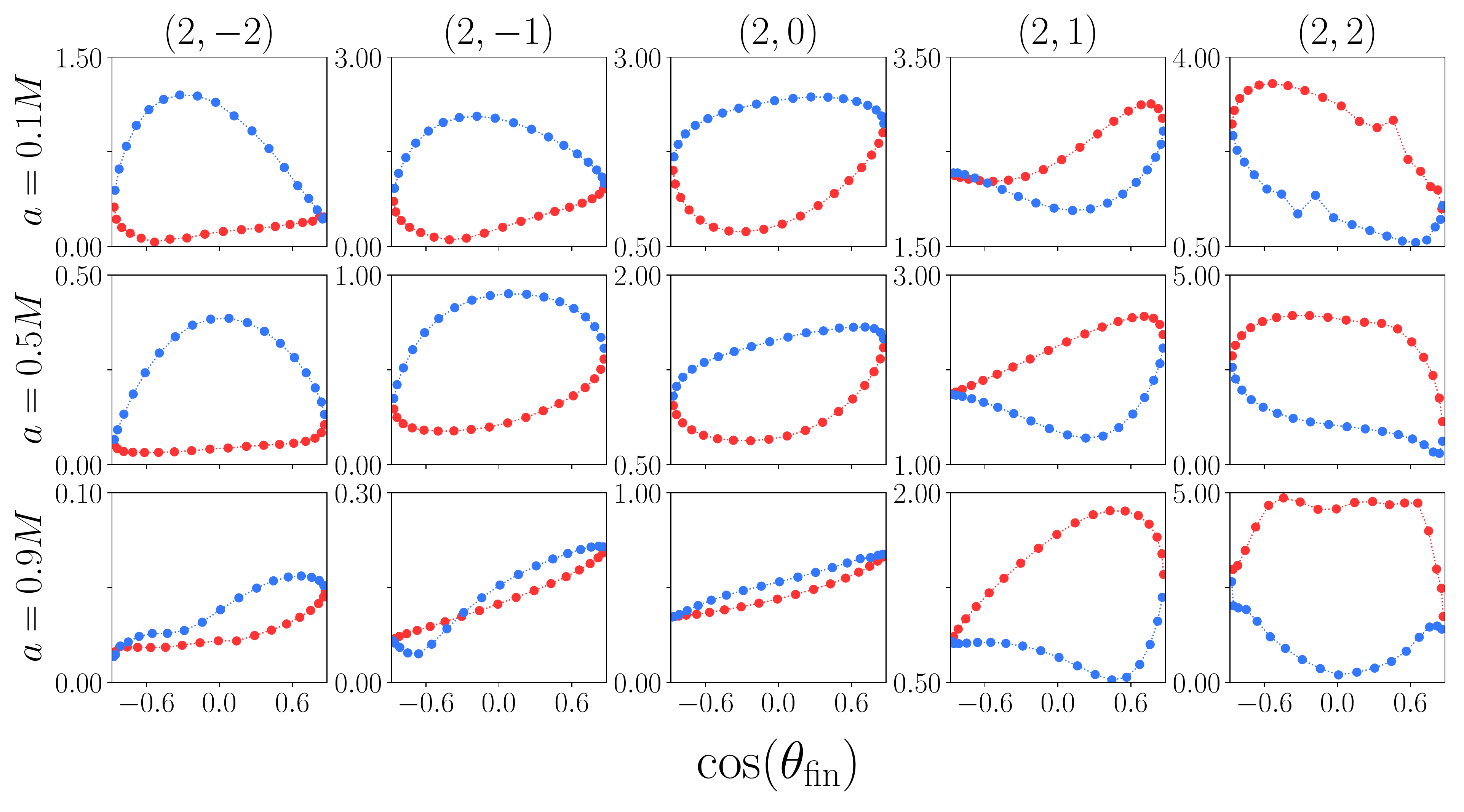}
\caption{Mode excitation magnitude ${\cal A}_{kmn}$ for spheroidal QNMs with $k = 2$, $m \in (-2,\ldots,2)$, $n = 0$ for inspiral and plunge with $I = 60^\circ$.  All other details are as in Fig.\ {\ref{fig:I020_l2}}.  At this high inclination we see more numerical noise, as evident in the ${\cal A}_{220}$ amplitudes.}
\label{fig:I060_l2}
\end{figure*}

\begin{figure*}[p]
\includegraphics[width = 1\textwidth]{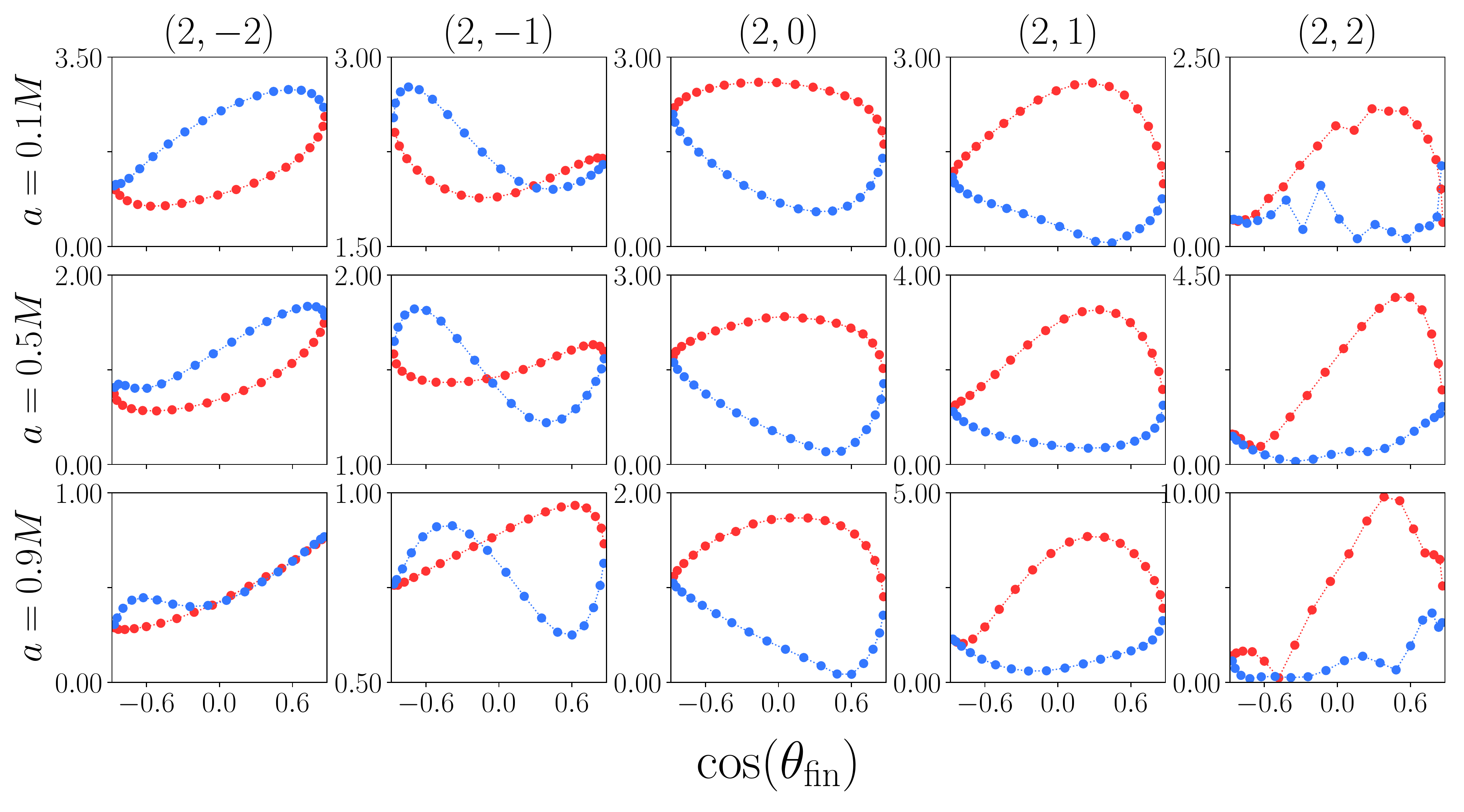}
\caption{Mode excitation magnitude ${\cal A}_{kmn}$ for spheroidal QNMs with $k = 2$, $m \in (-2,\ldots,2)$, $n = 0$ for inspiral and plunge with $I = 120^\circ$ (the retrograde complement of the cases shown in Fig.\ {\ref{fig:I060_l2}}.  All other details are as in Fig.\ {\ref{fig:I020_l2}}.  At this high inclination we see more numerical noise, as evident in the ${\cal A}_{220}$ amplitudes.}
\label{fig:I120_l2}
\end{figure*}

We begin by presenting a catalog of spheroidal modes with $k = 2$ for a range of black hole spins, $a \in (0.1M, 0.5M, 0.9M$), and for four values of orbital inclination, $I \in (20^\circ, 60^\circ, 120^\circ, 160^\circ)$.  The inclinations $20^\circ$ and $60^\circ$ represent low and high prograde values (i.e., inclinations for which $L_z$ is positive); the inclinations $160^\circ$ and $120^\circ$ represent low and high retrograde values (with $L_z < 0$).

Figures {\ref{fig:I020_l2}} -- {\ref{fig:I120_l2}} show the magnitude of the fundamental mode excitations ${\cal A}_{2m0}$ that we find in these 12 different cases.  In each plot, the top row shows results for $a = 0.1M$, the middle for $a = 0.5M$, and the bottom row for $a = 0.9M$.  Going from left to right, the columns present data for $m = -2$ through $m = 2$.  Each point in the panels shows ${\cal A}_{2m0}$ for a different value of $\theta_{\rm fin}$; red dots are for $\dot\theta_{\rm fin} > 0$, blue are for $\dot\theta_{\rm fin} < 0$.  Additional plots, presenting phases and additional values of $k$, are shown in Appendix {\ref{app:catalog_more}}.

\begin{figure*}[ht]

    \includegraphics[width=.31\textwidth]{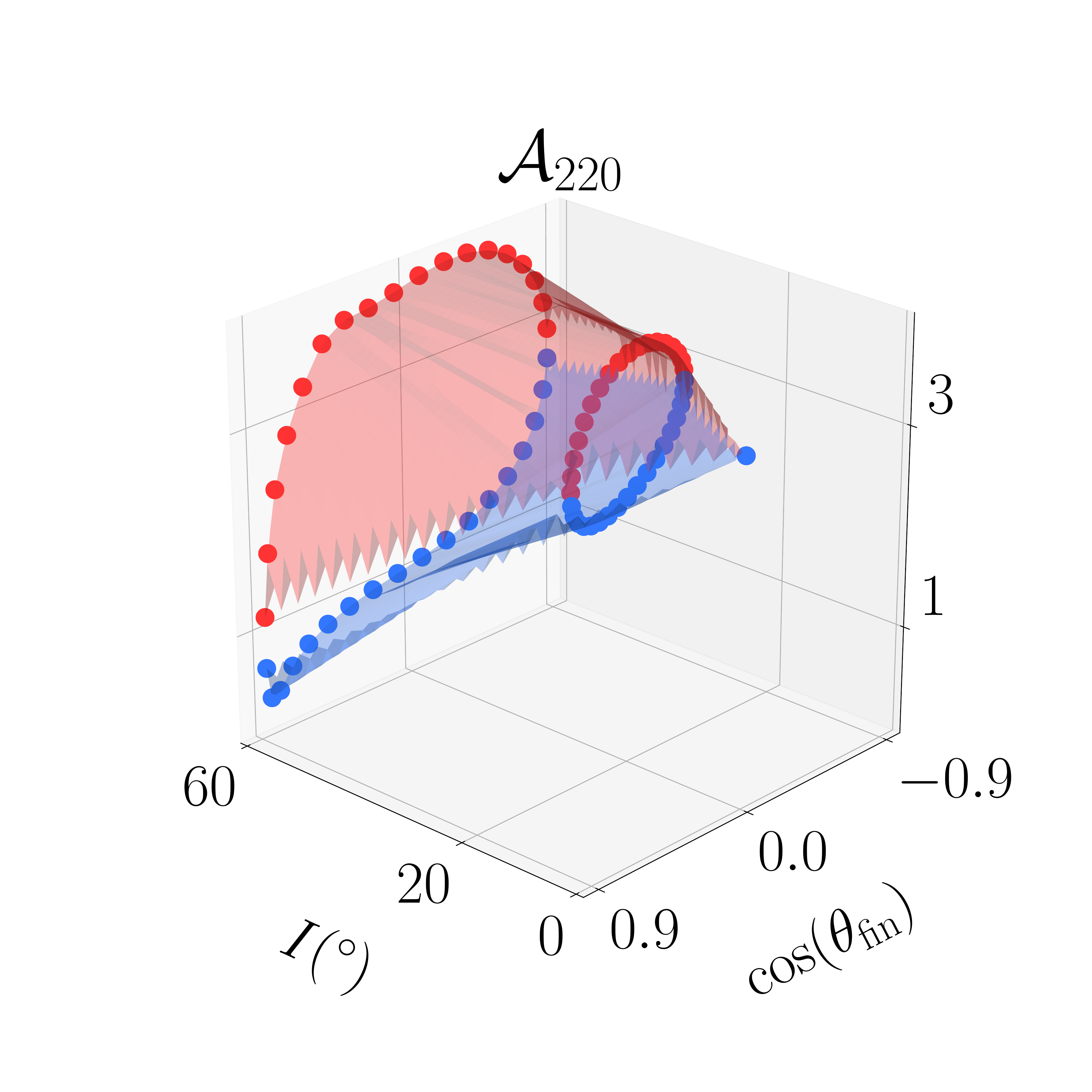}
    \includegraphics[width=.31\textwidth]{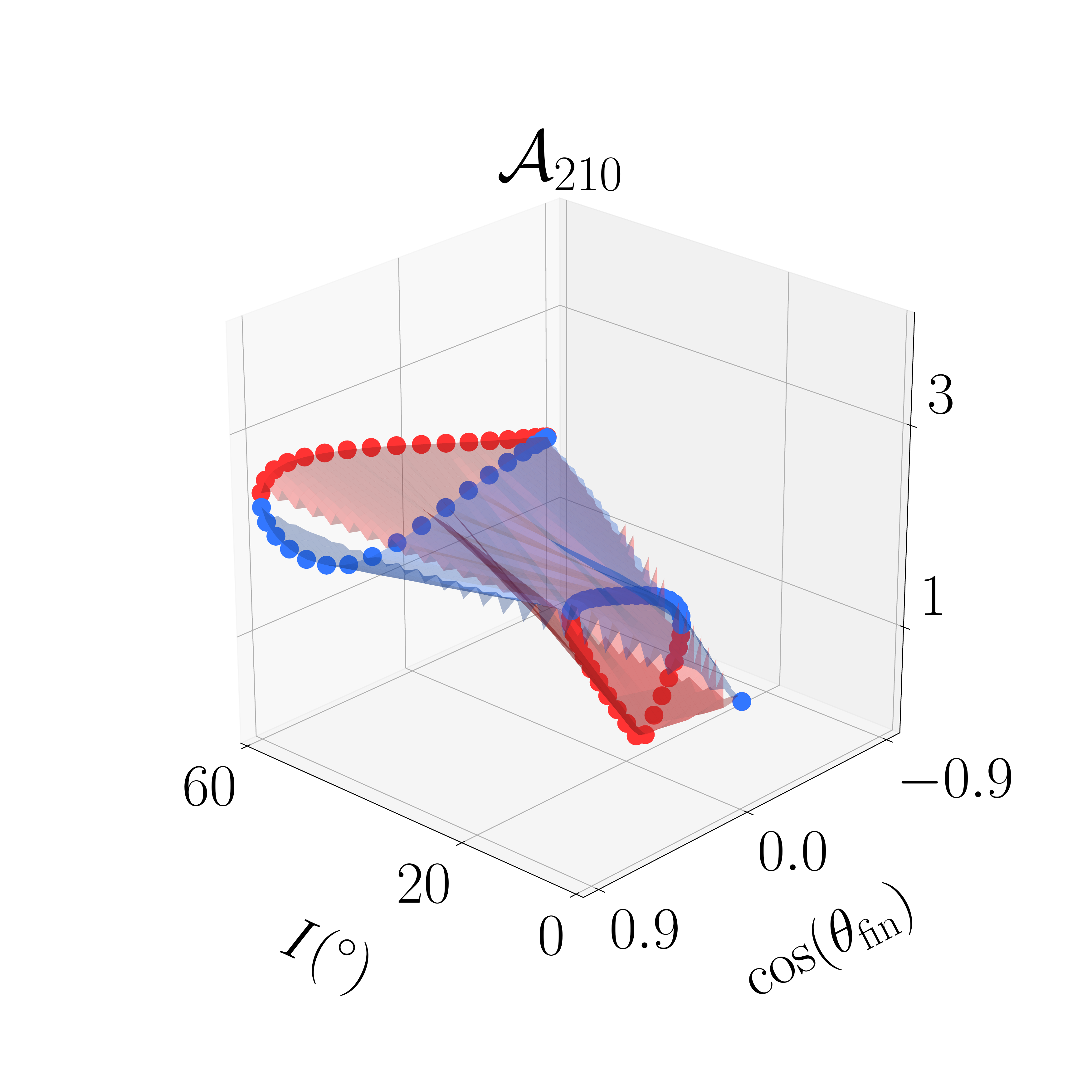}
    \includegraphics[width=.31\textwidth]{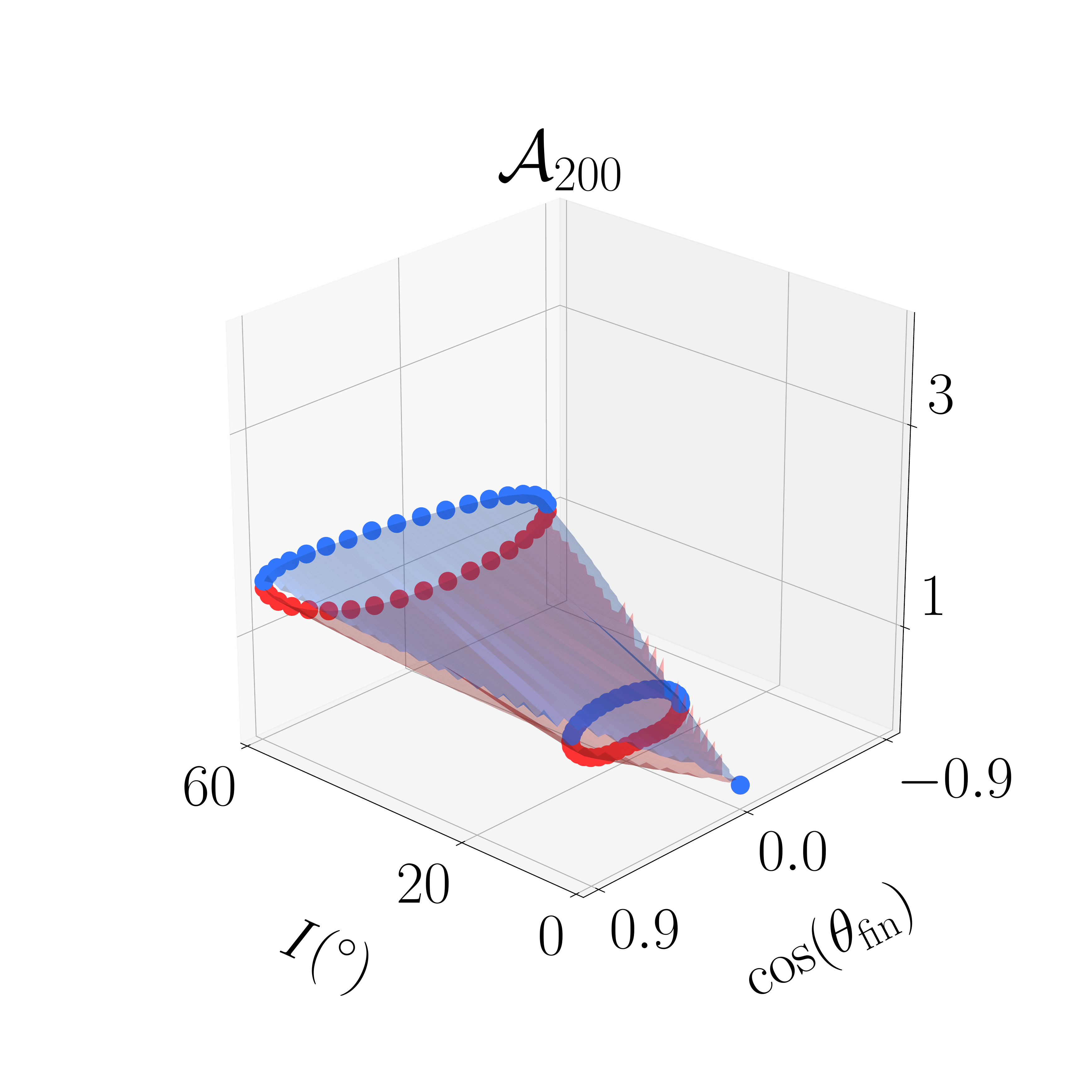}

    \includegraphics[width=.31\textwidth]{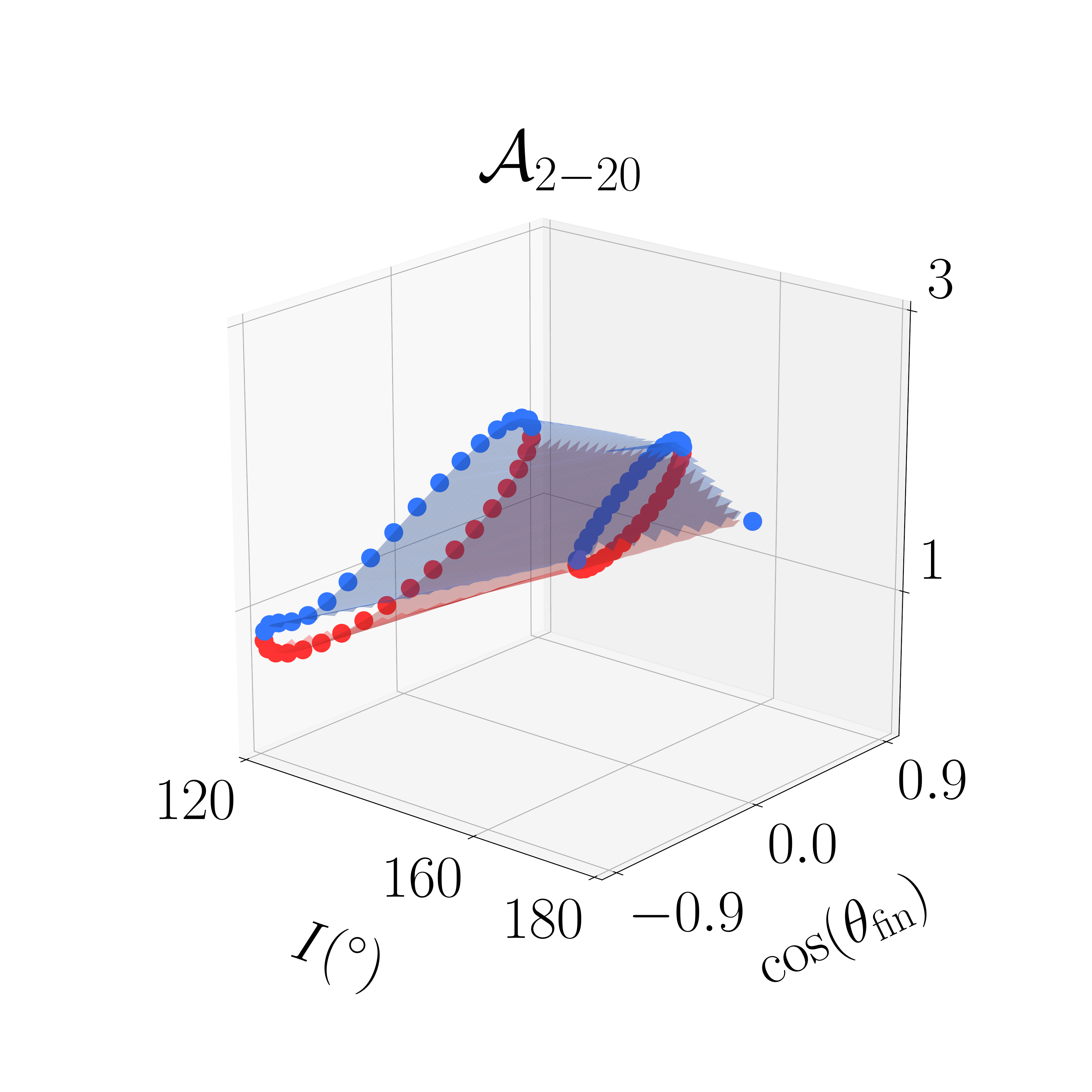}
    \includegraphics[width=.31\textwidth]{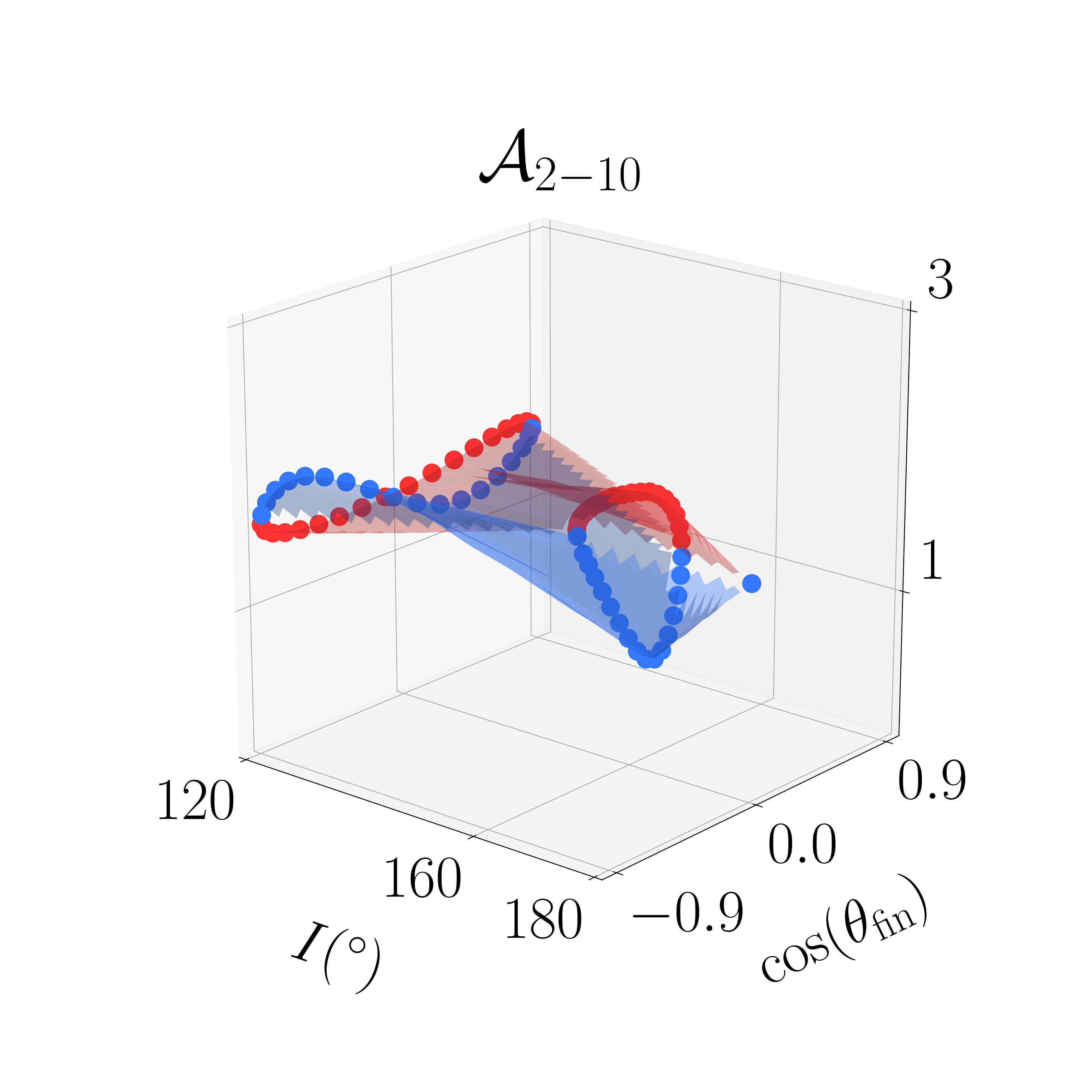}
    \includegraphics[width=.31\textwidth]{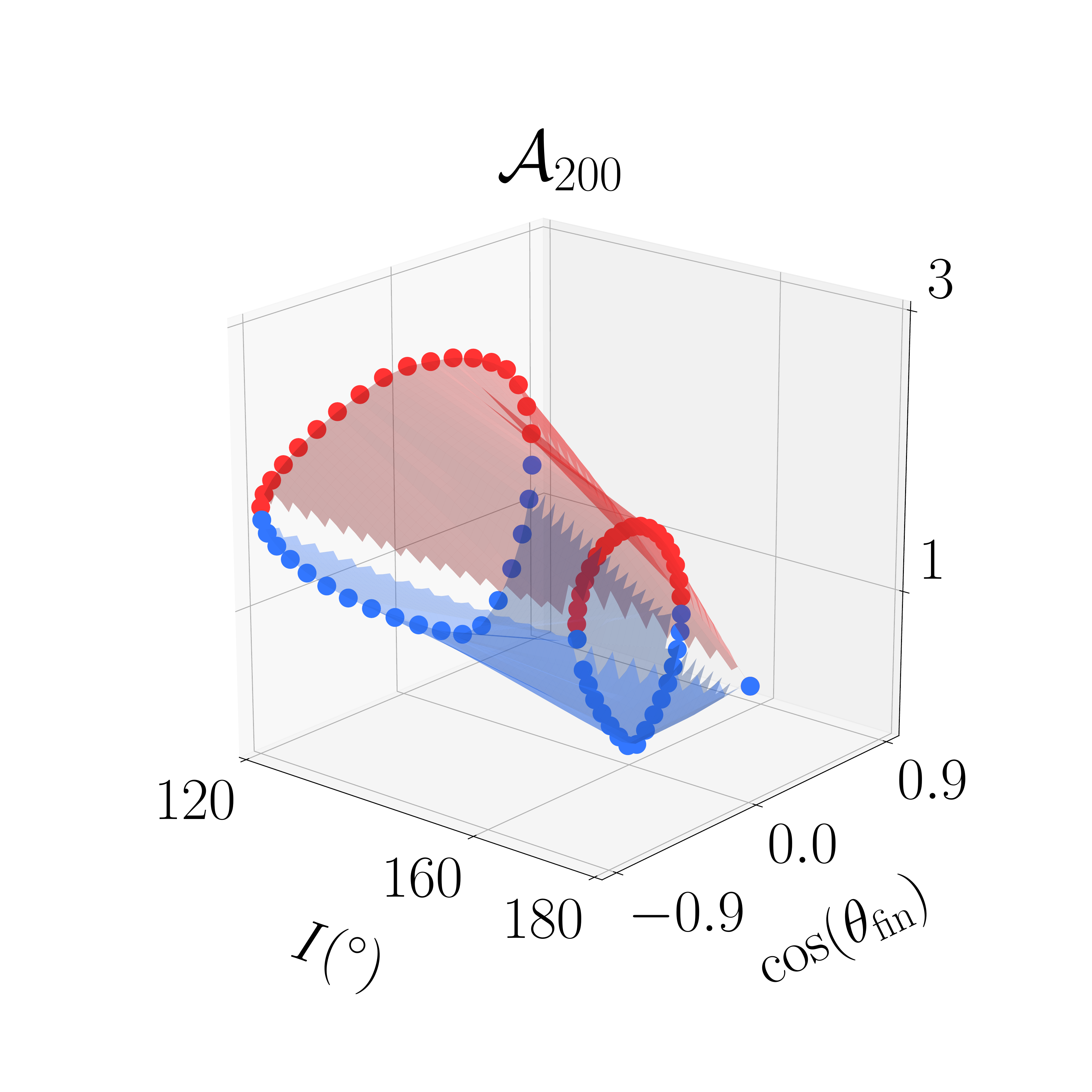}

    \caption{Mode excitation as a function of inclination $I$ and final polar angle $\theta_{\rm fin}$.  The top panels show only the inclination range $I \le 60^\circ$, so all orientations are prograde. The bottom panels show only the inclination range $120^\circ \leq I \leq 180^\circ$, so all orientations are retrograde. Each surface shows mode excitation for $a = 0.5M$; top left panel shows ${\cal A}_{220}$, top middle shows ${\cal A}_{210}$, and top right shows ${\cal A}_{200}$.  Red (blue) dots and surfaces are for surfaces that end with $\dot{\theta}_{\rm fin} > 0$ ($\dot{\theta}_{\rm fin} < 0$).  Notice that the spectral character evolves significantly with increasing inclination: some modes [e.g., $(k,m) = (2,0)$] are essentially absent at zero or small inclination, but are important at large $I$; other modes [e.g., $(k,m) = (2,\pm 1)$] show significant change in the dependence on $\cos\theta_{\rm fin}$.}
\label{fig:3d}
\end{figure*}

Several modes for the high inclination cases we examine ($I = 60^\circ$ and $I = 120^\circ$) appear to be affected by numerical noise.  We believe this is because in these cases the plunging body rapidly crosses multiple angular grid zones; the same effect led to the relatively large errors we find for the $I = 60^\circ$ Schwarzschild rotation test described in Sec.\ {\ref{sec:checks}}.  The shallow inclination cases ($I = 20^\circ$ and $160^\circ$) are substantially less affected by noise.  In Appendix\ {\ref{app:resolution_comparison}}, we estimate this numerical error by comparing results with a set of higher resolution waveforms --- calculated with a higher resolution integration of Teukolsky's equation.  The least reliable mode extractions appear to be ${\cal A}_{k20}, {\cal A}'_{k-20}$ for $I = 60^\circ$ and $I=120^\circ$.  Code enhancements to improve this behavior are under study right now. 

The key result we wish to illustrate is that each spheroidal fundamental mode $(k,m,0)$ is excited in a way that depends uniquely and predictably on the parameters $[a,I,\theta_{\rm fin}, \sgn(\dot\theta_{\rm fin})]$ characterizing its final plunge.  Figure {\ref{fig:3d}} shows another view of this, illustrating how mode excitation varies as a function of the angles $(I,\theta_{\rm fin})$ at $a = 0.5M$.  This figure illustrates how the spectral content of ringdown modes varies, in some cases significantly, as a function of orbit inclination: some modes, such as $(k,m) = (2,0)$ are absent or weak at small inclination, but are very strong for $I$ large; others, such as $(k,m) = (2,1)$ are present at all inclinations, but show large changes in how they depend on $\theta_{\rm fin}$ as $I$ increases.

At least in principle, the clean mapping between source geometry and mode excitation suggests that the inverse problem may be feasible: inferring the properties of the source geometry given knowledge of the excitation of multiple QNMs.  We discuss this further in our conclusions.

\subsection{Universal mode excitation for shallow inclination?}
\label{sec:universal}

In assembling this catalog, we have found intriguing trends in how certain modes are excited.  A particularly interesting one occurs at low inclination: for $I = 20^\circ$, we find that modes with the same $k - m$ are excited in largely the same manner, showing a nearly universal functional dependence on $\theta_{\rm fin}$; the same behavior is seen for $I = 160^\circ$ for modes with the same $k + m$.  This behavior is only weakly dependent on spin in the range that we have investigated.

\begin{figure*}[ht]
  \centering
  \includegraphics[width=.9\linewidth]{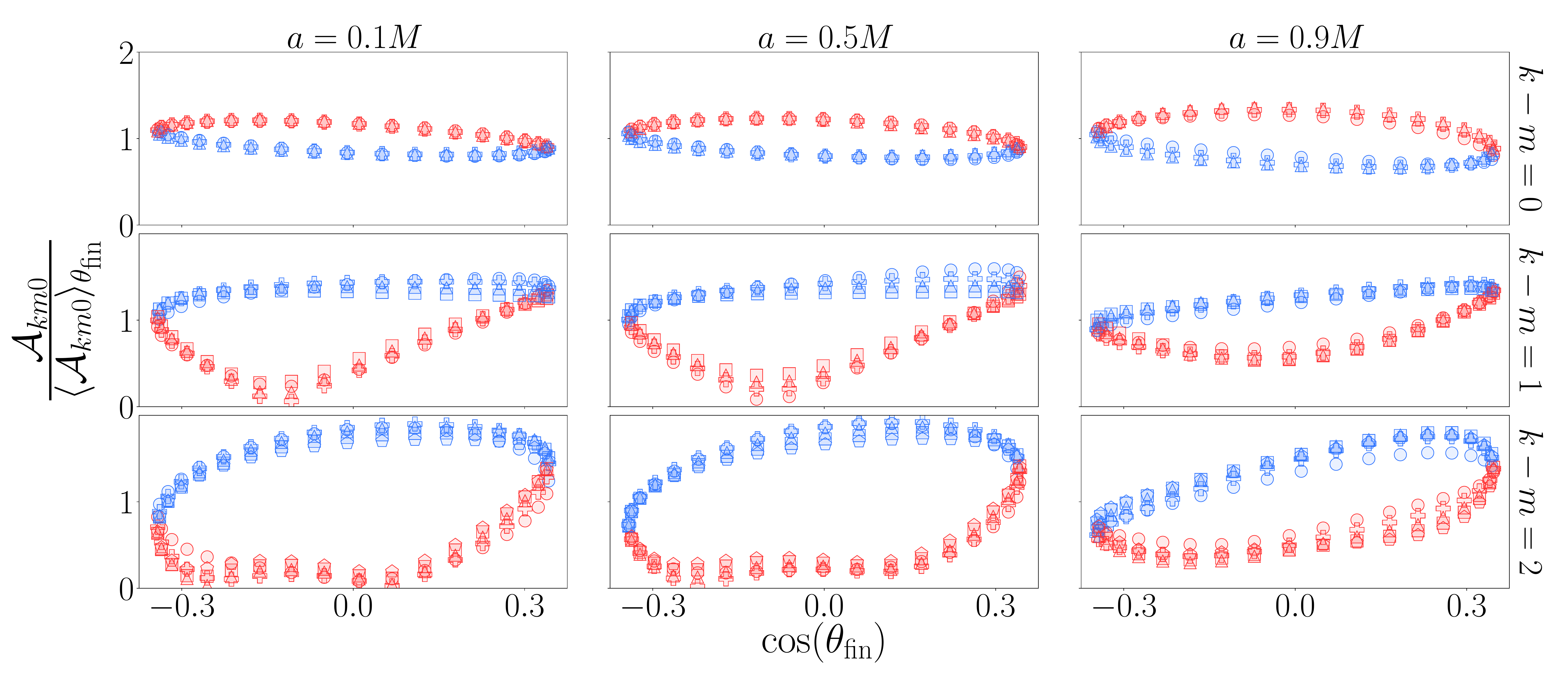}
  \includegraphics[width=.9\linewidth]{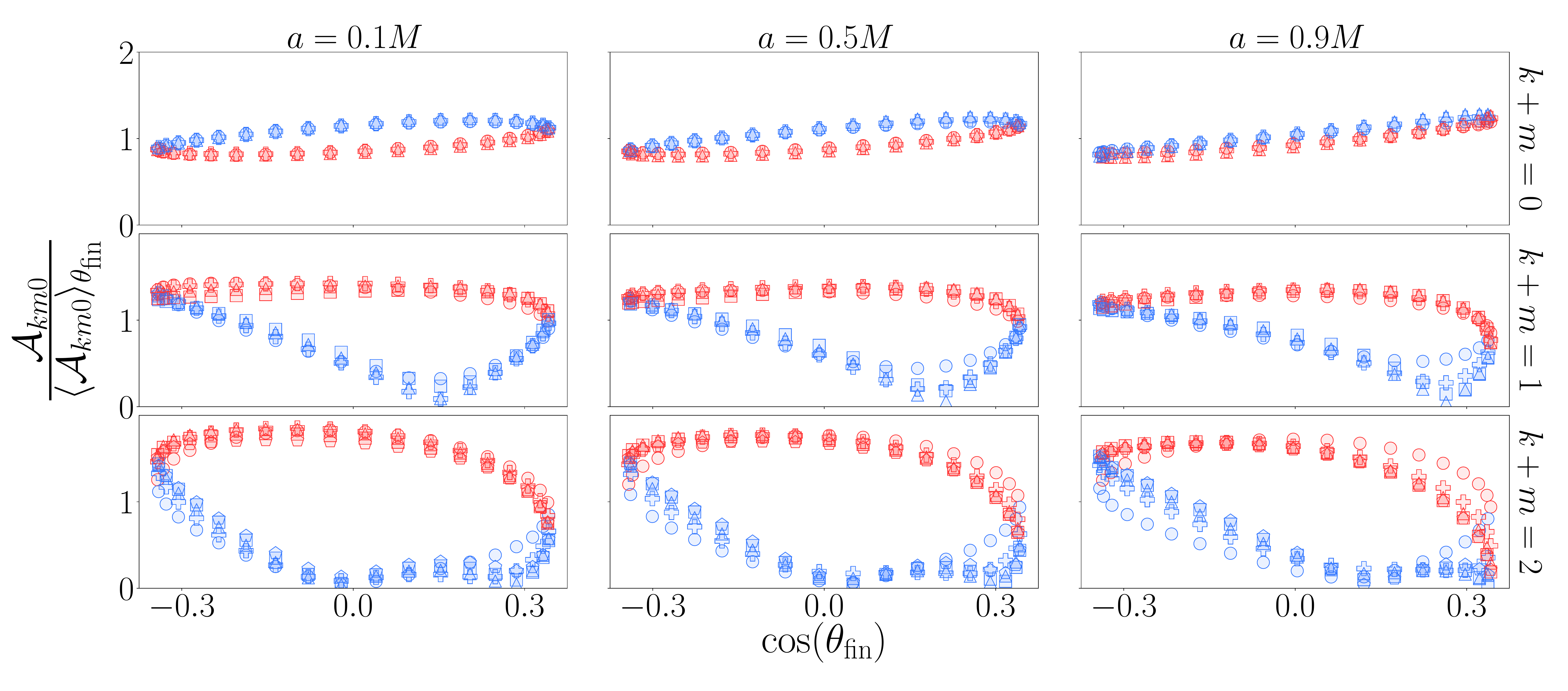}
  \caption{Nearly universal mode excitation as a function of $\cos\theta_{\rm fin}$ at low inclinations $I = 20^\circ$ (top panel) and $I=160^\circ$ (bottom bottom panel).  Each panel shows ${\cal A}_{km0}$ versus $\cos\theta_{\rm fin}$, normalized via Eq.\ (\ref{eq:thfinaverage}).  From left to right, columns range over spin from $a = 0.1M$ at the left to $a = 0.9M$ at the right; from top to bottom, rows range over $k - m$ ($I=20^\circ$) or $k+m$ ($I=160^\circ$) from 0 at the top to 2 at the bottom.  In each panel, modes of order $k = 2$, $3$, $4$, $5$, and $6$ are denoted by circles, crosses, triangles, squares, and pentagons, respectively.  For instance, the case $\mathcal{A}_{530}/\langle\mathcal{A}_{530}\rangle_{\theta_{\rm fin}}$ at $a = 0.5M$ is plotted in the top panel with pentagons in the third row, second column.}  
\label{fig:low_inc20_multiple}
\end{figure*}

This is illustrated in Fig.\ \ref{fig:low_inc20_multiple}.  Here, we overlay normalized mode amplitudes for several different values of $k$, grouping mode amplitudes with the same black hole spin and the same $k-m$ or $k+m$ values.  Amplitudes with given $(a,I,k,m)$ are normalized by averaging over $\cos\theta_{\rm fin}$:
\begin{eqnarray}
& &\left\langle \mathcal{A}_{kmn}\right\rangle_{\theta_{\rm fin}} \equiv
\nonumber\\  
& &\left(\frac{1}{\cos\theta_{\rm max} - \cos\theta_{\rm min}}\right) \int_{\cos\theta_{\rm min}}^{\cos\theta_{\rm max}} \mathcal{A}_{kmn}(\cos\theta_{\rm fin})\, d(\cos\theta_{\rm fin})\;.
\nonumber\\
\label{eq:thfinaverage}
\end{eqnarray}
The range $\theta_{\rm max/min}$ is given by Eq.\ (\ref{eq:polarrange}); for $I = 20^\circ$, $\theta_{\rm max} = 110^\circ$ and $\theta_{\rm min} = 70^\circ$.  The integral in Eq.\ (\ref{eq:thfinaverage}) is numerically evaluated by interpolating between mode amplitudes calculated at various $\theta_{\rm fin}$.

As Fig.\ {\ref{fig:low_inc20_multiple}} shows, for each value of spin and each value of $k - m$ or $k+m$, a nearly universal functional dependence emerges: In each panel, the normalized amplitude traces out a figure that is nearly the same at that spin for all spheroidal mode indices that have a given value of $k - m$ (for $I = 20^\circ$) or $k + m$ for ($I = 160^\circ$).  It is noteworthy that the dependence on spin is quite weak: the results at $a = 0.1M$ and $a = 0.5M$ are nearly identical, and are not dramatically different from the results at $a = 0.9M$.

These trends break down at higher inclinations $I = 60^\circ, 120^\circ$; presumably there is some maximum angle at which this nearly universal excitation form holds (although it should be noted that the highly inclined results are significantly more polluted by numerical noise).  As we discuss in the Conclusions, further work may show that it will be possible to exploit this behavior to better understand mode excitation for misaligned plunges.

\subsection{Overtones}
\label{sec:overtone_results}
\begin{figure*}[ht]
    \centering
    \includegraphics[width=.41\textwidth]{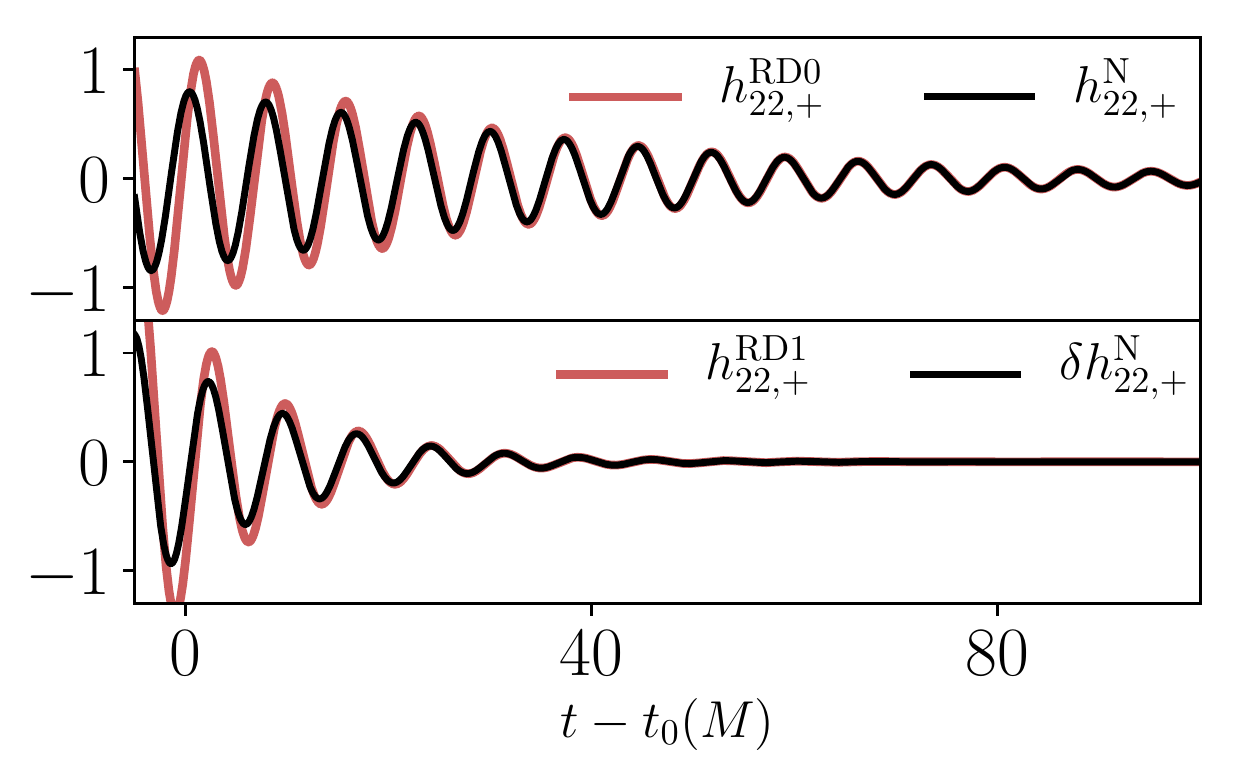}
    \includegraphics[width=.45\textwidth]{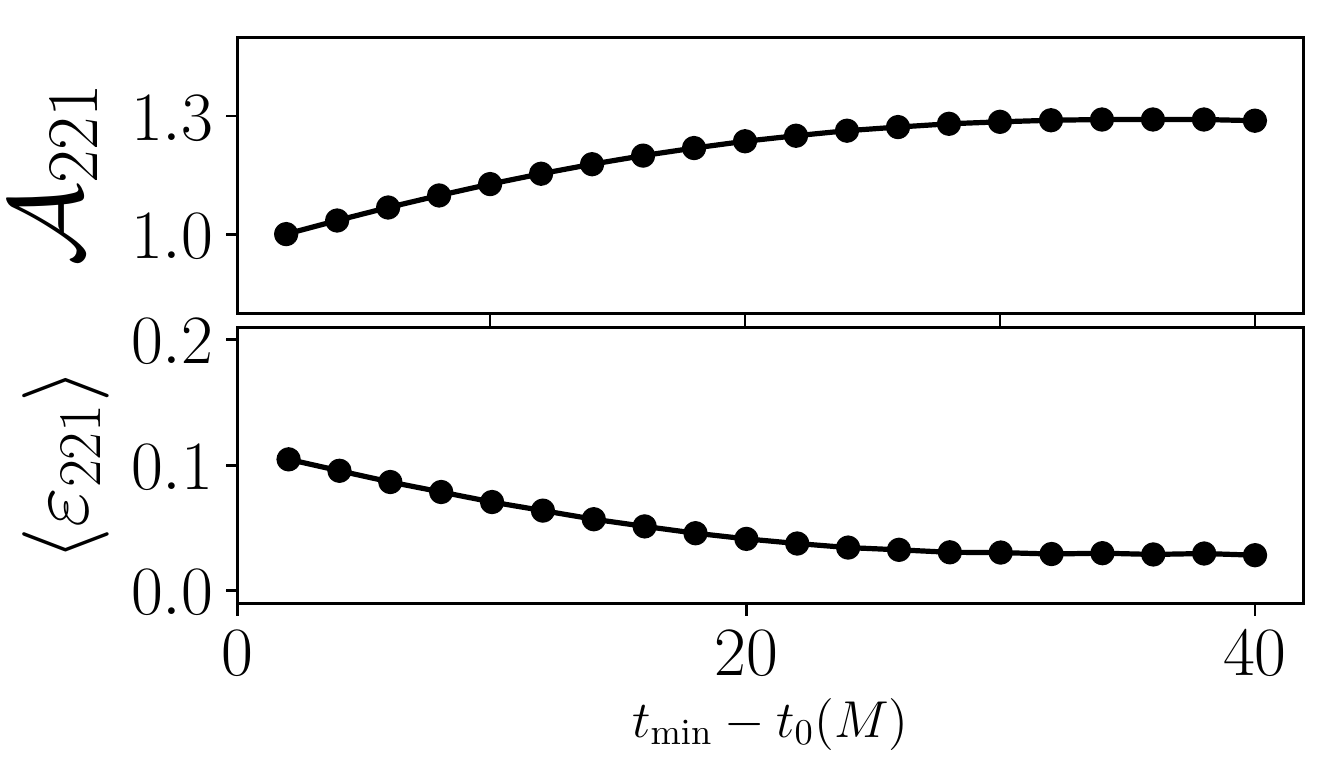}

    \caption{A high spin ($a=0.99M$) test case for extracting overtones. The top left panel shows the fundemantal model $h^{{\rm RD}0}_{22}$ used to describe the late ringdown. The bottom left panel shows the residual of the fundamental model, $\delta h^{N}_{22}$. We find that the residual can be fit with a single frequency $\sigma_{221}$ and amplitude $\mathcal{A}_{221} = 1.28$.  The right panel shows the relative fit error in modeling the early ringdown for various choices of $t_{\rm min}$.  When $t_{\rm min} \gtrsim 30 M$, the relative fit error $\langle \varepsilon_{221} \rangle$ and extracted overtone amplitude $\mathcal{A}_{221}$ stabilize to constant values of $\langle \varepsilon_{221} \rangle = 0.03$ and $\mathcal{A}_{221} = 1.28$.}
    \label{fig:overtone_error}
\end{figure*}
Another intriguing feature which we noticed in assembling our catalog of mode excitation is, possibly, the presence of overtone modes (i.e., modes with $n \ge 1$) excited by the final plunge and merger.  Here we briefly investigate this idea, and present evidence that we may be finding the first overtone in some cases.

Since the overtones are short-lived, amplitudes calculated using the algorithm described in Sec.\ {\ref{sec:modeextract}} do not stabilize to constant values as in the example in Fig.~{\ref{fig:modeextraction}}.  Therefore we conduct a Levenberg-Marquardt nonlinear least squares fit \cite{lmleastsquares} to model the residuals $\delta h^{\rm N}_{22}$ with overtones.  We expect this fit to work best for rapidly rotating black holes as excited overtones will decay more slowly.  To test this, we examined an equatorial $I=0^\circ$ inspiral and plunge for $a = 0.99 M$ and modeled the early ringdown waves with a single complex frequency $\sigma_{221}$ and amplitude $\mathcal{C}_{221}$.  While $t_0$ is still defined as the time at which the small body crosses the light ring, we still need to choose a time window over which to fit the model. To choose this the window we examine the following notion of relative fit error:
\begin{equation}
    \langle \varepsilon_{221} \rangle = \frac{ | \sum_{i} h^{{\rm RD}1}_{22}(t_i) - \delta\tensor*[]{h}{_{22}^{\rm N}}(t_i) |}{|\sum_{i} \delta\tensor*[]{h}{_{22}^{\rm N}}(t_i) |}\;,
    \label{eq:relativefit}
\end{equation}
where $t_i$ is sampled from $t_{\rm min} < t_i < t_0 + 70M$.  A similar equation defines the relative fit error $\langle\varepsilon_{331}\rangle$.

 The right panel of Fig. \ref{fig:overtone_error} shows the relative fit error $\langle\varepsilon_{221}\rangle$ for the $I=0^\circ$, $a = 0.99M$ test case.  Not surprisingly, the error is large for $t_{\rm min} - t_0$ near zero, since at these times the radiation is not yet fully described by QNMs.  The error decreases to a minimum when $t_{\rm min}-t_0 \gtrsim 30 M$; concomitant to this, the fitted amplitude ${\cal A}_{221}$ settles down to a constant.  The left panel of Fig. \ref{fig:overtone_error} shows the resulting waveform fits, where the top left panel shows the spherical $(2,2)$ mode waveform, along with the fundamental ringdown model constructed with the algorithm discussed in Sec.\ {\ref{sec:modeextract}}.  The bottom left panel of this figure shows the overtone model.

Encouraged by this test case, we applied this method to the cases in our ringdown catalog.  For $I=20^\circ$, we construct the overtone model $h^{{\rm RD}1}_{22}$ with two frequencies $\sigma_{221}, \sigma_{321}$ and amplitudes $\mathcal{C}_{221}, \mathcal{C}_{321}$.  The most promising results are for $n = 1$ overtones at low inclination and high spin, $(a,I) = (0.9M, 20^\circ)$.  In this case, we find that the residuals $\delta h_{22}^{\rm N}$ are well fit by a superposition of one or two overtones.  Interestingly, the $n = 1$ amplitudes we find are similar in form to the $n=0$ mode excitation.  In particular, the modes ${\cal A}_{221}$ and ${\cal A}_{331}$ depend on $\cos\theta_{\rm fin}$ in a manner that is reminiscent of the universal form we discussed in Sec.\ {\ref{sec:universal}}.

\begin{figure}[b]
    \centering
    \includegraphics[width=.46\textwidth]{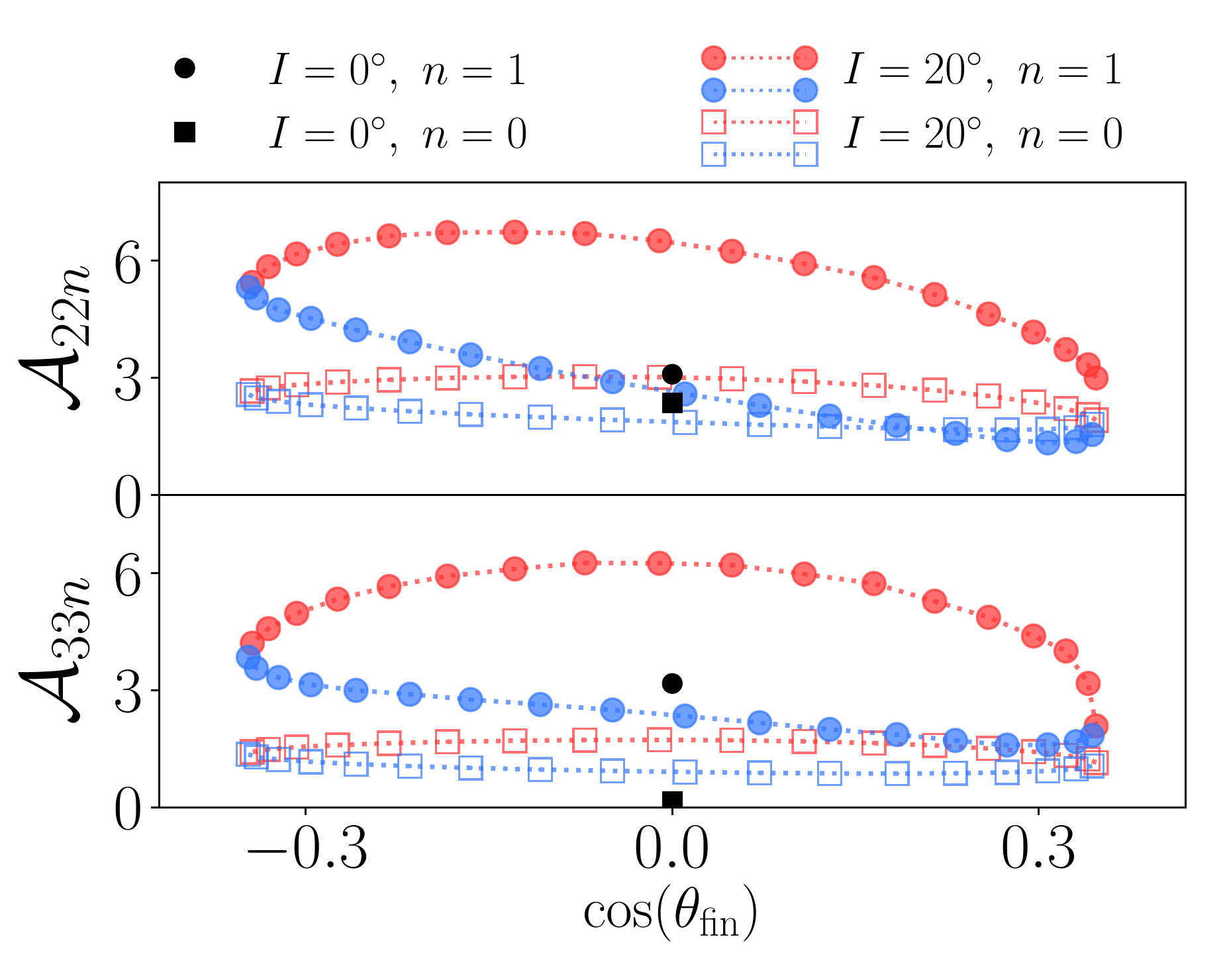}
    \caption{Mode excitation for inferred $n=1$ overtones when $a = 0.9M$ as functions of $\cos\theta_{\rm fin}$.  As with other figures, red (blue) points show data with $\dot\theta_{\rm fin} > 0$ ($\dot\theta_{\rm fin} < 0$).}
    \label{fig:overtone_amp}
\end{figure}

This iterative procedure of calculating overtone amplitudes relies on the assumption that the initial (fundamentals-only) model $\tensor*[]{h}{_{\ell m}^{\rm RD0}}$ is not biased by contaminating overtones. As a check on this assumption, in Appendix\ {\ref{app:overtone_test}}, we calculate the systematic error incurred when a single overtone is present using toy waveform data. The small systematic errors in the recovered $n=0$ and $n=1$ amplitudes provide further evidence of present overtones plotted in Fig.\ {\ref{fig:overtone_amp}}.

\section{Conclusion}
\label{sec:conclude}

We have calculated the black hole QNM excitation resulting from a plunging small body whose trajectory is calculated using a generalized Ori-Thorne algorithm.  Our method to extract the $n=0$ mode amplitudes does not involve choosing a fitting region, but instead determines when the ringdown model is self-consistent.  We find that the mode amplitudes are cleanly parameterized in terms of the black hole spin $a$, the orbital inclination $I$, the small body's final polar angle $\theta_{\rm fin}$ and its final angular direction $\sgn(\dot{\theta}_{\rm fin})$.  We have tabulated the results of our analysis and will provide a {\sc Mathematica} notebook that can be used to plot mode amplitudes as a function of different combinations of $\boldsymbol{(}a,I,\theta_{\rm fin}, \sgn(\dot{\theta}_{\rm fin})\boldsymbol{)}$ \cite{mathematica}.  Importantly, we find that this parameterization removes the influence of the final plunge on {\it ad hoc} characteristics of the inspiral and plunge model; this is discussed in detail in Appendix\ {\ref{app:worldline_robust}}.  As long as we use the parameter set $\boldsymbol{(}a,I,\theta_{\rm fin}, \sgn(\dot{\theta}_{\rm fin})\boldsymbol{)}$ to characterize the data, our conclusions about mode excitation appear to be robust.
\begin{figure}
    \centering
    \includegraphics[width=.45\textwidth]{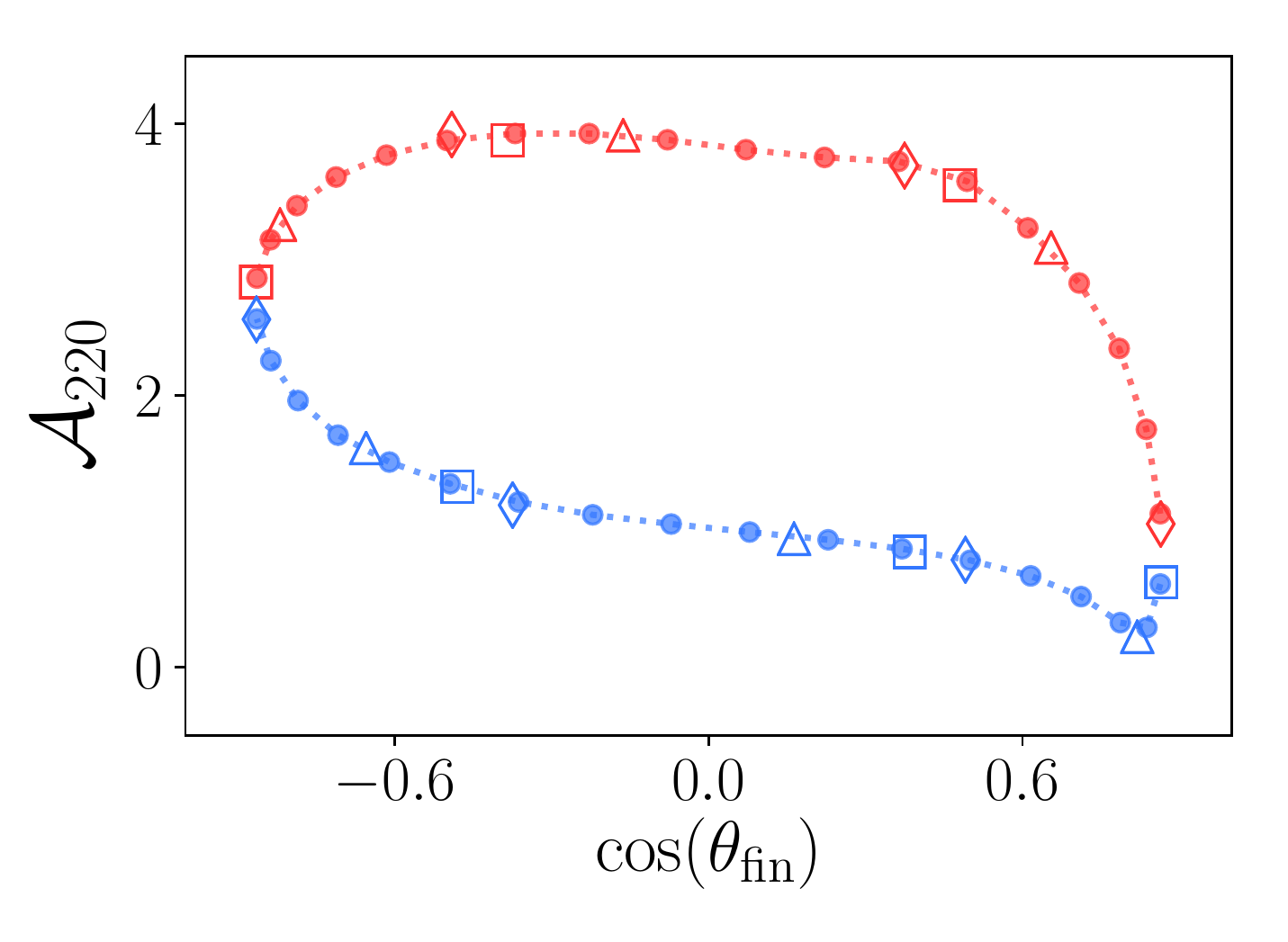}
    \caption{Effect of the choice of $L_f$ on the mode excitation. We calculate the mode amplitudes for a series of trajectories for $a = 0.5M$ and $I=60^\circ$ that share identical prescriptions for generating the inspiral and transition, but differ in transition end times $L_f$. Filled in circles plot the $(k,m)=(2,2)$ fundamental mode amplitude as excited from trajectories with $L_f=2.5$; hollowed squares, triangles, and diamonds plot amplitudes with $L_f = 2.5, 2.59, 2.83,$ and $3.07$, respectively. Red points indicate $\dot{\theta}_{\rm fin} > 0$ and blue points indicate $\dot{\theta}_{\rm fin} < 0$. Even though the small body freezes onto the horizon at different polar positions $\theta_{\rm fin}$, the functional dependence of the mode amplitude on $\theta_{\rm fin}$ is the same.}
    \label{fig:TEPB}
\end{figure}
At least in the large mass ratio limit, our results indicate that there is a clean map between ringdown mode excitation and the properties of the merging binary at plunge.  This supports our motivating idea that, by measuring multiple ringdown modes, it may be possible to learn about a binary's characteristics. By measuring a set of mode amplitudes, a ringdown measurement may provide useful information about the orbit inclination $I$, similar to how Refs.\ \cite{kamaretsos1,kamaretsos2} demonstrated that ringdown preserves the memory of a binary's mass ratio and aligned spin. For large-mass systems (which radiate relatively few inspiral cycles in band), and especially as the high- and mid-frequency sensitivity of ground-based detectors is improved, this could significantly increase what gravitational-wave observations can learn about spin-orbit alignment, a property that is particularly important for constraining the formation history of binary black holes. We refer to the reader to Ref.\ {\cite{mandelfarmer2018}} for overview discussion, {\cite{ligoastro2016}} and references therein for discussion of these observables and source astrophysics, and {\cite{gerosaetal2018}} for recent discussion.

Much work must be done to see whether these measurements can be done in practice.  First, we must determine to what extent inferences based on black hole perturbation theory can be trusted.  In the large-mass-ratio limit, the background spacetime has a well-defined spin which remains constant during the entire coalescence.  When the mass ratio is not so large, the spin of the merged remnant is dominated by the binary's orbital angular momentum at plunge, and information about the members' spins will be washed away.  In addition, studies must be done to determine how accurately detectors can measure the amplitude of ringdown modes.  Although there has been much work studying how well the modes' frequencies and damping times can be measured, less attention has been paid to date on the modes' amplitudes (though see Ref.\ {\cite{carullo2018}}).  We hope this work will motivate more research in this vein.

Our work has also uncovered interesting properties of mode excitation for misaligned coalescence.  An intriguing behavior we discussed in Sec.\ {\ref{sec:universal}} is that, at least for shallow inclination angle, the excitation of modes has a nearly universal form: the dependence of the fundamental mode excitation ${\cal A}_{km0}$ on $\cos\theta_{\rm fin}$ and $\sgn(\dot\theta_{\rm fin})$ is nearly identical across $k$ and $m$ for fixed $k - m$ (for prograde orbits) or $k+m$ (for retrograde orbits), for a wide range of spin $a$.  It may be possible to exploit this relation to make a ``quick and dirty'' assessment of mode excitation across a wide range of mode values and black hole spins.  If this universality includes the Schwarzschild limit, one might even be able to use Eq.\ (\ref{eq:moderotation}) in concert with a catalog of Schwarzschild {\it equatorial} mode excitations to rapidly estimate the QNM strength for a wide range of physically relevant parameters.

Finally, we see indications that it may be possible to extract information about overtone modes, at least when the black hole spin is large and the plunge inclination is shallow.  This is a somewhat delicate operation, so we are cautious about claims in this regime.  However, if it can be shown that there is a regime in which the overtones can be reliably understood, this may make possible additional consistency tests and ways of testing the nature of the Kerr metric using ringdown gravitational waves.

\section*{Acknowledgments}

Our work on this problem was supported at Massachusetts Institute of Technology (MIT) by National Science Foundation Grants No. PHY-1403261 and No. PHY-1707549; H.\ L.\ was in addition supported by an MIT Dean of Science Graduate Fellowship.  G.\ K.\ acknowledges research support from NSF Grant No. PHY-1701284 and Office of Naval Research/Defense University Research Instrumentation Program Grant No.\ N00014181255.  We thank Lionel London for helpful discussions and feedback on this work, Emanuele Berti for helpful discussions and for providing tools for computing ringdown mode properties, Michael Boyle for providing tools which enabled our rotation tests, and Gregory Cook for providing data (based on Ref.\ {\cite{cz14}}) that allowed us to validate our spherical-spheroidal mode mixing coefficients.

\appendix

\section{Robustness of mode excitation to the plunge worldline prescription}
\label{app:worldline_robust}

As we strongly emphasized in Paper I and re-emphasized in Sec.\ {\ref{sec:worldline}}, the generalized Ori-Thorne algorithm we use to construct the worldline followed by our inspiraling and plunging small body requires us to make three {\it ad hoc} choices.  Two of these (a time parameter $L_i$ which defines when we end ``inspiral'' and begin ``transition,'' plus the model we use to smooth the behavior of orbital constants during the transition) turn out to have very small or even negligible impact on the worldline.  The impact on the worldline of the third choice, the time $L_f$ at which ``transition'' ends and begin ``plunge'' begins, is not negligible.  Worldlines which start with the same initial conditions but use different choices for $L_f$ will end with different values of the final polar angle variable $\theta_{\rm fin}$.

Figure {\ref{fig:TEPB}} illustrates that the impact of this {\it ad hoc} choice on QNM excitation is nugatory.  Although changing the parameter $L_f$ for fixed initial conditions changes $\theta_{\rm fin}$, it does {\it not} affect the manner in which QNM excitation depends on $L_f$.  In Fig.\ {\ref{fig:TEPB}}, we show mode excitation for $(a,I) = (0.5M, 60^\circ)$ as a function of $\cos\theta_{\rm fin}$ for the $(k,m,n) = (2,2,0)$ spheroidal mode.  Data represented dots, squares, triangles, and diamonds were computed using $L_f = 2.5$, $2.59$, $2.83$, and $3.07$, respectively. The mode excitation shows no dependence on $L_f$, only on the value $\theta_{\rm fin}$ at which the plunge terminates.

We find the same effect for all other modes that we have examined.  Changing $L_f$ changes the relation between a worldline's initial conditions and $\theta_{\rm fin}$, but does not affect the fundamental manner in which mode excitation depends on a plunge's final geometry.  Although, as we have emphasized elsewhere, it would extremely useful to eliminate the need for the {\it ad hoc} parameters introduced by the generalized Ori-Thorne model, we nonetheless can make robust assessments of how mode excitation behaves as a function of parameters which characterize the final geometry of a coalescing binary.

\section{Resolution effects on mode amplitudes}
\label{app:resolution_comparison}
As discussed in Secs.\ {\ref{sec:modesymmetry} and \ref{sec:catalog}}, numerical errors dominate the accuracy of our mode extraction at high orbital inclination. Resolution effects are apparent when, for example, plotting the mode amplitudes and phases as a function of the plunge angle, $\cos(\theta_{\rm fin})$. In Figs. {\ref{fig:a0.5_I60_l3}} ($I=60^\circ$) and {\ref{fig:a0.5_I120_l3}} ($I=120^\circ$), there are small jumps in the $m=2$ mode amplitudes. In contrast, the mode amplitudes for low orbital inclination (Figs. {\ref{fig:a0.5_I20_l3}} and {\ref{fig:a0.5_I160_l3}}) do not exhibit such jumps.

\begin{figure*}
    \centering
    \includegraphics[width=.45\textwidth]{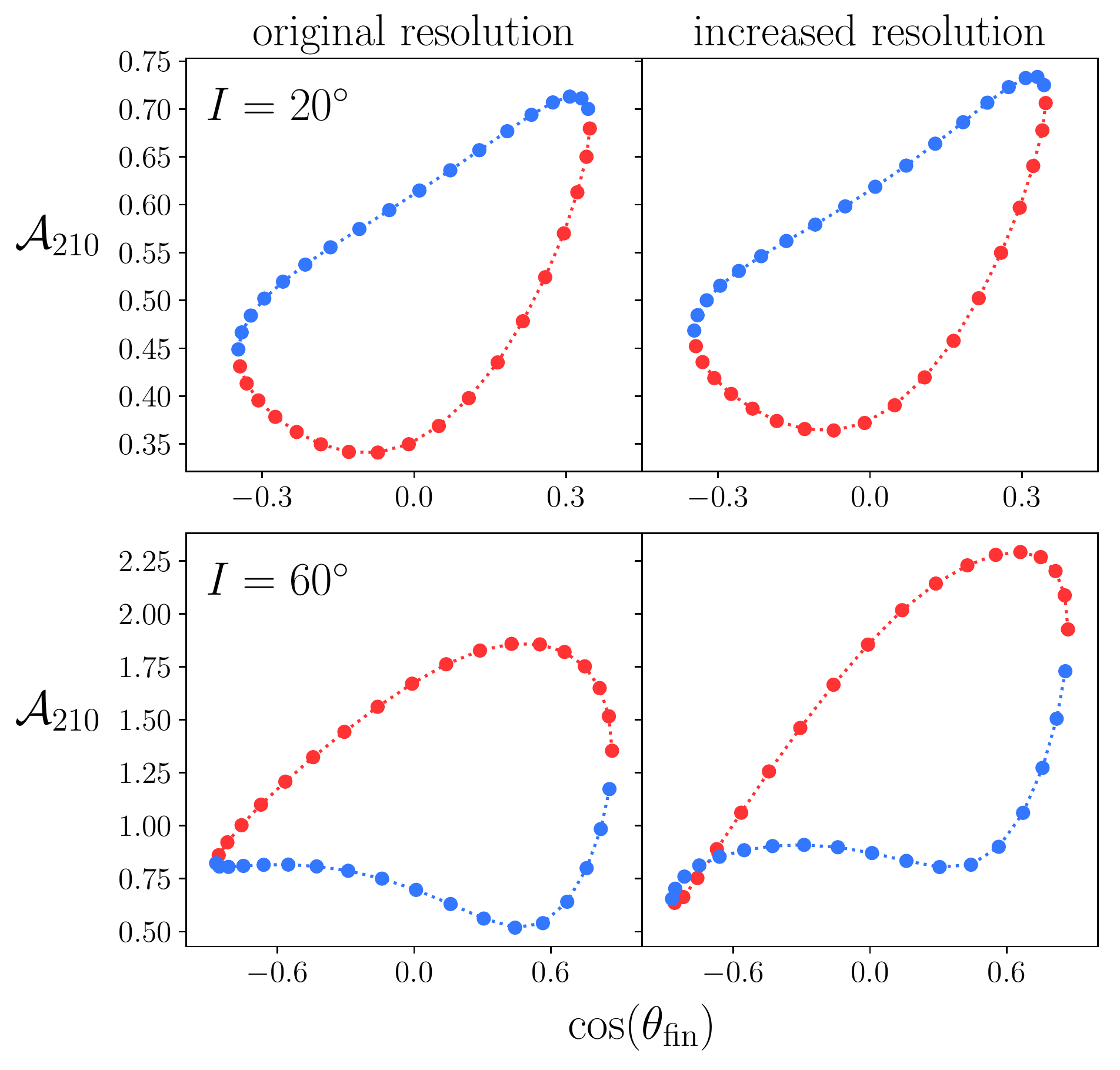}
    \includegraphics[width=.45\textwidth]{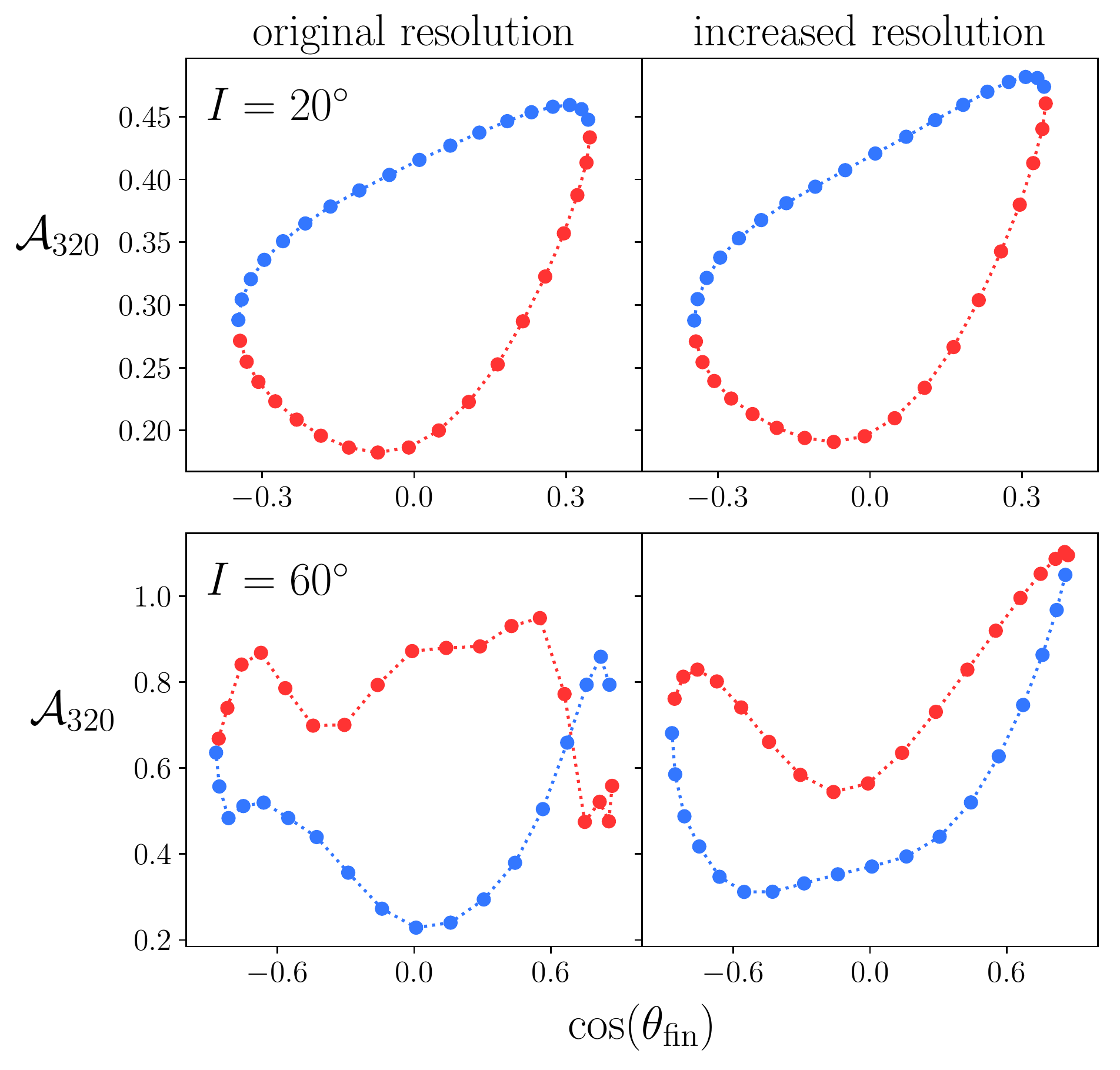}

    \caption{Effect of resolution on calculated mode amplitudes. The top (bottom) panels show the calculated mode amplitudes $\mathcal{A}_{210}$, $\mathcal{A}_{320}$ for low (high) inclination orbits. At low inclination ($I=20^\circ$) increasing the resolution of the waveforms has little effect on the calculated mode amplitudes. At high inclination ($I=60^\circ$) numerical errors dominate, leading to significant changes with increased resolution. }
\label{fig:resolution_test}
\end{figure*}

To further quantify our numerical error, we calculated the mode amplitudes from a set of higher resolution waveforms (Fig. {\ref{fig:resolution_test}}). With increased resolution, the low inclination results do not change much, indicating near convergence. However, the high inclination results are affected, indicating larger numerical errors.

\section{Estimating systematic errors from a toy model with overtones}
\label{app:overtone_test}
 At high spin, $a = 0.9M, 0.99M$, when overtones may be resolved, we employ an iterative process to extract the overtone amplitudes from the numerical data. Initially, we construct a ringdown model with only fundamental modes,  $\tensor*[]{h}{_{\ell m}^{\rm RD0}}$, as described in Sec. {\ref{sec:modeextract}}. Next, we model any remaining contributions to the ringdown with overtones (see Sec. {\ref{sec:OvertoneMethod}}). However, this iterative process relies on the assumption that the overtones will not bias the initial fundamentals-only model. If this assumption does not hold, then $\tensor*[]{h}{_{\ell m}^{\rm RD0}}$ may be biased, which can lead to systematic errors in extracting the overtone amplitudes.

\begin{figure}
    \centering
    \includegraphics[width=.5\textwidth]{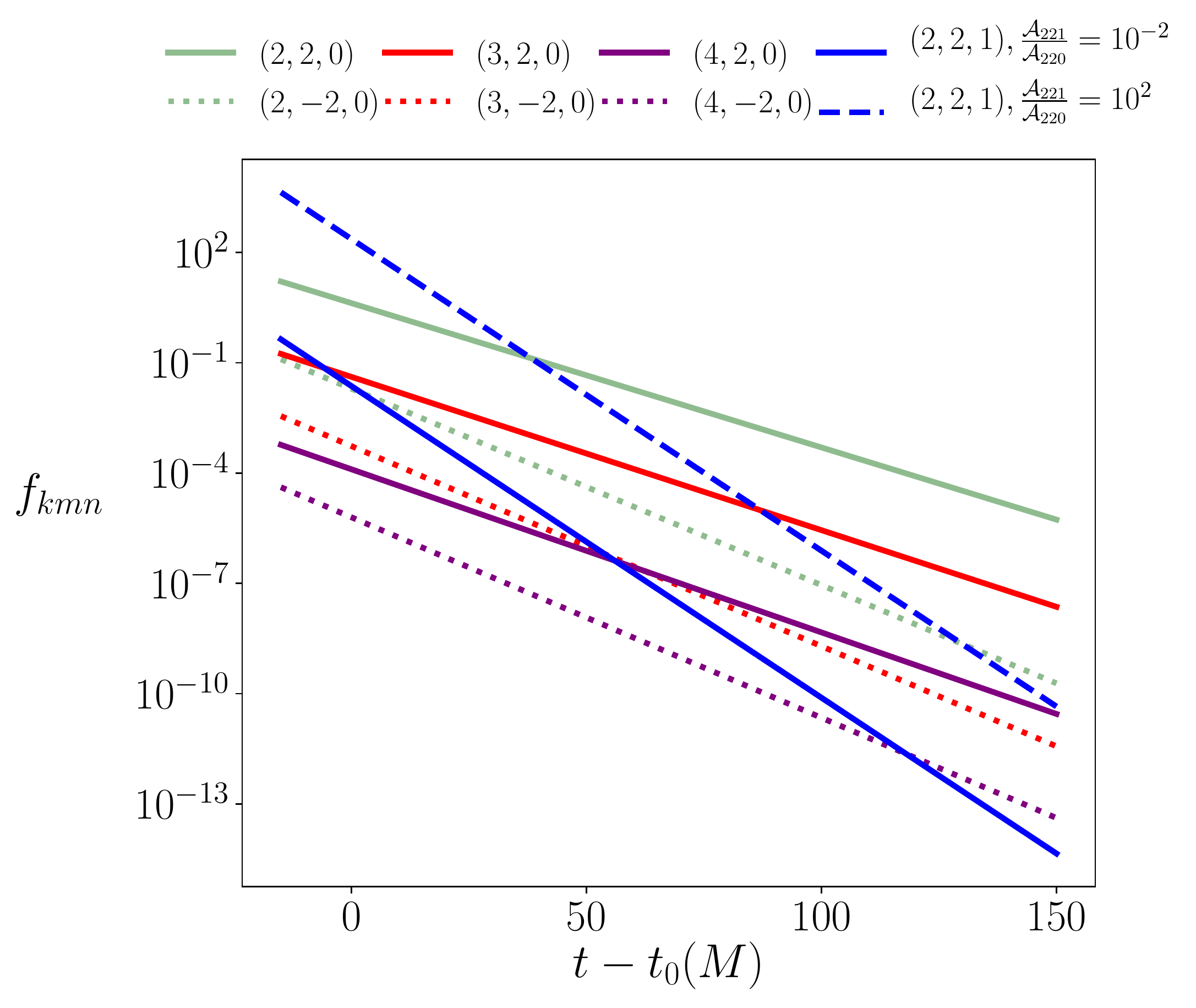}
    \caption{ Magnitude of injected QNMs present in the toy ringdown waveform. Each QNM --- labeled by the spheroidal indices $(k,m,n)$ --- contributes to the ringdown $\tensor*[]{h}{_{\ell 2}^{\rm N}}(t)$ with magnitude $f_{kmn} = |a_{m2 kn}(t)\ \mathcal{C}_{kmn}|$ for $m\geq0$ and $f_{kmn} = |a'_{-m2 kn}(t)\ \mathcal{C}'_{k-mn}|$ for $m\leq0$ (Eq. {\ref{eq:multipole2}}). We constructed 11 toy waveforms with the same injected $n=0$ mode amplitudes (show in green, red, and violet) but with varying $(2,2,1)$ mode amplitudes in the range $10^{-2} \leq  \frac{\mathcal{A}_{221}}{\mathcal{A}_{220}} \leq 10^2$, where we fix $\mathcal{A}_{220} = 2.35$. We plot the $\frac{\mathcal{A}_{221}}{\mathcal{A}_{220}} = 10^{-2}$ magnitude in solid blue; the mode extraction algorithm extracts the modes over the range $12.9M \leq t - t_0 \leq 52.9M$, chosen as the interval of least variance at fixed $\Delta t=40M$ (see Fig.~{\ref{fig:modeextraction}}). We also plot the $\frac{\mathcal{A}_{221}}{\mathcal{A}_{220}} = 10^{2}$ magnitude in dashed blue; since the overtone contamination is larger, the modes are extracted at later times during $76.1M \leq t - t_0 \leq 116.1M$.
    }
\label{fig:amplitude_decomp}
\end{figure}

To estimate the magnitude of such systematic errors, we analyzed toy waveform data --- consisting of only QNMs --- with known (injected) mode amplitudes. Since we would like the toy waveform data to resemble our real waveform data, we fix the injected $n=0$ mode amplitudes to some representative values. Here we have adopted the $n=0$ mode amplitudes for $a = 0.9M$, $I=0^\circ$, which were calculated using the fundamentals-only model. Before $n=1$ overtones are added, the leading order spherical mode $\tensor*[]{h}{_{22}^{\rm N}}$ comprises QNM contributions with magnitude $|a_{m2 k0}(t)\ \mathcal{C}_{k20}|$ (for QNMs with $m\geq0$) and $|a'_{-22 k0}(t)\ \mathcal{C}'_{k-20}|$ (for QNMs with $m\leq0$). We then add the $(k,m,n)=(2,2,1)$ overtone, and generate a 11 different waveforms each with various overtone amplitudes in the range $10^{-2}\leq \mathcal{A}_{221} / \mathcal{A}_{220} \leq 10^{2}$, while fixing $\mathcal{A}_{220}$ to $2.35$ (Fig.\ {\ref{fig:amplitude_decomp}}).

\begin{figure}
    \centering
    \includegraphics[width=.45\textwidth]{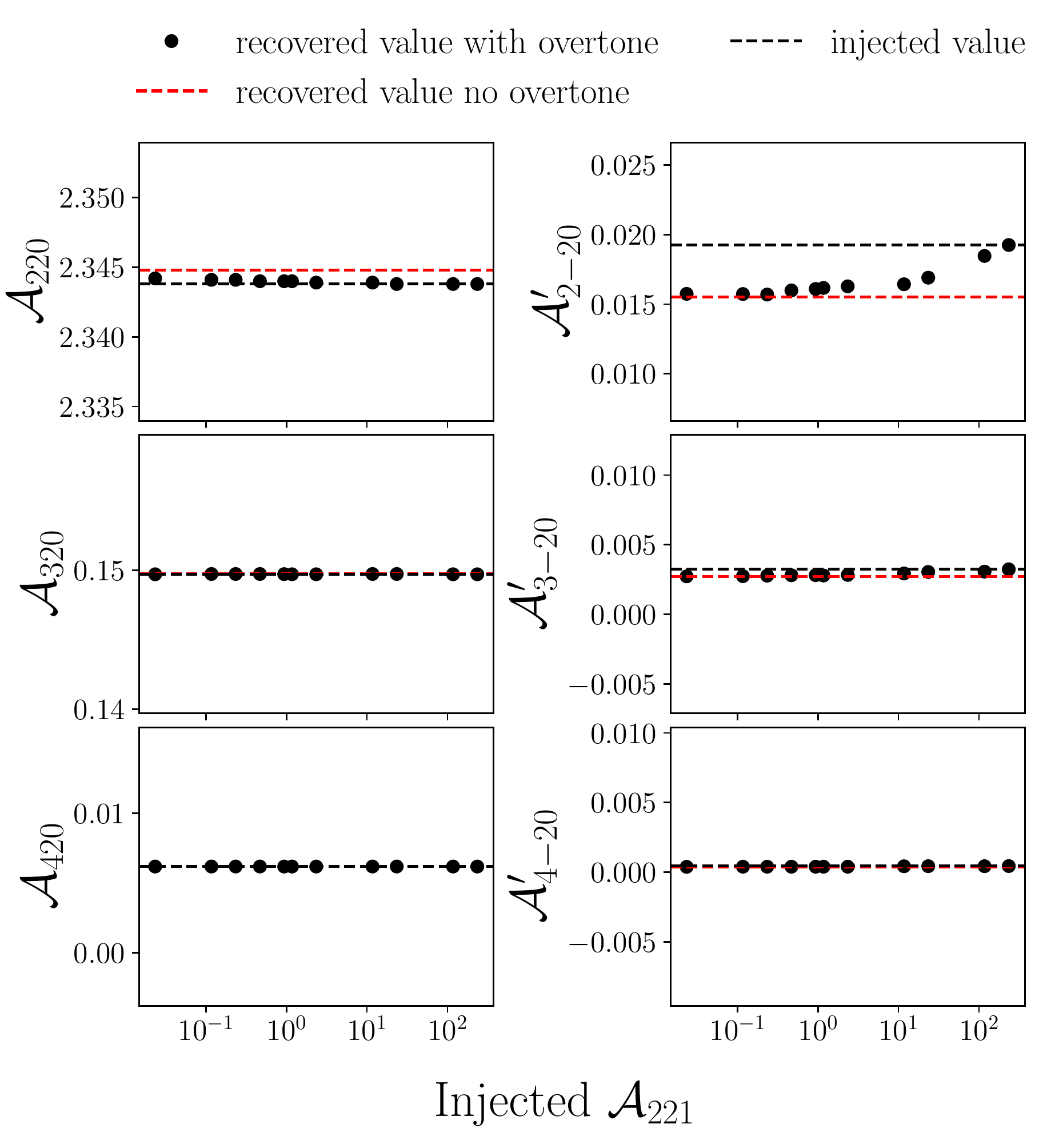}
    \caption{Recovery of $n=0$ mode amplitudes from toy waveform data $\tensor*[]{h}{_{\ell 2}^{\rm N}}$ in the presence of a single overtone $(k,m,n)=(2,2,1)$. Even when the overtone amplitude is large, the recovered mode amplitudes (plotted with black dots) are well within $0.01$ of their injected values (shown with dashed black lines). For comparison, we also performed a mode extraction of purely $n=0$ QNMs, which are shown with dashed red lines --- this shows the error in the limit when $\mathcal{A}_{221}\rightarrow 0$.}
\label{fig:overtone_test_error}
\end{figure}

\begin{figure}
    \centering
    \includegraphics[width=.45\textwidth]{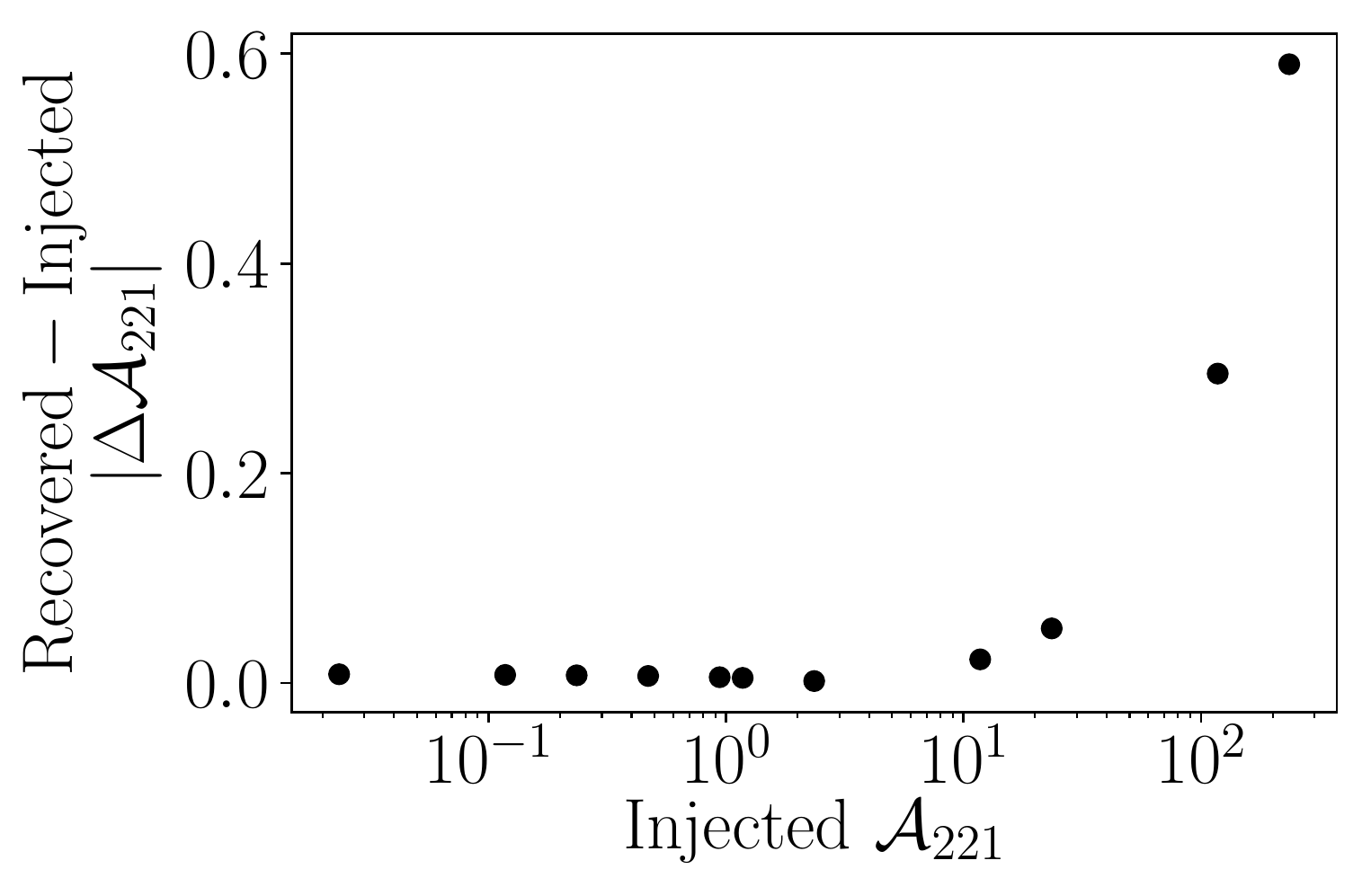}
    \caption{Systematic error in recovering the $n=1$ mode amplitude in toy waveform data using an iterative method. When the injected $(k,m,n)=(2,2,1)$ overtone amplitude is relatively large ($4\times 10^{-1} \leq \frac{\mathcal{A}_{221}}{\mathcal{A}_{220}} \leq 10^{2}$), the relative error is recovered amplitude $\frac{|\Delta\mathcal{A}_{221}|}{\mathcal{A}_{221}}$ does not exceed 1\%. For smaller overtone amplitudes ($10^{-2} \leq \frac{\mathcal{A}_{221}}{\mathcal{A}_{220}} \leq 2\times 10^{-1}$), the relative error is larger than 1\% but the absolute error is still within $|\Delta\mathcal{A}_{221}| < 0.01$. Note that we fix $\mathcal{A}_{220} = 2.35$ in the toy waveform data.  }

\label{fig:overtone_test_bias}
\end{figure}

Our results indicate that the calculated $n=0$ mode amplitudes are not significantly biased, and can be recovered well within $0.01$ even when the overtone amplitude is large (Fig.\ {\ref{fig:overtone_test_error}}). The $n=1$ mode amplitude is also faithfully recovered with the iterative procedure, incurring either a relative error under 1\% (for a loud overtone) or an absolute error under $0.01$ (for a weak overtone) (Fig.\ {\ref{fig:overtone_test_bias}}). This is primarily because the algorithm extracts the amplitudes at times when the fundamentals-only model is most self consistent with the data; as the injected overtone amplitude increases, the mode extraction algorithm compensates by adjusting the fitting interval to later times in the ringdown (see the caption of Fig.\ {\ref{fig:amplitude_decomp}}). Bias is also minimized by the averaging procedure (i.e. associating $\bar{\mathcal{C}}_{km0}$ and $\bar{\mathcal{C'}}_{km0}$ with the mode amplitudes), even when overtone contribution $|a_{22 21}(t)\ \mathcal{C}_{221}|$ is exceeds than that of subdominant higher-order modes.

\begin{figure*}[hp!]
\includegraphics[width = 1\textwidth]{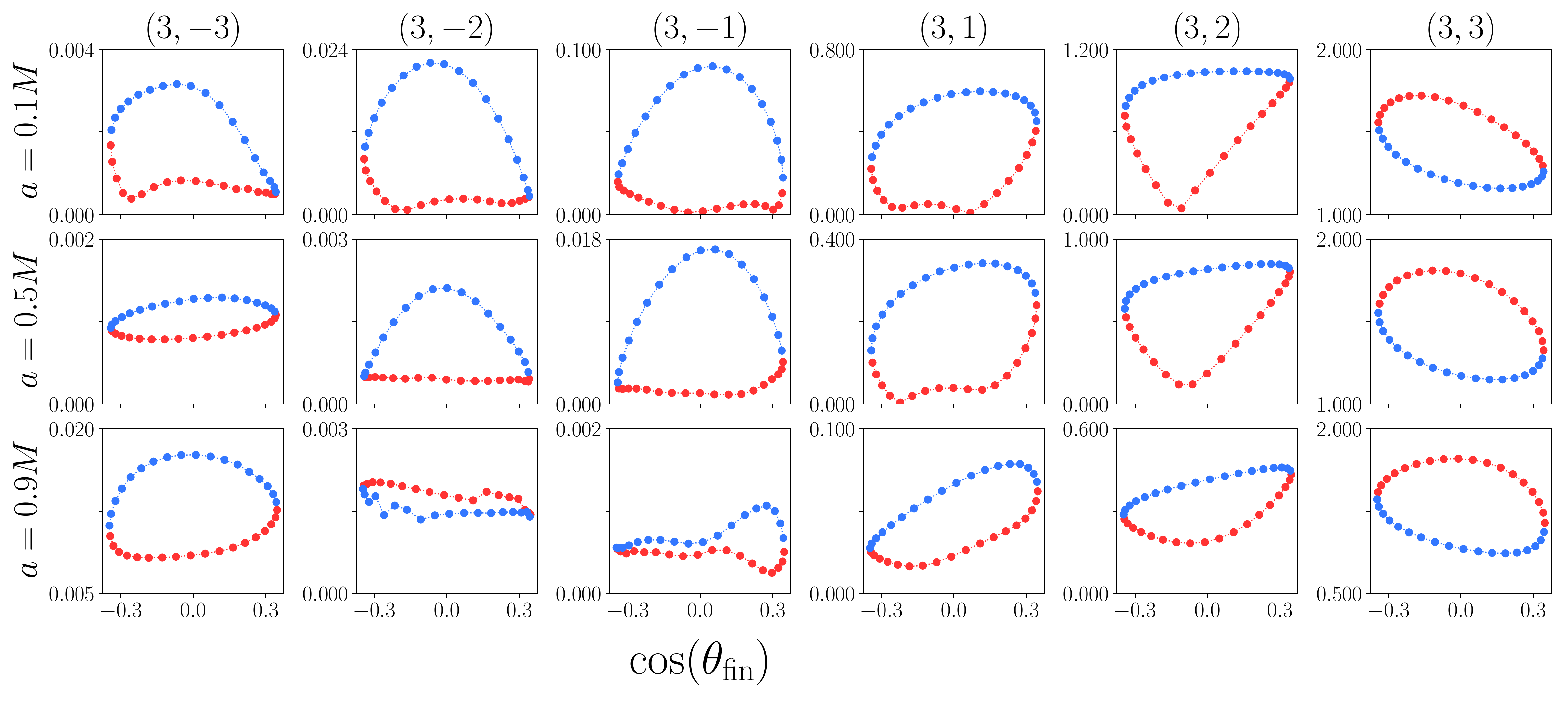}
\caption{Mode excitation magnitude ${\cal A}_{kmn}$ for spheroidal QNMs with $k = 3$, $m \in (-3,\ldots,3)$, $n = 0$ for inspiral and plunge with $I = 20^\circ$.  From top to bottom, the black hole spins vary from $a = 0.1M$ to $a = 0.5M$ to $a = 0.9M$; from left to right, $m$ varies from $3$ to $-3$.  Note that the $m = 0$ mode is not present here, as our numerical data achieved convergence for $m = 0$ using only the $k = 2$ spheroidal modes.  In each plot, red (blue) dots indicate $\dot\theta_{\rm fin} > 0$ ($\dot\theta_{\rm fin} < 0$).}
\label{fig:a0.5_I20_l3}
\end{figure*}

\begin{figure*}[hp!]
\includegraphics[width = 1\textwidth]{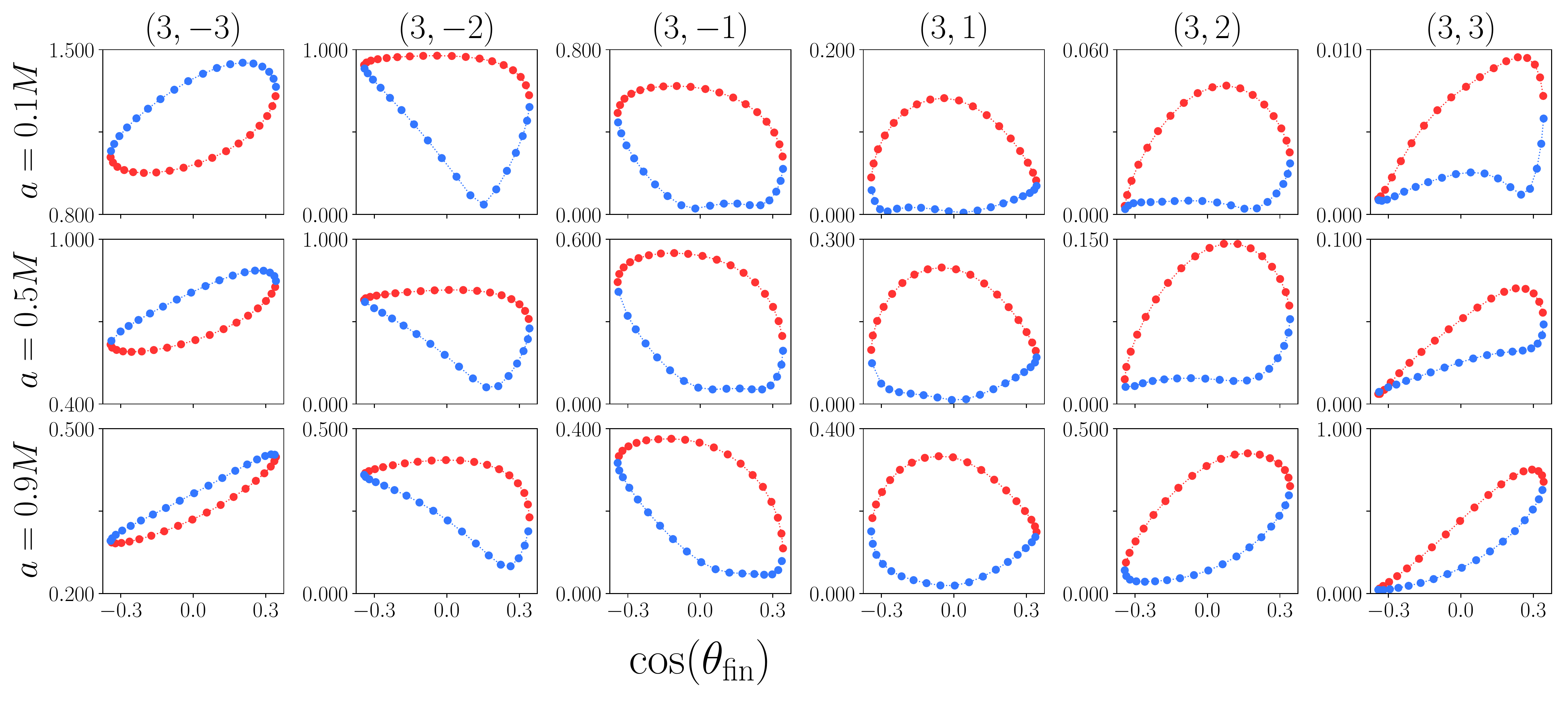}
\caption{Mode excitation magnitude ${\cal A}_{kmn}$ for spheroidal QNMs with $k = 3$, $m \in (-3,\ldots,3)$, $n = 0$ for inspiral and plunge with $I = 160^\circ$.  From top to bottom, the black hole spins vary from $a = 0.1M$ to $a = 0.5M$ to $a = 0.9M$; from left to right, $m$ varies from $3$ to $-3$.  Note that the $m = 0$ mode is not present here, as our numerical data achieved convergence for $m = 0$ using only the $k = 2$ spheroidal modes.  In each plot, red (blue) dots indicate $\dot\theta_{\rm fin} > 0$ ($\dot\theta_{\rm fin} < 0$).}
\label{fig:a0.5_I160_l3}
\end{figure*}

\begin{figure*}[hp!]
\includegraphics[width = 1\textwidth]{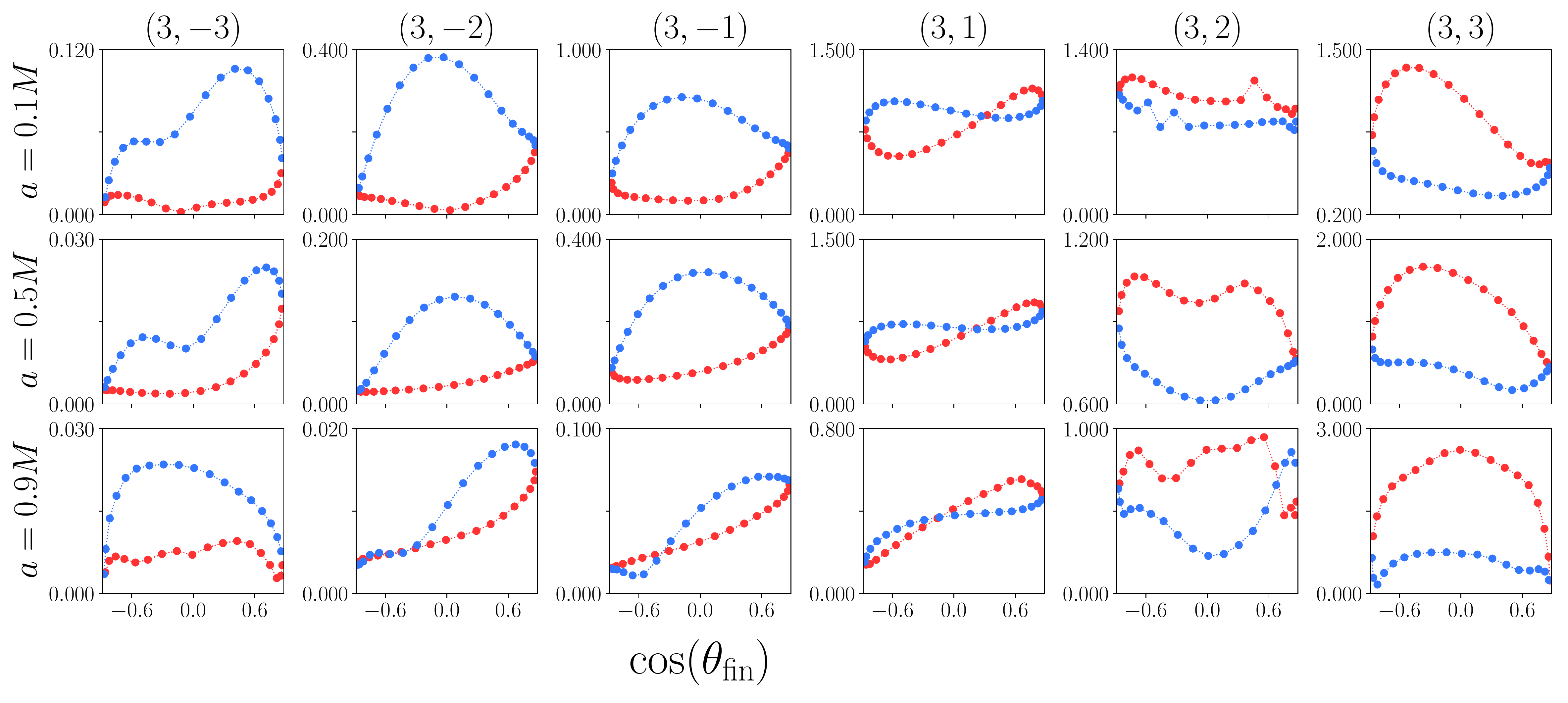}
\caption{Mode excitation magnitude ${\cal A}_{kmn}$ for spheroidal QNMs with $k = 3$, $m \in (-3,\ldots,3)$, $n = 0$ for inspiral and plunge with $I = 60^\circ$.  From top to bottom, the black hole spins vary from $a = 0.1M$ to $a = 0.5M$ to $a = 0.9M$; from left to right, $m$ varies from $3$ to $-3$.  Note that the $m = 0$ mode is not present here, as our numerical data achieved convergence for $m = 0$ using only the $k = 2$ spheroidal modes.  In each plot, red (blue) dots indicate $\dot\theta_{\rm fin} > 0$ ($\dot\theta_{\rm fin} < 0$).  At this high inclination we see more numerical noise as evident in the $m = 2$ amplitudes. }
\label{fig:a0.5_I60_l3}
\end{figure*}

\begin{figure*}[hp!]
\includegraphics[width = 1\textwidth]{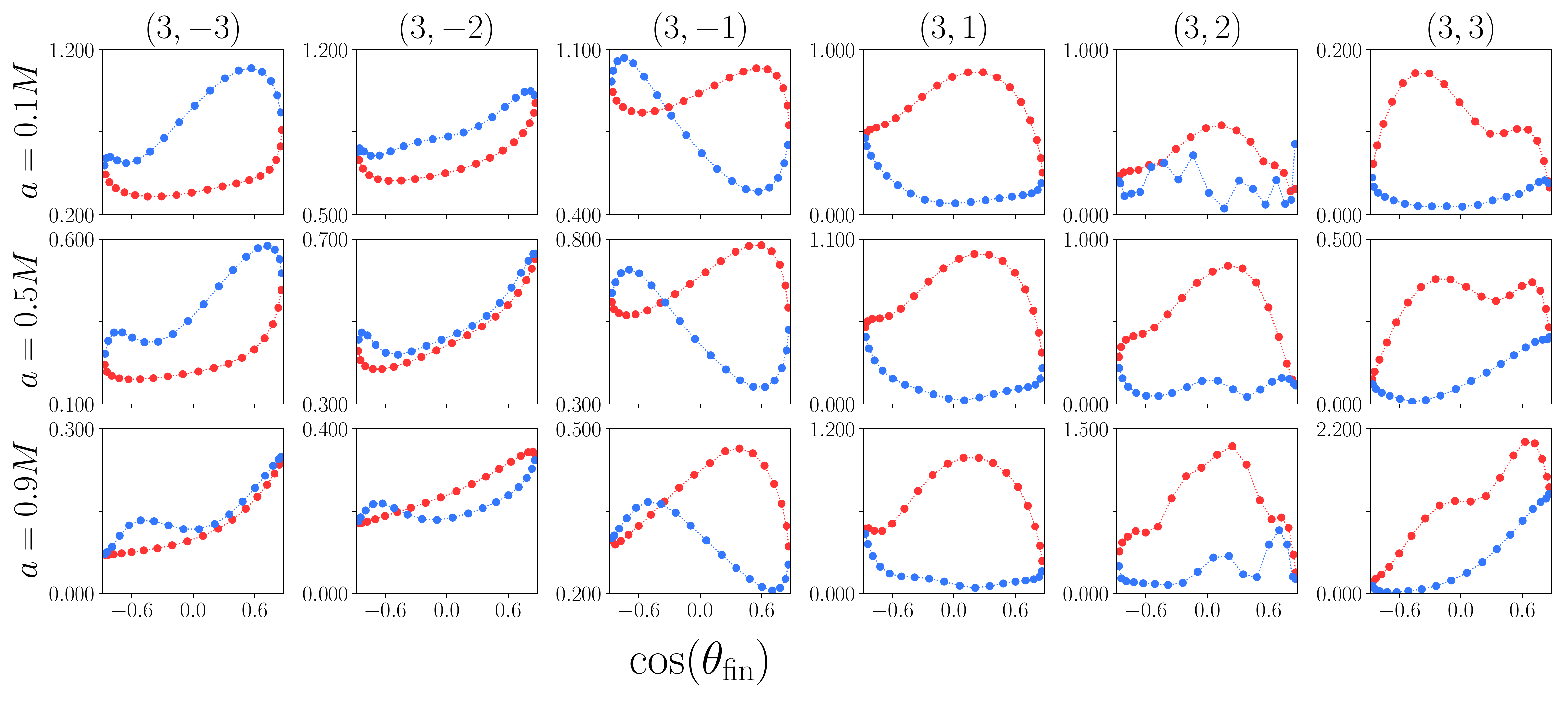}
\caption{Mode excitation magnitude ${\cal A}_{kmn}$ for spheroidal QNMs with $k = 3$, $m \in (-3,\ldots,3)$, $n = 0$ for inspiral and plunge with $I = 120^\circ$.  From top to bottom, the black hole spins vary from $a = 0.1M$ to $a = 0.5M$ to $a = 0.9M$; from left to right, $m$ varies from $3$ to $-3$.  Note that the $m = 0$ mode is not present here, as our numerical data achieved convergence for $m = 0$ using only the $k = 2$ spheroidal modes.  In each plot, red (blue) dots indicate $\dot\theta_{\rm fin} > 0$ ($\dot\theta_{\rm fin} < 0$).  At this high inclination we see more numerical noise as evident in the $m = 2$ amplitudes.}
\label{fig:a0.5_I120_l3}
\end{figure*}

\begin{widetext}

\section{Further mode excitation catalog entries}
\label{app:catalog_more}

Here we present additional catalog entries describing QNM excitation. In Fig.\ {\ref{fig:I060_l2_phase}} we show the mode excitation phase $\phi_{2m0}$ for the spins $a = 0.1M$, $0.5M$, and $0.9M$ for $I=60^\circ$. In Figs.\ {\ref{fig:a0.5_I20_l3}--\ref{fig:a0.5_I120_l3}} and we show the amplitude magnitude ${\cal A}_{km0}$ for $k = 3$.

\begin{figure*}[hp]
\includegraphics[width = .8\textwidth]{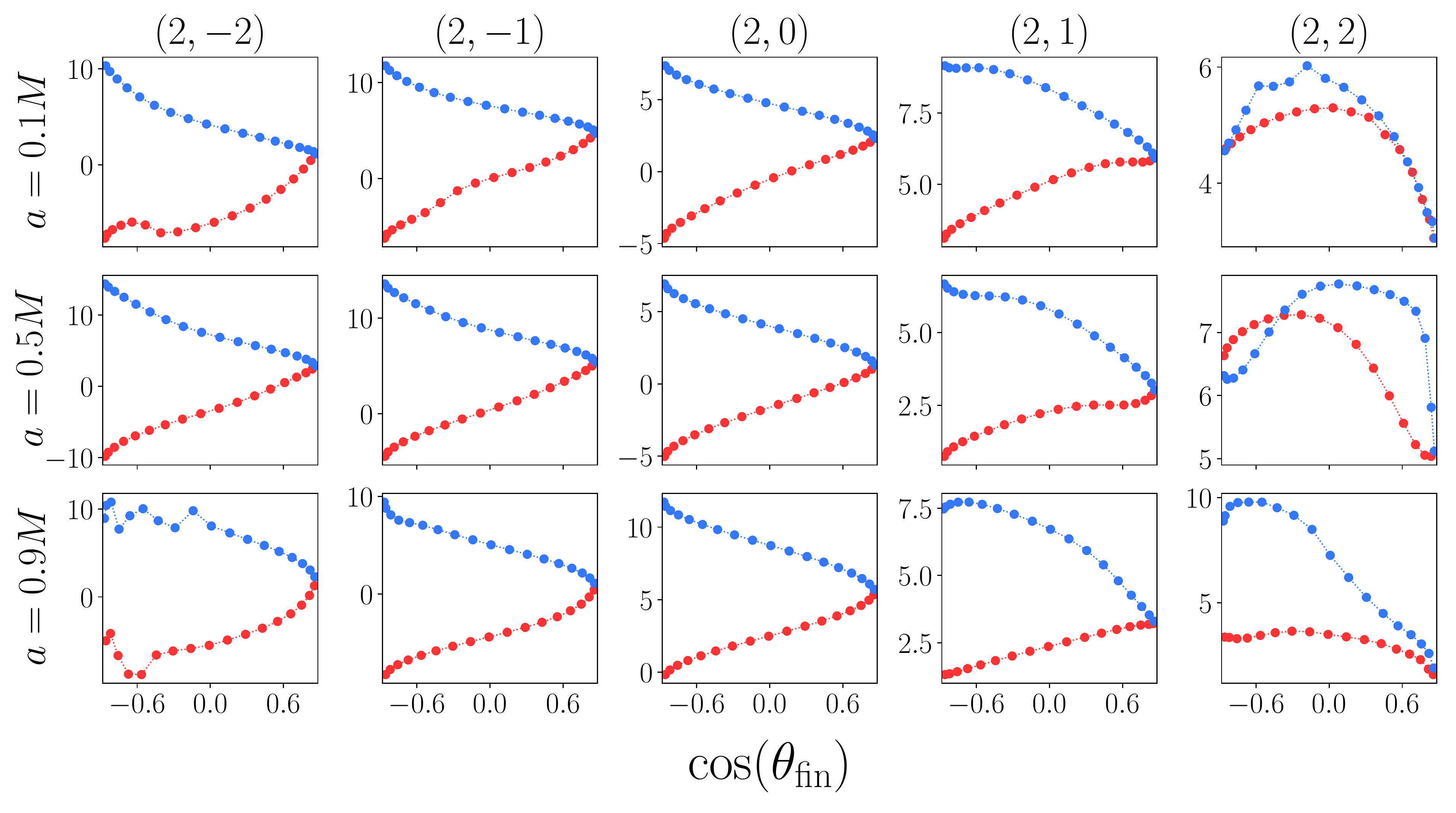}
\caption{Mode excitation phase $\phi_{kmn}$ for spheroidal QNMs with $k = 2$, $m \in (-2,\ldots,2)$, $n = 0$ for inspiral and plunge with $I = 60^\circ$. The corresponding mode excitation magnitudes ${\mathcal A}_{kmn}$ are shown in Fig.\ {\ref{fig:I060_l2}}.  At this high inclination we see more numerical noise.}
\label{fig:I060_l2_phase}
\end{figure*}

\end{widetext}




\bibliographystyle{apsrev4-1}
\bibliography{references}
\end{document}